\documentclass[twocolumn]{article}
\usepackage{multicol}
\usepackage{ragged2e}
\usepackage{graphicx,setspace}
\usepackage{natbib}
\usepackage{subfig}
\usepackage{fancyhdr,amsmath,amssymb, amsthm,mathtools}
\usepackage{etoolbox}

\usepackage[colorlinks,citecolor=blue,urlcolor=blue]{hyperref}
\usepackage{booktabs}
\usepackage{tikz}
\usetikzlibrary{matrix,positioning}
\usetikzlibrary{fit,positioning,calc}
\usepackage{mathrsfs}
\usepackage{amsfonts}
\usepackage{bbm,bm}
\usepackage{scrextend}
\usepackage{hyperref}
\usepackage{chngcntr}
\usepackage{apptools}
\usepackage{ulem}
\usepackage{caption}
\usepackage[T1]{fontenc}
\usepackage{tgtermes}
\usepackage[small]{titlesec}
\usepackage[export]{adjustbox}
\usepackage{multirow}
\RequirePackage{times}
\usepackage{newtxmath}
\usepackage{algorithm}
\usepackage{algpseudocode}
\usepackage{float}
\usepackage{stfloats}

\renewcommand{\P}{\mathbb{P}}
\newcommand{\Pcr}{\mathscr{P}}
\newcommand{\X}{\mathbb{X}}
\newcommand{\Xcr}{\mathscr{X}}
\newcommand{\N}{\mathbb{N}}

\newcommand{\E}{\mathbb{E}}
\newcommand{\R}{\mathbb{R}}

\newcommand{\bH}{ {\bf H} }
\newcommand{\bell}{\bm{\ell}}
\newcommand{\bt}{ {\bf t} }
\newcommand{\bn}{ {\bf n} }

\newcommand{\bZ}{ {\bf Z} }
\newcommand{\bx}{ {\bf x} }

\newcommand{\bS}{ {\bf S} }

\newcommand{\bX}{ {\bf X} }

\newcommand{\bY}{ {\bf Y} }

\newcommand{\bV}{ {\bf V} }

\newcommand{\bW}{ {\bf W} }

\newcommand{\bC}{ {\bf C} }

\newcommand{\bPsi}{ {\boldsymbol \Psi} }

\newcommand{\bz}{ {\bf z} }

\newcommand{\bw}{ {\bf w} }

\newcommand{\rpm}{\tilde{P}}
\newcommand{\e}{\bm{e}}
\newcommand{\q}{\bm{q}}
\newcommand{\indic}{\mathbbm{1}}
\def\simiid{\stackrel{\mbox{\scriptsize{iid}}}{\sim}}
\newcommand{\indep}{\perp\!\!\!\perp}
\newcommand{\cmean}[2]{\E\left[#1\left. \, | \, #2\right.\right]}
\newcommand{\ddr}{\mathrm{d}}

\newcommand{\prob}[1]{\P\left(#1\right)}
\newcommand{\Fcr}{\mathscr{F}}
\newcommand{\edr}{\mathrm{e}}
\newcommand{\cprob}[2]{\P\left(#1\left|#2\right.\right)}

\newtheorem{Proposition}{Proposition}
\newtheorem{Theorem}{Theorem}
\newtheorem{Corollary}{Corollary}
\newtheorem{Remark}{Remark}

\newtheorem{Definition}{Definition}

\newtheorem{prp}{Proposition}[section]
\newtheorem{thm}{Theorem}[section]
\newtheorem{cor}{Corollary}[section]
\newtheorem{defi}{Definition}[section]
\newtheorem{exe}{Example}[section]

\newcommand\marklessfootnote[1]{%
  \addtocounter{footnote}{1}
  \footnotetext{#1}
}

\begin{document}

\title{Partially Exchangeable Stochastic Block Models \\ for (Node-Colored) Multilayer Networks}
	\author{Daniele Durante$^{\mbox{\small a}}$,  Francesco Gaffi$^{\mbox{\small b}*}$, Antonio Lijoi$^{\mbox{\small a}}$ and Igor Pr\"unster$^{\mbox{\small a}}$ \\ {\small $^{\mbox{\small a}}$Bocconi Institute for Data Science and Analytics, Bocconi University, Milan, Italy}, \\ {\small $^{\mbox{\small b}}$Department of Mathematics, University of Maryland, College Park, US}, {\small$^*$corresponding author (fgaffi@umd.edu)}}
	\date{}

\twocolumn[
  {%
    \let\newpage\relax
    \let\clearpage\relax
	
	\maketitle

\vspace{5pt}

 \begin{abstract}
Multilayer networks generalize single-layered connectivity data in several directions. These generalizations include, among others, settings where  multiple types of edges are observed among the same set of nodes (edge-colored networks) or where a single notion of connectivity is measured between nodes belonging to different pre-specified layers (node-colored networks). While progress has been made in statistical modeling of edge-colored networks, principled approaches that flexibly account for both within and across layer block-connectivity structures while incorporating layer information through a rigorous probabilistic construction are still lacking for node-colored multilayer networks. We fill this gap by introducing a novel class of partially exchangeable stochastic block models specified in terms of a hierarchical random partition prior for the allocation of nodes to groups, whose number is learned by the model. This goal is achieved without jeopardizing probabilistic coherence, uncertainty quantification and derivation of closed-form predictive within- and across-layer co-clustering probabilities. Our approach facilitates prior elicitation, the understanding of theoretical properties and the development of yet-unexplored predictive strategies for both the connections and  the allocations of future incoming nodes. Posterior inference proceeds via a tractable collapsed Gibbs sampler, while performance is illustrated in simulations and in a real-world criminal network application. The notable gains achieved over competitors clarify the importance of developing general stochastic block models based on suitable node-exchangeability structures coherent with the type of multilayer network being analyzed.  
\end{abstract}
\vspace{5pt}

\noindent%
  }%

\vspace{27pt}

]

\section{Introduction}\label{sec_1}
\marklessfootnote{{\bf FUNDING:} This research was largely conducted while Francesco Gaffi was a Ph.D. student at Bocconi University, Italy. Daniele Durante is funded by the European Union (ERC, NEMESIS, project num.: 101116718). Views
and opinions expressed are however those of the author(s) only and do not necessarily reflect those of
the European Union or the European Research Council Executive Agency. Neither the European Union
nor the granting authority can be held responsible for them. Antonio Lijoi and Igor Pr\"unster were partially supported by the European Union – NextGenerationEU PRIN-PNRR (project P2022H5WZ9).} Modern network data encode elaborate connectivity information among a set of nodes. This growing complexity has motivated increasing efforts towards extending the wide literature on single-layered networks to the context of multilayer networks. Recalling the comprehensive review by \citet{kivela2014multilayer} (see also Section S1.1 in the supplementary materials), multilayer networks broadly refer to connectivity data that characterize edges among nodes in a multidimensional layering space, where each dimension denotes a feature of the edges, the nodes, or both. Remarkable special cases within this general class are edge-colored networks, where the layers encode connections among the same, or possibly varying, set of nodes w.r.t.\ different types of relationships, and node-colored networks, that encode a single notion of connectivity among nodes belonging to different pre-specified layers \citep[][Sect. 2.1--2.5]{kivela2014multilayer}.

\begin{figure}[t]
\centering
    \includegraphics[trim=0cm 0.8cm 0cm  0.5cm,clip,width=0.332\textwidth]{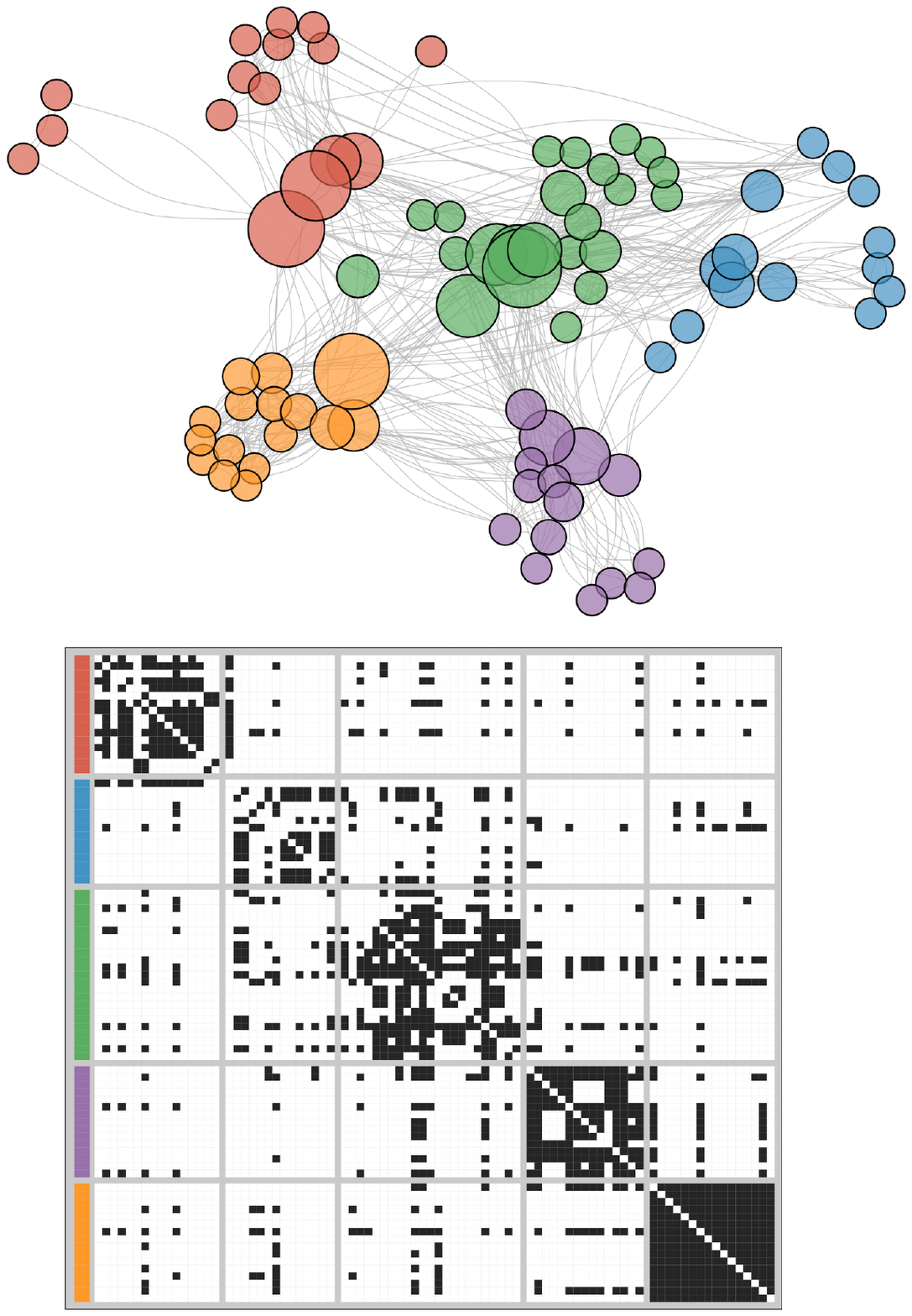}
\captionof{figure}{\footnotesize{Top: Graphical representation of the node-colored \textit{Infinito}  network \citep[][]{calderoni2017communities}.  Each node is a criminal and edges denote co-attendance to at least one of the summits of the Mafia organization. The size of each node is proportional to the corresponding betweenness, whereas the colors indicate layers, which, in this context, define the membership to different \textit{locali}, i.e., structural units administering crime in specific territories. Bottom: \textit{Supra}-adjacency matrix representation of the network. Rows and columns correspond to nodes, while entries measure the presence (black) or absence (white) of an edge for each pair of nodes. Side colors refer to layer (\textit{locali}) division. See also Section~\ref{sec_5}.}}
    \label{figure:1}
\end{figure}

In modeling the above data structures, attention has focused on a primary goal in network analysis, namely the identification of block-connectivity architectures based on the shared patterns of edge formation among the nodes \citep[see, e.g.,][]{fortunato2016community,abbe2017community,lee2019review}. This focus has led to several extensions of classical single-layered community detection algorithms \citep{girvan2002community,blondel2008fast}, spectral clustering methods \citep{von2007tutorial} and stochastic block models (SBMs) \citep{holland1983stochastic,nowicki2001estimation} to identify grouping structures among nodes in multilayer networks. Within this context, key advances have been achieved in edge-colored settings \citep[see, e.g.,][]{mucha2010community,stanley2016clustering,durante2017nonparametric,durante2017bayesian,wilson2017community,paul2020spectral,lei2020consistent,arroyo2021inference,jing2021community,gao2022testing,pensky2021clustering,noroozi2022sparse,amini2019hierarchical}. Albeit relevant, all these extensions do not naturally apply to node-colored networks, which possess a substantially different structure. As showcased within Figure~\ref{figure:1}, these connectivity data can be reconstructed from a single-layered network whenever the nodes are labelled with some ``type'' defining a natural division into subpopulations. Examples include connectivity networks among the brain regions belonging to different lobes \citep{bullmore2009complex}, bill co-sponsorship networks between the lawmakers of different parties \citep{briatte2016network}, social relationships among individuals from various sociodemographic groups \citep{handcock2007model}, and co-attendances to summits of criminals belonging to different territorial units \citep{calderoni2017communities}. To infer grouping structures among the nodes in these ubiquitous networks, it is fundamental to devise rigorous models that: (i) incorporate flexible block-connectivity structures, both within and across layers, via a principled representation; (ii) automatically learn the number of nodes' clusters from the data; (iii) provide formal uncertainty quantification; (iv) yield coherent projections as the size of the network grows; (v) allow for investigation of co-clustering properties; (vi) preserve computational tractability. 

Despite the importance of node-colored networks, a state-of-the-art formulation capable of incorporating all the these \textit{desiderata} is still missing. Leveraging the previously-mentioned direct connection with node-labelled network data, one possibility to partially address such a gap is to treat the layer membership information as a categorical attribute, and then rely on the available attribute-assisted methods for single-layered networks  \citep[see, e.g.,][]{tallberg2004bayesian,xu2012model,yang2013community,sweet2015incorporating,newman2016structure,zhang2016community,binkiewicz2017covariate,stanley2019stochastic,yan2021covariate,mele2022spectral,legramanti2022extended}. Although these solutions provide useful extensions of community detection algorithms, spectral clustering methods and SBMs, none of them jointly addresses all the above  \textit{desiderata}. In fact, these extensions are either based on algorithmic strategies, which fail to properly quantify uncertainty, or consider model-based approaches that, however, do not rely on consistent constructions, in the Kolmogorov sense, for the layer-dependent mechanism of nodes' allocations to the groups. This, in turn, undermines the development of models that preserve projectivity as the network size grows, a necessary condition for principled inference and prediction. As illustrated in Sections~\ref{sec_4}--\ref{sec_5}, achieving these objectives translates into remarkable empirical gains over state-of-the-art attribute-assisted models. 

In order to address the above \textit{desiderata} (i)-(vi), in Section~\ref{sec_2} we develop a novel  and general class of partially exchangeable stochastic block models (pEx-SBM) for node-colored networks. This class relies on a constructive and interpretable probabilistic representation, which crucially combines stochastic equivalence structures \citep[e.g.,][]{nowicki2001estimation} for the edges in the \textit{supra}-adjacency matrix of the node-colored network \citep[see, e.g.,][]{kivela2014multilayer} with a partial exchangeability assumption \citep{deF(38)} that allows to rigorously incorporate layers' division into the mechanism of groups formation. The latter assumption is motivated by the natural parallel between the division of statistical units into subpopulations, typical of a partially exchangeable framework, and the division of the nodes into layers. Such a perspective yields a Bayesian representation that assumes within- and across-layers edges to be conditionally independent realizations from  Bernoulli random variables whose probabilities only depend on the group allocation of the two involved nodes.  The prior distribution for these allocations is characterized by a partially exchangeable partition probability function (pEPPF) \citep{camerlenghi2019distribution}. 

The partially exchangeable regime behind pEPPFs encodes a more general distributional invariance than exchangeability: the observations are assumed to be drawn from different subpopulations, and the corresponding distribution is only invariant to permutations within the same subpopulation. Therefore, similarly to an exchangeable partition probability function (EPPF) \citep{pitman1995exchangeable} that identifies random partitions induced by the ties of an exchangeable sequence, a pEPPF characterizes the random partition induced by the ties of a partially exchangeable array, for which every row stores the observations from a different subpopulation. Crucially, such ties can occur also across  rows, that is, across different subpopulations. A concise account on partially exchangeable arrays is provided in the supplementary materials. The parallel between subpopulations and network layers is, therefore, natural: we suppose a partially exchangeable array of  latent attributes for the nodes within the subpopulations corresponding to layer membership, and then rely on a broad class of discrete nonparametric priors for such an array, namely \textit{hierarchical normalized completely random measures} (H-NRMIs) \citep{camerlenghi2019distribution}. Due to the discreteness of H-NRMIs, ties in the realizations of such auxiliary attribute may occur and nodes with the same attribute value are grouped together according to a probabilistic mechanism determined by a pEPPF. Moreover, the number of groups is random and learned from data. As is clear from this formulation, the auxiliary attribute is not object of inference. Rather, it represents an instrumental quantity that facilitates the constructive definition of general pEPPF priors.

Crucially, the above construction yields a \textit{Kolmogorov consistent} sequence of distributions $(P_V)_{V\geq1}$, with $P_V$ defined on the space of partitions of the $V$ nodes in the network. This means that the distribution of the partition of $V-1$ nodes obtained by deleting uniformly at random one node from a random partition distributed as $P_V$ is given by $P_{V-1}$, for any $V$. These sequences are known as \textit{partition structures} \citep{Kin(78)}, and guarantee theoretically-validated prediction and uncertainty quantification. We refer to the partition structures determined by a pEPPF as partially exchangeable partition (pExP) priors. Moreover, the pExP priors induced by H-NRMIs allow us to analytically investigate yet-unexplored clustering and co-clustering probabilities, both within and across layers. The corresponding expressions establish generalized notions of homophily  that favor the creation of within-layer clusters relative to groups spanning across multiple layers. As shown in Section~\ref{sec_3}, these results also yield closed-form urn schemes which allow for the derivation of a collapsed Gibbs sampler for posterior inference, and novel $k$-step-ahead predictive schemes for the allocations and edges of future incoming nodes. Both are achieved  preserving coherence between the updated and the full model, which is missing in the state-of-the-art alternatives. In fact, principled and accurate $k$-step-ahead predictive schemes as the one we develop within Section~\ref{sec_3.2} are lacking in the literature. Hence, our contribution along these lines represents an additional crucial advancement provided by pEx-SBM. To summarize, our pEx-SBM simultaneously meets the \textit{desiderata} (i)--(vi). More generally, although the focus is on node-colored networks, our results clarify how pairing the specific structure of the multilayer network analyzed with the most suitable notion of node-exchangeability can yield key foundational, methodological and practical gains, thus opening to future research in the broader class of multilayer networks (including edge-colored ones), beyond node-colored settings.

The simulation studies (see Section~\ref{sec_4}) and the application to a multilayer criminal network (see Section~\ref{sec_5}) showcase the practical gains in point estimation (including empirical evidence of frequentist  posterior consistency as $V$ grows), uncertainty quantification and prediction of pEx-SBMs, when compared to state-of-the-art competitors \citep{blondel2008fast,zhang2016community,binkiewicz2017covariate,come2021hierarchical,legramanti2022extended}. Finally, as highlighted in Section~\ref{sec_6}, although the focus is on  binary undirected node-colored networks, suitable adaptations of the novel modeling framework underlying pEx-SBMs facilitate the inclusion of other relevant network settings. Proofs, additional results and further empirical investigations are deferred to the supplementary materials. 

\section{Partially Exchangeable SBMs (pEx-SBM)}\label{sec_2}
Networks are typically represented through an adjacency matrix with rows and columns indexed by the $V$ nodes, and binary entries taking value $1$ if there is an edge between the corresponding pair of nodes and $0$ otherwise.  Similarly, a node-colored network can be represented via a \textit{supra}-adjacency (block) matrix with diagonal and off-diagonal blocks encoding connectivity structures within each layer and  across the pairs of layers, respectively;  see Figure~\ref{figure:1}. We consider here a generic undirected node-colored network with $V$ nodes divided into $d$ layers. Every layer $j$ contains $V_j$ nodes, with $\sum_{j=1}^d V_j=V$. The corresponding $V \times V$ binary and symmetric \textit{supra}-adjacency matrix is denoted by $\bY$. Since nodes are stacked consecutively in $\bY$, there is a one-to-one correspondence between the row-column indexes of the   \textit{supra}-adjacency matrix, namely $v=1, \ldots, V$, and the pairs $(j,i)$, for $j=1,\ldots,d$ and $i=1,\ldots,V_j$, denoting the $i$-th node in the $j$-th layer.  In the following we use $v$ or $(j,i)$ based on convenience.

We now derive the proposed partially exchangeable stochastic block model (pEx-SBM). The binary edges $y_{vu} =y_{uv}$, for $v=2, \ldots, V$, $u=1, \ldots, v-1$, in $\bY$ are assumed to be conditionally independent realizations from  Bernoulli random variables, given probabilities that depend solely on the group allocations of the two involved nodes. These allocations are, in turn, assigned a partially exchangeable partition (pExP) prior incorporating layer memberships. Notice that such a construction does not imply independence among edges and layers. In fact, marginalizing out the layer-informed pExP prior yields  a direct dependence among  layers and edges. Including  layer information also in block probabilities is possible. However, while in edge-colored settings this can be useful in modeling the different types of relationships, in node-colored networks encoding a single notion of connectivity such a choice would add redundant  flexibility, while complicating prior elicitation, posterior computation and interpretability.

\subsection{Stochastic Block Models}\label{sec_2.1}
Set $[n]:=\{1,\ldots,n\}$ for any generic integer $n \geq 1$. If each node is identified with the index $v\in[V]$, as in the \textit{supra}-adjacency matrix $\bY$, every partition of nodes into $H$ clusters can be represented by disjoint sets $(\mathbbm{V}_1, \ldots, \mathbbm{V}_H)$, where $\mathbbm{V}_h\subset[V]$ and $\bigcup_{h=1}^H\mathbbm{V}_h=[V]$. To make this representation unique, as customary, we label clusters in order of appearance, so that $\mathbbm{V}_1$ contains node $v=1$, $\mathbbm{V}_2$ contains node $v^*$, with $v^*=\min\{[V]\smallsetminus\mathbbm{V}_1\}$, etc. Define $\bz=(z_1, \ldots, z_V) \in [H]^V$ as the allocations vector corresponding to the node partition $(\mathbbm{V}_1, \ldots, \mathbbm{V}_H)$, such that $z_v=h$ if and only if $v \in \mathbbm{V}_h$. Recalling classical SBM representations for single-layered networks \citep[see, e.g.,][]{holland1983stochastic,nowicki2001estimation}, we assume that $(y_{vu} \, | \, z_v=h, z_u=h',\psi_{hh'}) \sim \mbox{Bern}(\psi_{hh'})$, independently for $v=2, \ldots, V$ and $u=1, \ldots, v-1$, with $\psi_{hh'} \in (0,1)$ the probability of an edge among the generic nodes in groups $h$ and $h'$, for $h,h'\in[H]$. Such a model for the \textit{supra}-adjacency matrix facilitates efficient inference on homogenous node groups, while allowing to incorporate a variety of heterogeneous block-interactions among groups via the $H \times H$ matrix $\bPsi$ with entries $\psi_{hh'}$ for $h,h'\in[H]$. These include any combination of community, core-periphery and disassortative structures, thus improving flexibility w.r.t.\ basic community detection algorithms \citep[e.g.,][]{fortunato2016community,lee2019review}.

Consistent with the above focus on inferring group structures among stochastically-equivalent nodes, we follow classical SBM implementations for single-layered networks \citep[e.g.,][]{schmidt_2013} by assuming independent $\mbox{beta}(a,b)$ priors for the block-probabilities in $\bPsi$, and then marginalize these quantities out  from the model to obtain the beta-binomial likelihood
 \begin{equation}
	p(\bY \, | \, \bz)= \prod_{h=1}^H \prod_{h'=1}^h \frac{\mbox{B}(a+m_{hh'},b+\overline{m}_{hh'})}{\mbox{B}(a,b)}, 
	\label{eq1}
\end{equation}
 where $m_{hh}$ and $\overline{m}_{hh'}$ are the total number of edges and non-edges among the nodes in groups $h$ and $h'$, respectively. Although inference on $\bPsi$ is also of interest, as clarified in Section~\ref{sec_3}, treating it as a nuisance parameter facilitates computation, inference and prediction for the main object of analysis, i.e., $\bz$. 

Importantly, the likelihood in \eqref{eq1} does not account for layer information. In fact, with such a distribution for the entries of the \textit{supra}-adjacency matrix $\bY$,  within- and across-layers edges are not disentangled. The proposed pEx-SBMs crucially address this shortcoming by  embedding the layer division information into the prior for the allocation vector $\bz$. Conversely, single-layered SBMs, which combine likelihood \eqref{eq1} with Dirichlet-multinomial \citep{nowicki2001estimation}, Dirichlet process \citep{kemp_2006}, mixture-of-finite-mixture \citep{geng_2019} or unsupervised Gibbs-type \citep{legramanti2022extended} priors for $\bz$, would be conceptually and practically suboptimal, as these priors are not designed to incorporate structure from layer division. Recalling Section~\ref{sec_1} and Figure~\ref{figure:1}, one expects nodes in the same layer to be more likely to exhibit similar connectivity patterns, with these patterns possibly varying within and across layers. This would translate into a prior for $\bz$ that reinforces the formation of one or multiple within-layer groups, while still allowing for the possibility of creating clusters that comprise nodes in different layers. The latter property is important since the exogenous layer division does not necessarily overlap with the endogenous stochastic equivalence structures in the network. 

\subsection{Partially Exchangeable Partition Priors}\label{sec_2.2}
Our goal is to embed the layer information into the model via the prior for the allocation vector. This translates into completing the likelihood \eqref{eq1} with a pExP prior for $\bz$ informed by the division in layers. Set $\bV=(V_1,\ldots,V_d)$, where we recall that $V_j$ stands for the number of nodes in the $j$-th layer for $j\in[d]$, and the total number of nodes is $V=\sum_{j=1}^dV_j$. Then, we assume
	\begin{equation}
		\bz\sim \mbox{pExP}(\bV),
		\label{eq2}
	\end{equation}
meaning that, given the partition structure defined by a pEPPF  \citep{camerlenghi2019distribution}, the distribution of $\bz$ is determined by this pEPPF evaluated at the vector of layer sizes $\bV$. Let $\bm{e}_j$ be the vector with all zero entries but the $j$-th which is $1$, for $j\in[d]$. Then, in view of our construction, marginalizing $\mbox{pExP}(\bV+\bm{e}_j)$ w.r.t.\ a node in the $j$-th group yields $\mbox{pExP}(\bV)$ for any $j\in[d]$.

\begin{Remark}
\normalfont The direct construction of a prior for a random partition is challenging even in the exchangeable case. Operationally, it is more convenient to start from an exchangeable sequence directed by a discrete random probability measure and then obtain the partition structure by marginalizing over the labels and the probability weights. Hence, \textit{a fortiori}, we follow this approach also in our more general setting and induce a pExP prior through a partially exchangeable array of latent auxiliary attributes of the nodes, directed by a large class of discrete random probability measures known as \textit{hierarchical normalized completely random measures} (H-NRMIs). See Definition~\ref{def:hnrmi}. As anticipated in Section~\ref{sec_1}, the array of latent auxiliary attributes is not of interest for inference.  Rather, it defines a purely instrumental quantity that is useful for constructively defining the pExP prior.
\end{Remark}

\subsubsection{Hierarchical construction}\label{sec_2.2.1}
The pExP prior in \eqref{eq2} allows for an appealing generative construction employing partial exchangeability. Let $\bX$ denote a random array with row $j$ having $V_j$ entries $x_{ji}\in\mathbb{X}$, for each $j\in[d]$ and $i\in[V_j]$. Every entry $x_{ji}$ acts as a latent auxiliary attribute corresponding to the generic node $v\in[V]$, according to the previously mentioned bijection $v\leftrightarrow(j,i)$. Partial exchangeability is enforced by assuming the rows of $\bX$ to be directed by discrete random probability measures $\tilde{P}_1, \ldots, \tilde{P}_d$. This setup is ideally suited for node-colored networks as it incorporates a generalized notion of homophily: if $(\tilde{P}_1, \ldots, \tilde{P}_d)$ are marginally identically distributed, the probability of a tie across the rows of $\bX$ is always less than or equal to the probability of a tie within the same row, with equality holding when the $\tilde P_j$'s coincide almost surely \citep[][]{bea}. Thus, nodes in the same layer are more likely to exhibit the same latent attribute \textit{a priori}, and therefore, to show a similar connectivity behavior. Since we aim to develop a projective representation whose construction and properties hold, coherently, for any finite vector $(V_1,\ldots,V_d)\in \N^d$, the natural invariance condition to assume is the partial exchangeability in Definition~\ref{def:pex} for the infinite array $\bX^\infty$.

\begin{Definition}\label{def:pex}
	 $\bX^\infty$ is partially exchangeable if, for any vector $(V_1,\ldots,V_d)\in\N^d$, it holds
	\begin{equation}\label{eq:parex}
(x_{ji}: j\in[d]; i\in[V_j] )\overset{\mbox{\normalfont \scriptsize d}}{=} (x_{j,\pi_j(i)}: j\in[d]; i\in[V_j] ),
	\end{equation}
for every family of index permutations $\{\pi_1, \ldots, \pi_d\}$, where $\overset{\mbox{\normalfont \scriptsize d}}{=} $ denotes equality in distribution.
\end{Definition}

Definition~\ref{def:pex}  implies that the joint distribution of the entries in $\bX^\infty$ is invariant w.r.t.\ within-layer permutations of the nodes, but not necessarily  across-layers ones. When this holds, de Finetti's representation theorem  \citep{deF(38)} ensures the existence of a vector of random probability measures $(\rpm_1,\dots,\rpm_d)$ such that, for any $j_1,\ldots,j_k\in[d]$ and $i_1,\ldots,i_k\ge 1$, one has
	\begin{equation}\label{eq:definetti}
	\begin{split}
	(x_{j_1i_1},\ldots,x_{j_ki_k})\, | \,(\rpm_1,\dots,\rpm_d)&\simiid\rpm_{j_1}\times \cdots \times \rpm_{j_k},\\ 
(\rpm_1,\dots,\rpm_d)&\sim Q,
	\end{split}
	\end{equation}
for some distribution $Q$, named \textit{de Finetti measure}, on the space of $d$-dimensional vectors of probability measures on $\mathbb{X}$. Note that, when all  $\rpm_j$'s coincide (or $d=1$), \eqref{eq:definetti} reduces to de Finetti's representation of an exchangeable sequence. Its extension to distributional invariance within layers, but not necessarily across, allows a probabilistically sound incorporation of the layer division in the distribution of $\bX^\infty$.

To complete \eqref{eq:definetti}, the prior $Q$ has to be specified. Since the final goal is to induce a prior on $\bz$, our Bayesian nonparametric approach focuses on priors $Q$ selecting, almost surely, vectors of discrete probability measures. With $V_1,\ldots,V_d$ being positive integers associated with the observed network $\bY$, such a choice implies that the entries $\{x_{ji}: j\in[d]; i\in [V_j]\}$ of the finite array $\bX$ can display ties and, therefore, it is possible, and natural, to induce a prior on node partitions by interpreting these ties as a co-clustering relationship. More specifically, let $\bx^*=(x^*_1,\dots,x^*_H)$ denote the unique values of $\bX$, with $H$ the number of clusters, which can be random. Then, the node allocations $z_{ji}$, $j\in[d]$, $i\in[V_j]$, in the random array $\bZ$ can be defined as
	\begin{equation}\label{eq:partition} 
	z_{ji}=h \ \mbox{ if and only if }  \ x_{j i}=x^*_{h}, \quad  \text{for }\, h\in[H],
	\end{equation}
thus inducing a prior distribution on the space of random partitions of $\{1, \ldots, V\}$ from the ties in $\bX$. Clearly, its clustering and co-clustering properties depend on the choice of $Q$. We opt for $Q$ belonging to the broad class of H-NRMIs \citep{camerlenghi2019distribution}, which includes the hierarchical Dirichlet process (H-DP) \citep{Teh_06} and the hierarchical normalized stable process (H-NSP) as noteworthy special cases. The choice of H-NRMIs is motivated by three main reasons: (i) the nodes can be clustered both within and across layers; (ii) the number of clusters in the network is random and learned from the data; (iii) mathematical and computational tractability. Extending the NRMIs construction in the supplementary materials to the hierarchical setting,  it is  possible to define H-NRMIs as follows.

\begin{Definition}\label{def:hnrmi}
	$(\rpm_1,\dots,\rpm_d)$ is a vector of \mbox{\normalfont H-NRMIs} on $\mathbb{X}$ with parameters $(\rho,\rho_0,c,c_0,P_0)$ if
	\begin{equation}\label{eq:hnrmi}
		\begin{split}
			\rpm_1,\dots,\rpm_d \, | \, \rpm_0&\ {\normalfont \simiid} \ \mbox{\normalfont NRMI}(\rho,c,\rpm_0), \\
			\rpm_0&\sim\mbox{\normalfont NRMI}(\rho_0,c_0,P_0),
		\end{split}
	\end{equation}
where $\rho$ and $\rho_0$ are the jump components of the underlying L\'evy intensities of $\rpm_j\, | \,\rpm_0$  for every $j\in[d]$ and $\rpm_0$, respectively, with $c,c_0 \in \mathbb{R^+}$, while $P_0$ is a diffuse probability measure on $\mathbb{X}$.
\end{Definition}

The role of the parameters $(\rho,\rho_0,c,c_0)$ characterizing specific H-NRMIs is described in the supplementary materials.  As desired, \eqref{eq:definetti} combined with \eqref{eq:hnrmi} generates ties among latent node attributes both within and across layers, the latter being induced by the almost sure discreteness of  $\tilde{P}_0$ at the root of the hierarchy. The distribution of the allocation vector associated to the unique values in $\bx^*$ is specified as
	\begin{equation}\label{eq:z}
		\bz\sim\mbox{pExP}(\bV;\rho,\rho_0,c,c_0).
	\end{equation}
The prior induced on this grouping structure arising from $\bX$ is characterized by its pEPPF. This means that the prior probability mass function for $\bz$ (or alternatively $\bZ$), can be derived from the distribution induced by the H-NRMI construction on the array of positive integers $(\bn_1, \ldots, \bn_d)$, where each $\bn_j=(n_{j1}, \ldots, n_{jH})$ represents the allocation frequencies to $H$ different groups of the $V_j$ nodes in layer $j$, for each $j\in[d]$. As shown in \citet{camerlenghi2019distribution}, this distribution is equal to
\begin{equation}
\begin{split}
	&p^{(\bV)}_{H}(\bn_1, \ldots, \bn_d)=\sum_{\bell}\sum_{\bm{q}}\bigg[\Phi_{H,0}^{(|\bell|)}(\ell_{\cdot 1},\dots,\ell_{\cdot H})\\
	& \quad \times \prod_{j=1}^d\prod_{h=1}^H\frac{1}{\ell_{jh}!}\binom{n_{jh}}{q_{jh1},\dots,q_{jh\ell_{jh}}}\Phi_{\ell_{j\cdot},\,j}^{(V_j)}(\bm{q}_{j1},\dots,\bm{q}_{jH}) \bigg],
	\end{split}
	\label{eq:peppf1}
\end{equation}
where  $n_{jh}$ is the total number of nodes within layer $j$ allocated to group $h$, $\bell$ is an array with generic element $\ell_{jh} \in [n_{jh}]$, for $j\in[d]$ and $h\in[H]$, whose meaning will be clarified in detail later, while $\ell_{j\cdot}=\sum_{h=1}^H \ell_{jh}$, $\ell_{\cdot h}=\sum_{j=1}^d\ell_{jh}$ and $|\bell|=\sum_{j=1}^d\ell_{j\cdot}$. Finally, $\bm{q}_{jh}=(q_{jh1},\dots,q_{jh\ell_{jh}})$ is a vector of positive integers summing up to $n_{jh}$. 

In \eqref{eq:peppf1}, the functions $\Phi_{\ell_{j\cdot},j}^{(V_j)}$ and $\Phi_{H,0}^{(|\bell|)}$ are the EPPFs associated to NRMIs with parameters $(\rho,c)$ and $(\rho_0,c_0)$, respectively. Such EPPFs characterize the probability distribution of the exchangeable random partitions of $V_j$ nodes in $\ell_{j\cdot}$ groups and $|\bell |$ elements in $H$ clusters, respectively; see \cite{jlp} for the closed-form expressions of such functions.

To help intuition with the pExP prior and the quantities involved in \eqref{eq:peppf1}, we recast the \textit{Chinese restaurant franchise} (CRF) metaphor \citep{Teh_06} within the node-colored network setting for the whole H-NRMIs class. According to this metaphor, in every layer the nodes are first allocated to within-layer subgroups, and then each subgroup is assigned a sociability profile from a single list which is common across layers. Nodes in subgroups with the same sociability profile are, therefore, naturally characterized by similar connectivity patterns in the multilayer network and, hence, have the same group allocation in $\bz$. Notice that the  within-layer division in subgroups  is only a latent quantity, which is not of interest for inference, but rather provides an intermediate clustering that leads, under the hierarchical construction, to the formation of the grouping structure among nodes which we aim to infer (i.e., the one defined by the sociability profiles). Nonetheless, such a within-layer partition nested in the sociability profiles further clarifies how layer information is effectively incorporated in the formation  of node groups. In fact, two nodes in the same layer have equal group indicator if they are allocated to the same subgroup or, when they belong to different subgroups, the same sociability profile is assigned to these two subgroups. Conversely, nodes in different layers belong to a same final group only if the corresponding subgroups have the same sociability profile. This clarifies the role of layer division in reinforcing the formation of within-layer groups without ruling out across-layer clusters.

Consistent with the above metaphor, each $\rpm_j$ in \eqref{eq:hnrmi} allocates nodes to subgroups within each layer, whereas $\rpm_0$ assigns to subgroups the sociability profiles from a list shared across layers. Due to the discreteness of $\rpm_0$, the same sociability profile can be assigned to multiple subgroups both within and across layers. This facilitates also the interpretation of the arrays $\bell$ and $\bm{q}$ in \eqref{eq:peppf1}. More specifically, $\ell_{jh}$ is the number of subgroups in layer $j$ with sociability profile $h$ and $q_{jht}$ is the number of nodes in layer $j$ assigned to the $t$-th subgroup with sociability profile $h$; consequently, $\ell_{j\cdot}$ is the number of subgroups in layer $j$, $\ell_{\cdot h}$ is the total number of subgroups with sociability profile $h$, and $n_{jh}$ denotes the number of nodes in layer $j$ having sociability profile $h$. The metaphor also clarifies the role of the EPPFs $\Phi_{\ell_{j\cdot},j}^{(V_j)}$ and $\Phi_{H,0}^{(|\bell|)}$ in the sampling of the two nested partitions. The former regulates, for every $j\in[d]$, the distribution of the division in subgroups within each layer, whereas the latter drives the sociability profile assignment to the subgroups in the different layers.

The above discussion also clarifies the process through which the auxiliary attribute values in $\bX$ are generated. As already discussed, $\bX$ is treated in our construction as a latent quantity useful to define, study and manage the pExP prior for $\bz$. While $\bX$ is not of interest for inference, its generative process is, however, useful for posterior sampling and prediction under the pEx-SBM.
	\begin{enumerate}
		\item[1. ] For each layer $j\in[d]$, sample the entries within the vector $\bt_j=(t_{j1},\dots,t_{jV_j})$ of subgroup labels via
\begin{equation*}		
		t_{j1},\dots,t_{jV_j}\, | \,\tilde{Q}_j\simiid\tilde{Q}_j, 
		\end{equation*}
where $\tilde{Q}_j\sim\text{NRMI}(\rho,c,G_j)$, for some diffuse probability measure $G_j$.
		\item[2. ] For each  $j\in[d]$, allocate the nodes to subgroups according to the partition induced by the ties in $\bt_j$. This yields the subgroup allocation array $\bW$ with entries  
		\begin{equation}\label{eq:subpartition}w_{ji}=\uptau \ \mbox{ if and only if }  \  t_{ji}=t^*_{j\uptau},\end{equation}
for $\uptau\in[\ell_{j\cdot}]$ with $\bt_j^*=(t^*_{j1},\dots,t^*_{j\ell_{j\cdot}})$ the unique values of $\bt_j$ in order of occurrence.
		\item[3. ] Sample the array of sociability profile labels $\bS$, which has an entry $s_{j\tau}$ for each subgroup drawn from
		\begin{equation}\label{eq:s}
			s_{j\uptau}\, | \,\tilde{P}_{0}\simiid\tilde{P}_{0},
		\end{equation}
for $\uptau\in[\ell_{j\cdot}]$, $j\in[d]$, with $\tilde P_0\sim\mbox{NRMI}(\rho_0,c_0,P_0)$.
		\item[4. ] Obtain $\bX$ by assigning to each node the sociability profile label of its subgroup, that is $x_{ji}:=s_{j\uptau}$ whenever $w_{ji}=\uptau$.
	\end{enumerate}

Recalling \eqref{eq:partition}, the ties in $\bX$ yield the final clustering structure encoded in $\bz$ (or alternatively $\bZ$), which comprises the sociability profiles of the $V$ nodes. As $\bZ$ takes values in $[H]$, the $j$-th row of $\bW$ takes values in $[\ell_{j\cdot}]$, for every $j\in[d]$. We will represent the sociability profiles and the subgroups with these indices. Note that, since we are considering ties only for clustering, $G_j$ in step 1.\ and $P_0$ in step 3.\ are arbitrary, as long as they are diffuse. As discussed previously, the subgroup allocation array $\bW$  should be interpreted as a data augmentation scheme. This quantity is not object of inference, but is important to recover tractable full conditionals for computation and prediction on $\bz$ and $\bY$.

A crucial  feature of pExP priors is that \textit{Kolmogorov consistency} holds by construction.  This property implies that the joint distribution of any finite-dimensional sequence of node allocations can be obtained by marginalizing one of higher dimension defined through the same generative scheme. Consequently, our formulation remains coherent for every network size and preserves projectivity to growing number of nodes. Such a property, which is not met, e.g., by the supervised approach of \citet{legramanti2022extended}, is not only crucial in deriving novel and principled predictive strategies, but is also conceptually desirable since, in practice, not all the nodes in a network are observed. Hence, it is important that the model postulated for a subset of the whole network remains coherent also at the population level.

\begin{Remark} 
	\normalfont  Despite the potential for multilayer network modeling of the pExP priors induced by H-NRMIs, to the best of our knowledge, the only contribution moving in a related direction is the one by 
	\citet{amini2019hierarchical} which, however, focuses on edge-colored (multiplex) networks, rather than node-colored ones, and considers only the specific case of H-DP priors. Such a contribution can be recovered as a special case of our framework by imposing a block-diagonal \textit{supra}-adjacency matrix and selecting the H-DP from the general H-NRMIs class. However, although \citet{amini2019hierarchical} provide a valuable contribution in the substantially-different edge-colored setting, the partial exchangeability assumption behind the H-DP and its induced clustering and co-clustering mechanisms do not align naturally with the intrinsic structure of edge-colored networks. In fact, when the same set of nodes is replicated across the layers, as in edge-colored settings, such an assumption forces the model to ignore this key node-identity information across layers. As a result, it is also unclear, from a network perspective, how to interpret groups shared across multiple layers, which may include copies of the same nodes. Similar comments apply to the recent contribution by \citet{josephs2023nested} which relies on the nested Dirichlet process (also a partially exchangeable prior) and still focuses mainly on edge-colored, rather than node-colored, networks. As highlighted in Section~\ref{sec_6}, an assumption of separate exchangeability \citep[e.g.,][]{lin2021separate} would be more coherent with the structure of edge-colored networks.

Notice that the use of specific discrete hierarchical structures can be found also in \citet{dempsey2022hierarchical}, but with a focus on edge-exchangeable, rather than node-exchangeable, models. Such a perspective is desirable in situations where edge-specific structures are of interest, regardless of the  specific identities of the nodes among which these edges exist. This prevents from inferring group structures among nodes via SBMs, possibly informed by layer partitions. When this is the focus of inference, one has to identify nodes (and not edges) as statistical units and, hence, enforce notions of node exchangeability.  In fact, in \citet[][Remark 3.1]{dempsey2022hierarchical},  the hierarchical structure is used to generate, and infer, the interaction process itself, rather than as a prior on nodes' partitions.
\end{Remark}

\subsubsection{Clustering and co-clustering properties}\label{sec_2.2.2}
Clustering and co-clustering properties play a key role in SBMs. Nonetheless analytical results are missing even in the more general H-NRMI literature  \citep{camerlenghi2019distribution}. Here we fill this gap by deriving novel explicit expressions for clustering and co-clustering probabilities from a predictive perspective. In addition to providing insights into the model's behavior, such results are at the basis of the sampling algorithm and the prediction schemes developed in Section~\ref{sec_3}.
	
In the sequel we will consider the vectorized forms $\bz$ and $\bw$ for the sociability profiles and the subgroups allocations in $\bZ$ and $\bW$, respectively, defined as in \eqref{eq:partition} and \eqref{eq:subpartition}. With $\bz^{-v}$ and $\bw^{-v}$ we instead indicate the $(V-1)$-dimensional vectors obtained from $\bz$ and $\bw$ respectively, upon removing node $v$ and suitably rearranging the labels in order of appearance. Similarly, $\bell^{-v}$ and $\bm{q}^{-v}$ also stand for the arrays with node $v$ removed from the corresponding counts. Moreover, define the set 
\begin{equation}\label{eq:taus}
		\mathbbm{T}_{jh}^{-v}:=\{\uptau\in[\ell_{j\cdot}^{-v}]: s^{-v}_{j\uptau}=x^{*-v}_h\},
	\end{equation}
where $s^{-v}_{j\uptau}$ is an entry of the sociability array $\bS^{-v}$ obtained as in \eqref{eq:s}, and $\bx^{*-v}$ displays the unique sociability labels, as in \eqref{eq:partition}, but disregarding node $v$. The indices within $\mathbbm{T}^{-v}_{jh}$ are associated to the subgroups in layer $j$ with profile $h$. Indeed, $|\mathbbm{T}^{-v}_{jh}|=\ell_{jh}^{-v}$.  With these settings we can state the following result. 

\begin{Proposition}\label{prp:pred_hnrmi}
		Let $\bz$ denote a random allocation vector such that  $\bz\sim \mbox{\normalfont pExP}(\bV;\rho,\rho_0,c,c_0)$ as in \eqref{eq:z}, and $j$ be the layer of the generic node $v\in[V]$. Then, with $\mathbbm{T}_{jh}^{-v}$ defined as in \eqref{eq:taus} and $\uptau_{\text{new}}:=\ell^{-v}_{j\cdot}+1$ standing for a new subgroup in layer $j$, we have
		\begin{equation}\label{eq:pred}
		\begin{split}
			&\mathbb{P}(z_v=h \, | \, \bz^{-v}, \bw^{-v})=\mathbb{P}(v \rightsquigarrow \mathbb{T}^{-v}_{jh})\\
			&\qquad \quad +\mathbb{P}(v \rightsquigarrow \uptau_{\text{new}}\, | \, \uptau_{\text{new}} \leftsquigarrow h)\times \mathbb{P}(\uptau_{\text{new}} \leftsquigarrow h),
			\end{split}
		\end{equation}
for any  $h\in[H^{-v}+1]$, with $ \rightsquigarrow$ and $\leftsquigarrow$ meaning ``assigned to'', and
		\begin{equation}
			\begin{split}
				&\mathbb{P}(v \rightsquigarrow \mathbbm{T}^{-v}_{jh})=\frac{{\textstyle \sum_{t=1}^{\ell_{jh}^{-v}}}\Phi_{\ell_{j\cdot}^{-v},j}^{(V_j)}(\q_{j1}^{-v},\dots,\q^{-v}_{jh}+\e_t,\dots,\q^{-v}_{jH^{-v}})}{\Phi_{\ell_{j\cdot}^{-v},j}^{(V_j-1)}(\q_{j1}^{-v},\dots,\q^{-v}_{jH^{-v}})},\\
				&\mathbb{P}(v \rightsquigarrow \uptau_{\text{new}}\, | \, \uptau_{\text{new}} \leftsquigarrow h)\\
				& \qquad \qquad=\frac{\Phi_{\ell_{j\cdot}^{-v}+1,j}^{(V_j)}(\q_{j1}^{-v},\dots,(\q^{-v}_{jh},1),\dots,\q^{-v}_{jH_h^{-v}})}{\Phi_{\ell_{j\cdot}^{-v},\,j}^{(V_j-1)}(\q_{j1}^{-v},\dots,\q^{-v}_{jH^{-v}})},\\
				&\mathbb{P}(\uptau_{\text{new}} \leftsquigarrow h)=\frac{\Phi_{H_h^{-v},0}^{(|\bell^{-v}|+1)}(\ell_{\cdot 1}^{-v},\dots,\ell^{-v}_{\cdot h}+1,\dots,\ell^{-v}_{\cdot H_h^{-v}})}{\Phi_{H^{-v},0}^{(|\bell^{-v}|)}(\ell_{\cdot 1}^{-v},\dots,\ell^{-v}_{\cdot H^{-v}})},
			\end{split}
			\label{eq:pred1}
		\end{equation}
		where $\e_t$ is the $t$-th vector of the $\ell_{jh}^{-v}$-dimensional canonical basis, and $H_h^{-v}:=h\vee H^{-v}$.
	\end{Proposition}

\begin{Remark}
	{\normalfont The expression for $\mathbb{P}(z_v=h \, | \, \bz^{-v}, \bw^{-v})$ in Proposition~\ref{prp:pred_hnrmi} can be  interpreted in terms of the previously-introduced metaphor. The first summand is the probability of node $v$ being allocated to any of the already-occupied subgroups in $j$ with sociability profile $h$, while the second is the probability of being allocated to a new subgroup, which has been assigned sociability profile $h$. If profile $h$ is not present in layer $j$, then $\ell_{jh}^{-v}=0$, the first summand disappears and $h$ can be assigned to $v$ only by creating a new subgroup. Moreover, since $\bz^{-v}$ is arranged according to the order of occurrence in $\bx^{-v}$, $h=H^{-v}+1$ represents the case of a sociability profile new to the whole network.}
\end{Remark}

Corollary~\ref{ex1} specializes the results in Proposition~\ref{prp:pred_hnrmi} to the popular H-DP case, thereby re-obtaining its known urn scheme \citep{Teh_06}. An analogous result for the H-NSP is given in the supplementary materials.

\begin{Corollary}\label{ex1}
	Let $\bz\sim \mbox{\normalfont pExP}(\bV; \theta,\theta_0)$, where $\mbox{\normalfont pExP}(\bV;\theta,\theta_0)$ is the partition structure induced by a $\mbox{\normalfont H-DP}$ having concentration parameters $\theta$ and $\theta_0$. Then, for each $h\in[H^{-v}+1]$,
	\begin{equation}\label{eq:pred_dir}
	\begin{split}
		&\mathbb{P}(z_v=h \, | \, \bz^{-v}, \bw^{-v}) =\indic_{\{\ell^{-v}_{\cdot h}=0\}}\frac{\theta_0}{(\theta_0+|\bm{\ell}^{-v}|)}\frac{\theta}{(\theta+V_j-1)}\\
		&\qquad+\indic_{\{\ell^{-v}_{\cdot h}\neq0\}}\left[\frac{\ell^{-v}_{\cdot h}}{(\theta_0+|\bm{\ell}^{-v}|)}\frac{\theta}{(\theta+V_j-1)}+\frac{n_{jh}^{-v}}{\theta+V_j-1}\right].
		\end{split}
		\end{equation}
	\end{Corollary}
	
In Corollary \ref{ex1} the two scenarios are more explicit. If $h$ is a new sociability profile for the whole network, $\ell^{-v}_{\cdot h}=0$ and the probability of node $v$ receiving $h$ coincides with the probability of being assigned a new profile at a new subgroup. Conversely, if $\ell^{-v}_{\cdot h}\neq0$, then $h$ is not new to the network and we sum two terms, namely the probability of assigning $h$ to a new subgroup and the probability of being allocated to an already-occupied subgroup among those with profile $h$. The latter is $0$ if the sociability profile $h$ is new for the layer of node $v$, since $n_{jh}^{-v}=0$.

Besides clarifying the generative nature of $\bz$, as highlighted in Section~\ref{sec_3}, the results in Proposition~\ref{prp:pred_hnrmi} and Corollary~\ref{ex1} are a key to develop tractable collapsed Gibbs sampling schemes for  inference on the posterior distribution $p(\bz \, | \, \bY) \propto p(\bz)p(\bY \, | \, \bz)$ induced by the model defined in \eqref{eq1} and \eqref{eq:z}. The co-clustering probabilities derived in Theorem~\ref{coclust_hnrmi} below are instead crucial for prior elicitation and to devise rigorous predictive strategies for  both the allocations and edges of future incoming nodes. The apex $^{-vu}$ denotes all the previously-defined quantities, evaluated disregarding nodes $v,u\in[V]$.

\begin{Theorem}\label{coclust_hnrmi}
		Let $\bz$ denote a random allocation vector such that $\bz\sim \text{\normalfont pExP}(\bV;\rho,\rho_0,c,c_0)$. Moreover, let $j$ and $j'$ be the layers of any two distinct nodes $v,u\in[V]$, respectively. Then, with $\uptau_{\text{new}}$ and $\uptau_{\text{new}}'$ indicating new subgroups in layers $j$ and $j'$, respectively, the following results hold.
		
	 \text{(1)} If $j=j'$, we have that
		\begin{equation}\label{eq:monster1}
			\begin{split}
				&\mathbb{P}(z_v=z_u \, | \, \bz^{-vu}, \bw^{-vu})= \sum\nolimits_{h=1}^{H^{-vu}+1}\Big[  \mathbb{P}(\{v,u\} \rightsquigarrow \mathbbm{T}_{jh}^{-vu})\\
				&+  \mathbb{P}(\uptau_{\text{new}} \leftsquigarrow h) \times \Big\{\mathbb{P}(\{v,u\} \rightsquigarrow \uptau_{\text{new}}\, | \, \uptau_{\text{new}} \leftsquigarrow h)\\
				&\qquad+2\,\mathbb{P}(v  \rightsquigarrow \uptau_{\text{new}}, u \rightsquigarrow\mathbbm{T}_{jh}^{-vu}\, | \, \uptau_{\text{new}} \leftsquigarrow h)\Big\}\\
				&+   \mathbb{P}(\uptau_{\text{new}} \leftsquigarrow h, \uptau_{\text{new}}' \leftsquigarrow h)  \\
				&\qquad \times \mathbb{P}(v  \rightsquigarrow \uptau_{\text{new}}, u \rightsquigarrow \uptau_{\text{new}}'\, | \, \uptau_{\text{new}} \leftsquigarrow h, \uptau_{\text{new}}' \leftsquigarrow h) \Big],
			\end{split}
		\end{equation}
		with
		\begin{equation}\label{eq:monster11}
			\begin{split}
				&\mathbb{P}(\{v,u\} \rightsquigarrow \mathbbm{T}_{jh}^{-vu})\\
				&=\frac{\sum_{t=1}^{\ell_{jh}^{-vu}}\Phi_{\ell^{-vu}_{j\cdot},j}^{(V_j)}(\q_{j1}^{-vu},\dots,\q_{jh}^{-vu}+2\e_t,\dots,\q_{jH^{-vu}}^{-vu})}{\Phi_{\ell^{-vu}_{j\cdot},j}^{(V_j-2)}(\q_{j1}^{-vu},\dots,\q_{jH^{-vu}}^{-vu})} \\
				& \quad+\frac{\sum_{A\in C_{\ell_{jh}^{-vu},\,2}}2\cdot\Phi_{\ell^{-vu}_{j\cdot},j}^{(V_j)}(\q_{j1}^{-vu},\dots,\q_{jh}^{-vu}+\e_A,\dots,\q_{jH^{-vu}}^{-vu})}{\Phi_{\ell^{-vu}_{j\cdot},j}^{(V_j-2)}(\q_{j1}^{-vu},\dots,\q_{jH^{-vu}}^{-vu})},\\
				&\mathbb{P}(\uptau_{\text{new}} \leftsquigarrow h) = \frac{\Phi_{H^{-vu}_h,0}^{(|\bell^{-vu}|+1)}(\ell_{\cdot 1}^{-vu},\dots,\ell^{-vu}_{\cdot h}+1,\dots,\ell^{-vu}_{\cdot H^{-vu}_h})}{\Phi_{H^{-vu},0}^{(|\bell^{-vu}|)}(\ell_{\cdot 1}^{-vu},\dots,\ell^{-vu}_{\cdot H^{-vu}})},\\
				&\mathbb{P}(\{v,u\} \rightsquigarrow \uptau_{\text{new}}\, | \, \uptau_{\text{new}} \leftsquigarrow h)\\
				&\quad =\frac{\Phi_{\ell_{j\cdot}^{-vu}+1,\,j}^{(V_j)}(\q_{j1}^{-vu},\dots,(\q^{-vu}_{jh},2),\dots,\q^{-vu}_{jH^{-vu}_h})}{\Phi_{\ell_{j\cdot}^{-vu},\,j}^{(V_j-2)}(\q_{j1}^{-vu},\dots,\q^{-vu}_{jH^{-vu}})},\\
								&\mathbb{P}(v  \rightsquigarrow \uptau_{\text{new}}, u \rightsquigarrow\mathbbm{T}_{jh}^{-vu}\, | \, \uptau_{\text{new}} \leftsquigarrow h)\\
								&\quad =\frac{\sum_{t=1}^{\ell_{jh}^{-vu}}\Phi_{\ell_{j\cdot}^{-vu}+1,j}^{(V_j)}(\q_{j1}^{-vu},\dots,(\q^{-vu}_{jh}+\e_t,\,1),\dots,\q^{-vu}_{jH^{-vu}})}{\Phi_{\ell_{j\cdot}^{-vu},j}^{(V_j-2)}(\q_{j1}^{-vu},\dots,\q^{-vu}_{jH^{-vu}})},\\	
				&\mathbb{P}(\uptau_{\text{new}} \leftsquigarrow h, \uptau_{\text{new}}' \leftsquigarrow h) \\
				&\quad =\frac{\Phi_{H^{-vu}_h,0}^{(|\bell^{-vu}|+2)}(\ell_{\cdot 1}^{-vu},\dots,\ell^{-vu}_{\cdot h}+2,\dots,\ell^{-vu}_{\cdot H^{-vu}_h})}{\Phi_{H^{-vu},0}^{(|\bell^{-vu}|)}(\ell_{\cdot 1}^{-vu},\dots,\ell^{-vu}_{\cdot H^{-vu}})},\\
				&\mathbb{P}(v  \rightsquigarrow \uptau_{\text{new}}, u \rightsquigarrow \uptau_{\text{new}}'\, | \, \uptau_{\text{new}} \leftsquigarrow h, \uptau_{\text{new}}' \leftsquigarrow h)\\
				& \quad =\frac{\Phi_{\ell_{j\cdot}^{-vu}+2,j}^{(V_j)}(\q_{j1}^{-vu},\dots,(\q^{-vu}_{jh},1,1),\dots,\q^{-vu}_{jH^{-vu}_h})}{\Phi_{\ell_{j\cdot}^{-vu},j}^{(V_j-2)}(\q_{j1}^{-vu},\dots,\q^{-vu}_{jH^{-vu}})},
\end{split}
\end{equation}
for any $h\in[H^{-vu}+1]$, where $C_{\ell,2}$ is the set of $2$-combinations of indices in $[\ell]$, whereas, for $A\in C_{\ell,2}$, $\e_A$ is a $\ell$-dimensional vector with $\e_i=1$ if $i\in A$ and $0$ otherwise. Finally $H^{-vu}_h=H^{-vu}\vee h$. 
	
\noindent \text{(2)}  If, instead, $j \neq j'$, we have
		\begin{equation}\label{eq:monster2}
			\begin{split}
				&\mathbb{P}(z_v=z_u \, | \, \bz^{-vu}, \bw^{-vu})\\
				&= \sum\nolimits_{h=1}^{H^{-vu}+1}\Big[  \mathbb{P}(v  \rightsquigarrow \mathbbm{T}_{jh}^{-v})\times \mathbb{P}(u  \rightsquigarrow \mathbbm{T}_{j'h}^{-u})  \\
				& \quad +  \mathbb{P}(\uptau_{\text{new}} \leftsquigarrow h) \times \mathbb{P}(v\rightsquigarrow \uptau_{\text{new}}\, | \, \uptau_{\text{new}} \leftsquigarrow h) \times \mathbb{P}(u  \rightsquigarrow \mathbbm{T}_{j'h}^{-u})  \\
				& \quad +  \mathbb{P}(\uptau_{\text{new}}' \leftsquigarrow h) \times \mathbb{P}(u\rightsquigarrow \uptau_{\text{new}}'\, | \, \uptau_{\text{new}}' \leftsquigarrow h) \times \mathbb{P}(v  \rightsquigarrow \mathbbm{T}_{jh}^{-v})  \\
				& \quad +  \{ \mathbb{P}(\uptau_{\text{new}} \leftsquigarrow h, \uptau_{\text{new}}' \leftsquigarrow h)  \times \mathbb{P}(v  \rightsquigarrow \uptau_{\text{new}}\, | \, \uptau_{\text{new}} \leftsquigarrow h)\\
				& \qquad \times \mathbb{P}(u \rightsquigarrow \uptau_{\text{new}}'\, | \,\uptau_{\text{new}}'\leftsquigarrow h)\} \Big],
			\end{split}
		\end{equation}
		with
		\begin{equation}\label{eq:monster22}
			\begin{split}
				&\mathbb{P}(v  \rightsquigarrow \mathbbm{T}_{jh}^{-v}) \\
				&\ \ = \frac{\sum_{t=1}^{\ell_{jh}^{-vu}}\Phi_{\ell^{-vu}_{j\cdot},j}^{(V_{j})}(\q_{j1}^{-vu},\dots,\q_{jh}^{-vu}+\e_t,\dots,\q_{jH^{-vu}}^{-vu})}{\Phi_{\ell^{-vu}_{j\cdot},j}^{(V_{j}-1)}(\q_{j1}^{-vu},\dots,\q_{jH^{-vu}}^{-vu})}, \\
				&\mathbb{P}(u  \rightsquigarrow \mathbbm{T}_{j'h}^{-u}) \\
				&\ \ =\frac{\sum_{t=1}^{\ell^{-vu}_{j'h}}\Phi_{\ell^{-vu}_{j'\cdot},j'}^{(V_{j'})}(\q_{j'1}^{-vu},\dots,\q_{j'h}^{-vu}+\e_t,\dots,\q_{j'H^{-vu}}^{-vu})}{\Phi_{\ell^{-vu}_{j'\cdot},j'}^{(V_{j'}-1)}(\q_{j'1}^{-vu},\dots,\q_{j'H^{-vu}}^{-vu})}, \\
				&  \mathbb{P}(\uptau_{\text{new}} \leftsquigarrow h)= \mathbb{P}(\uptau_{\text{new}}' \leftsquigarrow h)\\
				&\ \ = \frac{\Phi_{H^{-vu}_h,0}^{(|\bell^{-vu}|+1)}(\ell_{\cdot 1}^{-vu},\dots,\ell^{-vu}_{\cdot h}+1,\dots,\ell^{-vu}_{\cdot H^{-vu}_h})}{\Phi_{H^{-vu},0}^{(|\bell^{-vu}|)}(\ell_{\cdot 1}^{-vu},\dots,\ell^{-vu}_{\cdot H^{-vu}})}, \\
				& \mathbb{P}(v\rightsquigarrow \uptau_{\text{new}}\, | \, \uptau_{\text{new}} \leftsquigarrow h)\\
				&\ \  = \frac{\Phi_{\ell_{j\cdot}^{-vu}+1,j}^{(V_{j})}(\q_{j1}^{-vu},\dots,(\q^{-vu}_{jh},1),\dots,\q^{-vu}_{jH^{-vu}_h})}{\Phi_{\ell_{j\cdot}^{-vu},j}^{(V_{j}-1)}(\q_{j1}^{-vu},\dots,\q^{-vu}_{jH^{-vu}})}, \\
				& \mathbb{P}(u\rightsquigarrow \uptau_{\text{new}}'\, | \, \uptau_{\text{new}}' \leftsquigarrow h)\\
				&\ \ =\frac{\Phi_{\ell_{j'\cdot}^{-vu}+1,\,j'}^{(V_{j'})}(\q_{j'1}^{-vu},\dots,(\q^{-vu}_{j'h},1),\dots,\q^{-vu}_{j'H_h^{-vu}})}{\Phi_{\ell_{j'\cdot}^{-vu},\,j'}^{(V_{j'}-1)}(\q_{j'1}^{-vu},\dots,\q^{-vu}_{j'H^{-vu}})}, \\
				& \mathbb{P}(\uptau_{\text{new}} \leftsquigarrow h, \uptau_{\text{new}}' \leftsquigarrow h)\\
				&\ \ =\frac{\Phi_{H^{-vu}_h,0}^{(|\bell^{-vu}|+2)}(\ell_{\cdot 1}^{-vu},\dots,\ell^{-vu}_{\cdot h}+2,\dots,\ell^{-vu}_{\cdot H^{-vu}_h})}{\Phi_{H^{-vu},0}^{(|\bell^{-vu}|)}(\ell_{\cdot 1}^{-vu},\dots,\ell^{-vu}_{\cdot H^{-vu}})},\\
			\end{split}
		\end{equation}
		for any $h\in[H^{-vu}+1]$. 
\end{Theorem}

\begin{Remark}
	{\normalfont As for Proposition~\ref{prp:pred_hnrmi}, the components in \eqref{eq:monster1} and \eqref{eq:monster2} can be interpreted in terms of our metaphor. More specifically, in \eqref{eq:monster1} we have the sum of: (i) the probability of both nodes $v$ and $u$ being allocated to an already-occupied subgroup, either the same or two different ones but still with same sociability profile (these two scenarios are accounted for by the two summands in \eqref{eq:monster11}); (ii) the probability of creating a new subgroup with either both nodes assigned to that subgroup or one node assigned to the new subgroup and the other to a previously-occupied subgroup having the same sociability profile; (iii) the probability of being allocated to two new subgroups with equal sociability profile. In \eqref{eq:monster2}, since $v$ and $u$ are in two different layers, we have the sum of: (i) the probability of being allocated to two different already-occupied subgroups having the same sociability profile; (ii) the probabilities of creating, and occupying, a new subgroup in one of the two layers, while the other node is assigned to an already-occupied subgroup with the same sociability profile; (iii) the probability of being allocated to a new subgroup in both layers, with each new subgroup having the same sociability profile.}
\end{Remark}

If one specializes the pEPPF to the H-DP case, the following novel result is obtained (an analogous one for the H-NSP can be found in the supplementary materials).

\begin{Corollary}\label{cor:coclust_hdp}
	Let $\bz\sim \mbox{\normalfont pExP}(\bV;\theta,\theta_0)$. Then, if both $v$ and $u$ are in the same layer $j$, we have
	\begin{equation}
	\begin{split} \label{eq:coclust_hdp1}
		&\mathbb{P}(z_v=z_u \, | \, \bz^{-vu}, \bw^{-vu})=\frac{1}{(\theta+V_{j}-2)(\theta+V_j-1)}\\
		&\times \left[\sum_{h=1}^{H^{-vu}} n_{jh}^{-vu}(n_{jh}^{-vu}+1)+\theta\left(1+\frac{2}{\theta_0+|\bm{\ell}^{-vu}|}\sum_{h=1}^{H^{-vu}}n_{jh}^{-vu}\ell_{\cdot h}^{-vu}\right)\right.\\
		&\left.\ \ \ +\frac{\theta^2}{(\theta_0+|\bm{\ell}^{-vu}|)(\theta_0+|\bm{\ell}^{-vu}|+1)}\left(\sum_{h=1}^{H^{-vu}}\ell_{\cdot h}^{-vu}(\ell_{\cdot h}^{-vu}+1)+\theta_0\right)\right],
		\end{split}
	\end{equation}
	whereas, if node $v$ is in layer $j$ and node $u$ is in layer $j'$, with $j\neq j'$, it follows that
	\begin{equation}
	\begin{split}\label{eq:coclust_hdp2}
		&\mathbb{P}(z_v=z_u \, | \, \bz^{-vu}, \bw^{-vu}) =\frac{1}{(\theta+V_{j}-1)(\theta+V_{j'}-1)}\\
		&\times \left[\sum_{h=1}^{H^{-vu}} n_{jh}^{-vu}n_{j'h}^{-vu}+\frac{\theta}{\theta_0+|\bm{\ell}^{-vu}|}\sum_{h=1}^{H^{-vu}}\ell_{\cdot h}^{-vu}\left(n_{jh}^{-vu}+n_{j'h}^{-vu}\right)\right.\\
		&\left.\ \ \ +\frac{\theta^2}{(\theta_0+|\bm{\ell}^{-vu}|)(\theta_0+|\bm{\ell}^{-vu}|+1)}\left(\sum_{h=1}^{H^{-vu}}\ell_{\cdot h}^{-vu}(\ell_{\cdot h}^{-vu}+1)+\theta_0\right)\right].
\end{split}
	\end{equation}
\end{Corollary}

The previous expressions are also useful for prior elicitation. According to our extended notion of homophily one expects the probability of a node being assigned to a sociability profile (i.e., a group) already present in its layer to be larger than the one of being allocated to a sociability profile that is new to its layer. As it can be argued from, e.g., \eqref{eq:pred_dir}, the probability of allocating node $v$ to a group comprising nodes from its layer $j$ or in a cluster having only nodes from other layers depends on the proportion $p_v=(1/|\bm{\ell}^{-v}|)\sum_{h\in \bH_j^{-v}}\ell_{\cdot h}^{-v}$ where $ \bH_j^{-v}$ is the set of unique sociability profiles for the nodes in layer $j$ upon removing node $v$. In fact, allocation of nodes to within-layer groups are favored. These occur by either assigning a node to an already-occupied subgroup or by creating a new one with a sociability profile  that is already present in the layer; conversely, only the latter option is possible to cluster strictly across layers. Nonetheless, if sociability profiles new to layer $j$ are popular (in the sense of several subgroups in other layers displaying them), then $p_v$, which measures the popularity of the sociability profiles already observed in $v$'s layer, is reduced. Consequently, clustering strictly across layers becomes increasingly probable. Albeit possible in some networks, such a situation points in an opposite direction relative to the concept of homophily and, hence, it is natural to elicit priors which exclude this possibility. Leveraging the previously-derived results, Proposition~\ref{prp:cond_dir}  provides conditions on the H-DP hyperparameters which eliminate the dependence on the proportion $p_v$, thus enforcing the general notion of homophily. As clarified in the supplementary materials, an analogous result holds for the H-NSP case.

\begin{Proposition}\label{prp:cond_dir}
	Let $\bz\sim \mbox{\normalfont pExP}(\bV;\theta,\theta_0)$. Then, for any generic node $v$ in layer $j$, we have
	\begin{equation}\label{eq:elicit1}
	\begin{split}
	&\theta\leq (V_j-1)\left(\theta_0/|\bm{\ell}^{-v}|+1\right)\\
	&\Longrightarrow\mathbb{P}(z_v\in \bH_j^{-v}\, | \,\bz^{-v},\bw^{-v})\geq\mathbb{P}(z_v\in \bH^{-v}\smallsetminus  \bH_j^{-v}\, | \,\bz^{-v},\bw^{-v}),
	\end{split}
	\end{equation}
	for each $j\in[d]$, where $\bH^{-v}$ denotes the set of already-observed sociability profiles in all layers. Moreover
	\begin{equation}\label{eq:elicit2}
		\begin{split}
		&\theta\leq V_j-1\\
		&\Longrightarrow
		\mathbb{P}(z_v\in \bH_j^{-v}\, | \,\bz^{-v},\bw^{-v})\geq\mathbb{P}(z_v\notin \bH_j^{-v}\, | \,\bz^{-v},\bw^{-v}),
			\end{split}
	\end{equation}
	for every node $v\in[V]$ and corresponding layer $j\in[d]$.
\end{Proposition}

\begin{Remark}\label{rmk_ho}
	{\normalfont  Both conditions in Proposition \ref{prp:cond_dir} force the grouping structure to be more adherent, \textit{a priori}, to the layer division, by making within-layer clusters more probable than across-layer ones. Such a generalized notion of homophily is sensible, since one would naturally expect that, in practice, it is more likely to share connectivity behavior (i.e., stochastic equivalence) within the layers than across layers. Condition \eqref{eq:elicit2} implies \eqref{eq:elicit1}, since $\{\bz_v \notin \bH_j^{-v}\}$ includes $z_v$ being assigned to a new sociability profile (i.e., a new cluster). For large enough networks (or in large enough layers), \eqref{eq:elicit2} is always attained, whereas for smaller networks (or in layers with few nodes), our results have the merit to suggest that this key property can be still enforced by tuning the prior parameters in such a way that \eqref{eq:elicit2} is satisfied.}
\end{Remark}

\section{Inference and Prediction}\label{sec_3}

The pEx-SBM allows for tractable posterior inference on $\bz$ given $\bY$, and prediction for the allocation and edges of future  nodes. In Section~\ref{sec_3.1} we accomplish the former via a tractable collapsed Gibbs sampler which exploits the urn scheme of the pExP prior in \eqref{eq:z}. Innovative predictive schemes, which leverage the projectivity properties and analytic results for the proposed pExP prior, are derived in Section~\ref{sec_3.2}.

\subsection{Collapsed Gibbs Sampler}\label{sec_3.1}
Unlike classical MCMC strategies for single-layer SBMs \citep[e.g.,][]{schmidt_2013}, our pExP prior in the pEx-SBM relies on a two-level grouping structure which first assigns nodes to layer-specific subgroups and then clusters these subgroups w.r.t.\ common sociability profiles that form the final grouping structure in $\bz$; see the generative process for the auxiliary node attributes $\bX$ within Section~\ref{sec_2.2.1}. To this end, rather than devising an MCMC scheme targeting the posterior $p(\bz \, | \, \bY)$, we propose a Gibbs sampler for the augmented posterior $p(\bz, \bw \, | \, \bY)$. This perspective facilitates the derivation of tractable full-conditional distributions $p(z_v, w_v \, | \, \bY, \bz^{-v}, \bw^{-v})$ for each node $v\in[V]$, while still allowing inference on  $p(\bz \, | \, \bY)$ by simply retaining only the samples for $\bz$. Neither slice-sampling steps nor truncations are required for inferring the total number of occupied groups, in contrast to, e.g., \citet{amini2019hierarchical}. Moreover, our Gibbs sampler covers the whole H-NRMI class.

By noting that \eqref{eq1} implies $(\bY\, | \,\bz)\indep\bw$, a direct application of the Bayes rule yields the following full-conditional distributions for the generic node $v$
\begin{equation}\label{eq:full1}
\begin{split}
	&\mathbb{P}(z_v=h,w_v=\uptau \, | \, \bY,\,\bz^{-v},\bw^{-v})\\
	&\propto \mathbb{P}(z_v=h,w_v=\uptau \, | \, \bz^{-v},\bw^{-v})\frac{p(\bY\, | \, z_v=h,\,\bz^{-v})}{p(\bY^{-v}\mid \bz^{-v})},
	\end{split}
\end{equation}
for any $h\in[H^{-v}+1]$ and $\uptau\in[\ell^{-v}_{j\cdot}+1]$, where $j$ is the layer of node $v$, while $\bY^{-v}$ denotes the $(V-1) \times (V-1)$ \textit{supra}-adjacency matrix without the rows and columns corresponding to node $v$. Recalling e.g.,  \citet{legramanti2022extended}, under the collapsed likelihood in \eqref{eq1} the term $p(\bY\, | \, z_v=h,\,\bz^{-v})/p(\bY^{-v}\, | \, \bz^{-v})$ in \eqref{eq:full1} can be evaluated in closed-form as
\begin{equation}
\begin{split}
	&\frac{p(\bY \, | \, z_v=h,\,\bz^{-v})}{p(\bY^{-v}\, | \, \bz^{-v})}\\
	&\qquad \quad =\prod_{h'=1}^{H^{-v}}\frac{\mbox{B}(a+m_{hh'}^{-v}+r_{vh'},b+\overline{m}_{hh'}^{-v}+\overline{r}_{vh'})}{\mbox{B}(a+m_{hh'}^{-v},b+\overline{m}_{hh'}^{-v})}, 
	\end{split}
	\label{eq_like}
\end{equation}
where $m_{hh'}^{-v}$ and $\overline{m}_{hh'}^{-v}$ are the number of edges and non-edges between  groups $h$ and $h'$, after excluding node $v$, while $r_{vh'}$ and $\overline{r}_{vh'}$ denote the number edges and non-edges among $v$ and the nodes in  $h'$. As for the prior factor, a closed-form and tractable expression for $\mathbb{P}(z_v=h,w_v=\uptau \, | \, \bz^{-v},\bw^{-v})$ in \eqref{eq:full1} is given in Proposition~\ref{prp:nuova}.

\begin{Proposition}\label{prp:nuova}
		Let $\bz$ denotes a random allocation vector such that $\bz\sim\mbox{\normalfont pExP}(\bV;\rho,\rho_0,c,c_0)$. Then for any node $v$ within layer $j$  the following results hold.
		
 \text{(1)} If $h\in\bH_j^{-v}$ and $\uptau=\uptau^{-v}_{jht}$ with $t\in[\ell_{jh}^{-v}]$:
		\begin{equation}\label{eq:jointpred1}
		\begin{split}
			&\mathbb{P}(z_v=h,w_v=\uptau^{-v}_{jht} \, | \, \bz^{-v},\bw^{-v})\\
			&\qquad =\frac{\Phi_{\ell_{j\cdot}^{-v},j}^{(V_j)}(\bm{q}_{j1}^{-v},\dots,\bm{q}^{-v}_{jh}+\e_t,\dots,\bm{q}^{-v}_{jH^{-v}})}{\Phi_{\ell^{-v}_{j\cdot},j}^{(V_{j}-1)}(\bm{q}^{-v}_{j1},\dots,\bm{q}^{-v}_{jH^{-v}})},
			\end{split}
		\end{equation}
		where $\uptau^{-v}_{jht}$ is the $t$-th entry of the vector $\bm{\uptau}_{jh}^{-v}$ obtained by ordering the set $\mathbbm{T}^{-v}_{jh}$ in \eqref{eq:taus}. 
		
 \text{(2)}	If $\uptau=\ell_{j\cdot}^{-v}+1$ and $h\in[H^{-v}+1]$:
		\begin{equation}\label{eq:jointpred2}
		\begin{split}
			&\mathbb{P}(z_v=h,w_v=\ell_{j\cdot}^{-v}+1 \,|\, \bz^{-v},\bw^{-v})\\
			&\quad =\frac{\Phi_{H_h^{-v},0}^{(|\bell^{-v}|+1)}(\ell_{\cdot 1}^{-v},\dots,\ell^{-v}_{\cdot h}+1,\dots,\ell^{-v}_{\cdot H_h^{-v}})}{\Phi_{H^{-v},0}^{(|\bell^{-v}|)}(\ell^{-v}_{\cdot 1},\dots,\ell^{-v}_{\cdot H^{-v}})}\\
			& \qquad \quad \times \frac{\Phi_{\ell_{j\cdot}^{-v}+1,j}^{(V_j)}(\q_{j1}^{-v},\dots,(\q^{-v}_{jh},1),\dots,\q^{-v}_{jH_h^{-v}})}{\Phi_{\ell^{-v}_{j\cdot},j}^{(V_{j}-1)}(\bm{q}^{-v}_{j1},\dots,\bm{q}^{-v}_{jH^{-v}})},
			\end{split}
		\end{equation}
		where $H^{-v}_h=H^{-v}\vee h$.
		
	\text{(3)} For any other choice of $(h,\uptau)\in[H^{-v}+1]\times[\ell_{j\cdot}^{-v}+1]$ the probability is $0$.
	\end{Proposition}

	Corollary~\ref{exe:jpred_hdp}  covers the H-DP case, whereas expressions for the H-NSP case are provided in the supplementary materials.

\begin{Corollary}\label{exe:jpred_hdp}
		If $\bz\sim\text{\normalfont pExP}(\bV;\theta,\theta_0)$, then, for any generic node $v$ in layer $j$, we have
		\begin{equation}
			\begin{split}
				\label{eq:jpred_hdp}
				&\mathbb{P}(z_v=h,w_v=\uptau\,|\, \bz^{-v},\bw^{-v})=\indic_{\left\{\uptau=\uptau^{-v}_{jht}\right\}}\frac{q_{jht}^{-v}}{\theta+V_j-1}\\
				&\ +\indic_{\left\{\uptau=\ell_{j\cdot}^{-v}+1\right\}}\frac{\theta}{\theta+V_j-1}\left[\frac{\ell_{\cdot h}^{-v}}{\theta_0+|\bell^{-v}|}+\indic_{\left\{h=H^{-v}+1\right\}}\frac{\theta_0}{\theta_0+|\bell^{-v}|}\right],
			\end{split}
		\end{equation}
 for any $h\in[H^{-v}+1]$, $\uptau\in[\ell^{-v}_{j\cdot}+1]$, where $\bm{\uptau}^{-v}_{jh}$ is defined as in Proposition \ref{prp:nuova}.
\end{Corollary}

\begin{algorithm}[b!]
	\caption{Gibbs sampler for pEx-SBM}\label{alg:gibbs}
	\begin{algorithmic}
		\State Initialize coherently $\bz$ and $\bw$ \Comment{Default: $V$ different sociability profiles and subgroups (one for each node)}
		\For{$s=1, \ldots, n_{\text{iter}}$ ($n_{\text{iter}}$: number of Gibbs iterations)}
		\For{$v=1 \ldots, V$ ($V$: number of nodes in the network)}
		\State 1. $j\gets$ layer of node $v$  
		\State 2. remove node $v$
		\State \hspace{3mm} 2.1 reorder the labels in $\bz^{-v}$ and $\bw^{-v}$ so that only 
		\State \hspace{8mm} sociability profiles $h\in[H^{-v}]$  and subgroups
		\State \hspace{8mm} $\uptau\in[\ell^{-v}_{j\cdot}]$ in layer $j$ are non-empty	
		\State \hspace{3mm} 2.2 compute $\bell^{-v}, \bm{q}^{-v}, {\bf m}^{-v},\overline{\bf m}^{-v},{\bf r}_v,\bar{{\bf r}}_v$
		\For{$h\in[H^{-v}+1]$ and $\uptau\in[\ell^{-v}_{j\cdot}+1]$}
		\State 3. compute $\mathbb{P}(z_v=h,w_v=\uptau \,|\, \bY,\,\bz^{-v},\bw^{-v})$  (up to
		\State \hspace{3mm} a proportionality constant) via \eqref{eq:full1}--\eqref{eq:jointpred2}
		\EndFor
		\State 4. sample $(z_v,w_v)$ from the bivariate discrete random
		\State  \hspace{3mm} variable with probabilities obtained by normalizing
		\State  \hspace{3mm}   those computed in step 3. 
		\State 5. update ${\bf m},\,\bell,\,\bm{q}$ based on the new allocation vectors 
		\State  \hspace{3mm}  $(z_v,\bz^{-v})$ and $(w_v,\bw^{-v})$
		\EndFor
		\State 6. save the sampled vector $\bz^{(s)}$ and $\bw^{(s)}$
		\EndFor
	\end{algorithmic}
\end{algorithm}    

Another key result, which can be deduced from \eqref{eq:jointpred1}--\eqref{eq:jointpred2}, is the following.

\begin{Corollary}\label{cor:suff}
	Let $\bz$ be a random allocation vector distributed as in \eqref{eq:z}, and  indicate by $\bw$ the corresponding subgroup allocation vector. Then $(\bell,\bm{q})$ is a predictive sufficient statistics for $(\bz,\bw)$, i.e.,
\begin{equation*}
\mathbb{P}(z_v=h,w_v=\uptau \, | \, \bz^{-v},\bw^{-v})=\mathbb{P}(z_v=h,w_v=\uptau  \, | \, \bell^{-v},\bm{q}^{-v})
\end{equation*}	
 for any  $h\in[H^{-v}+1]$, $\uptau\in[\ell^{-v}_{j\cdot}+1]$, and node $v\in[V]$ in layer $j\in[d]$.
\end{Corollary}

\begin{Remark}
	\normalfont  
	The result in Corollary~\ref{cor:suff} can be further refined in the H-DP case by stating that the vector of the column-wise sums $(\ell_{\cdot 1},\dots,\ell_{\cdot H^{-v}})$ of the matrix $\bell$, i.e., the frequencies of subgroups assigned to each already-observed sociability profile in the whole network, and the array $\bm{q}_j$, i.e., the frequencies of nodes assigned to each already-observed subgroup for each already-observed sociability profile in layer $j$, are predictive sufficient statistics.
\end{Remark}

Corollary~\ref{cor:suff}, combined with \eqref{eq:full1}--\eqref{eq:jointpred2}, facilitates the implementation of a  collapsed Gibbs sampler, which iteratively simulates values from the full-conditional allocation probabilities $p(z_v, w_v \,|\, \bY, \bz^{-v}, \bw^{-v})$  of each node $v\in[V]$ via simple updating of the quantities $(\bell, \bm{q})$ for the prior component in \eqref{eq:jointpred1}--\eqref{eq:jointpred2}, and $({\bf m}^{-v},\overline{\bf m}^{-v},{\bf r}_v,\bar{{\bf r}}_v)$ for the collapsed likelihood  in \eqref{eq_like}. The latter quantities denote, respectively, the two matrices counting the edges and non-edges between groups, after excluding node $v$, and the vectors  with the number of edges and non-edges between $v$ and the nodes in the different groups. The pseudo-code for the proposed computational strategy is outlined in Algorithm~\ref{alg:gibbs}.

Notice that, although Step 3 of Algorithm~\ref{alg:gibbs} requires the computation of an $(H^{-v}+1) \times ( \ell^{-v}_{j\cdot}+1)$-dimensional probability table, this matrix is highly sparse. In fact, recalling Proposition~\ref{prp:nuova}, the probability of being assigned $h$ at subgroup $\uptau$ is non-zero only if $\uptau$ is a previously-occupied subgroup with profile $h$ or if $\uptau$ is a new subgroup. Thus, each column of the probability table, except the last, has a single non-zero entry.

Leveraging on the samples of $\bz$ produced by Algorithm~\ref{alg:gibbs}, we conduct posterior inference on the nodes' grouping structures via the variation of information (VI) approach set forth in \citet{wade2018} for Bayesian clustering.  The VI defines distances among partitions through a comparison of individual and joint entropies \citep{meilua2007comparing}, thus providing a metric which allows for point estimation and uncertainty quantification directly in the space of partitions. In this framework, a point estimate for $\bz$ is defined as $\hat{\bz}= \mbox{argmin}_{\bz'} \cmean{\mbox{VI}(\bz,\bz')}{\bY}$, whereas a $(1-\alpha)$-credible ball around $\hat{\bz}$ is obtained by collecting all partitions with VI distance from $\hat{\bz}$ less than a pre-specified threshold defined to ensure the ball to contain at least  $1\!-\!\alpha$ posterior mass, while being of minimum size. These tasks can be performed via the \texttt{R} library \texttt{mcclust.ext} \citep{wade2018}, which requires the calculation of the $V \times V$ \textit{posterior similarity} (or \textit{co-clustering}) matrix $\bC$, whose generic entry $c_{vu}$ yields an estimate of $\mathbb{P}(z_v=z_u \, |\, \bY)$ by computing the relative frequency of MCMC samples in which $z_v^{(s)}=z_u^{(s)}$.

Model assessment and comparison is performed by leveraging the WAIC criterion, which has  also  a direct connection with Bayesian leave-one-out cross-validation \citep[][]{watanabe2010asymptotic,gelman2014understanding}. This criterion is further useful to select, in a data-driven manner, specific priors in the H-NRMI class and tune the corresponding parameters. Alternatively, prior elicitation can proceed by studying the expected number of non-empty clusters and the growth of $H$ as a function of the network size. For node-colored networks with  $V_1=\cdots V_d=V/d$, such a growth is $\mathcal{O}(\log \log V)$ under $\mbox{H-DP}(\theta,\theta_0)$ and $\mathcal{O}(V^{\sigma \sigma_0})$ for $\mbox{H-NSP}(\sigma, \sigma_0)$ \citep[e.g.,][]{camerlenghi2019distribution}. Instead, within each layer $j$, this growth is $\mathcal{O}(\log V_j)$ and $\mathcal{O}(V_j^\sigma)$, respectively. A final option is to specify hyperpriors for the parameters of the selected H-NRMI. This requires simply adding a step in  Algorithm~\ref{alg:gibbs} to sample from the full-conditionals of these parameters. As clarified in the supplementary materials, in the H-DP case with gamma hyperpriors for $\theta$ and $\theta_0$, closed-form and tractable full conditionals can be straightforwardly derived by extending results of \citet{escobar1995bayesian} from the DP to the H-DP.

\subsection{$k$-Step-Ahead Prediction}\label{sec_3.2}
The overarching focus of predictive strategies for SBMs has been on prediction of edges among the observed nodes, or forecasting the group allocations for new incoming nodes given $\bY$ and the observed connections between these new  nodes and the previously observed ones \citep[see, e.g.,][]{lee2019review}. These strategies fail to address the more challenging and realistic problem of predicting the allocations of new incoming nodes, conditioned only on $\bY$, and, most importantly, the edges among these new nodes and the previously-observed ones. In fact, in practice one expects that for  $k$ new incoming nodes only the corresponding layers are observed, and no information is available on the associated edges and group allocations. By inheriting the \textit{Kolmogorov consistency} from the pExP prior, the  pEx-SBM opens the avenues for addressing these predictive tasks through innovative methods that are not yet available in current literature. In particular, the sequential construction of the nonparametric priors employed in pEx-SBM allows for principled prediction of an arbitrary number $k$ of new nodes, from any layer.

Let us direct our attention to predicting the allocation $\bz_{\text{new}}=(z_{V+1}, \ldots, z_{V+k})$ and suballocation $\bw_{\text{new}}=(w_{V+1}, \ldots, w_{V+k})$ vectors for the $k$ incoming nodes conditioned only on the network $\bY$ among the previously-observed nodes.  Since the new allocations are conditionally independent from $\bY$, given $\bz$ and $\bw$, for any $v_{\text{new}} =V+1,\dots,V+k$ and $u=1, \ldots, v_{\text{new}}-1$ we have
	\begin{equation}\label{eq:prediction}
	\begin{split}
		\mathbb{P}(z_{v_{\text{new}}} =z_u \, | \, \bY)&= \int\mathbb{P}(z_{v_{\text{new}}} =z_u \, | \,  \bY, \bz, \bw)\,\mathcal{L}_{\bY}(\ddr\bz,\ddr\bw)\\
		&= \int\mathbb{P}(z_{v_{\text{new}}} =z_u \, | \,  \bz, \bw)\,\mathcal{L}_{\bY}(\ddr\bz,\ddr\bw),
		\end{split}
	\end{equation}
where $\mathcal{L}_{\bY}$ denotes the joint posterior law of $(\bz,\bw)$.

By combining \eqref{eq:prediction} with the results in Proposition~\ref{prp:pred_hnrmi} and Theorem~\ref{coclust_hnrmi}, $\mathbb{P}(z_{v_{\text{new}}} =z_u \, | \, \bY)$ can be estimated via Monte Carlo as 
\begin{equation*}
\hat{\mathbb{P}}(z_{v_{\text{new}}} =z_u \, | \, \bY)=\frac{1}{n_{\text{tot}}}\sum_{s=n_{\text{burn}}+1}^{n_{\text{iter}}}\mathbb{P}(z_{v_{\text{new}}} =z_u \, | \,  \bz^{(s)}, \bw^{(s)}),
\end{equation*}
where $n_{\text{burn}}$ corresponds to the number of discarded burn-in samples, and $n_{\text{tot}}=n_{\text{iter}}-n_{\text{burn}}$ is the number of subsequent MCMC draws from Algorithm~\ref{alg:gibbs} used for averaging.  If $u$ is a new node, $\mathbb{P}(z_{v_{\text{new}}} =z_u  \, | \,  \bz^{(s)}, \bw^{(s)})$ can be directly evaluated by Theorem~\ref{coclust_hnrmi}. Conversely, when $u$ denotes a previously-observed node, then $\mathbb{P}(z_{v_{\text{new}}} =z_u  \, | \,  \bz^{(s)}, \bw^{(s)})=\mathbb{P}(z_{v_{\text{new}}} =h  \, | \,  \bz^{-u(s)}, z^{(s)}_u=h, \bw^{(s)})$, which is computed via Proposition~\ref{prp:pred_hnrmi}.

The above procedure yields a Monte Carlo estimate of predictive co-clustering probabilities among the $k$  new incoming nodes as well as between such  nodes and the previously-observed ones. Combing these with the quantities computed for the in-sample nodes from the MCMC yields an augmented $(V+k) \times (V+k)$ posterior similarity matrix $\bC_{\text{aug}}$ which quantifies uncertainty for both in-sample and predictive co-clustering probabilities. A VI point estimate for $\bz_{\text{new}}$ can therefore be obtained via the  \texttt{R} library \texttt{mcclust.ext}. In contrast to other approaches  \citep[e.g.,][]{legramanti2022extended}, knowledge on the edges of the new incoming nodes is not required. Only layer information is employed. 

Let us now focus on predicting the edges $y_{v_{\text{new}},u}$ for these new incoming nodes. More specifically, the goal is to derive the predictive  $p(\bY_{\text{new}}\, | \, \bY)$, where $\bY_{\text{new}}$ comprises the unobserved edges $y_{v_{\text{new}},u}$ with $v_{\text{new}} =V+1,\dots,V+k$ and $u=1, \ldots, v_{\text{new}}-1$. 
	Let $\overline{\bz}:=(\bz,\bz_{\text{new}})$ be the complete allocation vector. Then 
	\begin{equation}\label{eq:pred3}
		p(\bY_{\text{new}}\, | \,  \bY)=\int p(\bY_{\text{new}}\, | \,  \bY, \overline{\bz})\,\mathcal{L}_{\bY}(\ddr \overline{\bz}),
	\end{equation}	
	with $\mathcal{L}_{\bY}$ the posterior law of $\overline{\bz}$. 
	
	Notice that $p(\bY \, | \,  \overline{\bz})=p(\bY \!\mid \bz)$, since $(\bY\, | \,  \bz) \indep \bz_{\text{new}}$. Hence, by Bayes' rule and \eqref{eq1}, $p(\bY_{\text{new}}\, | \,  \bY, \overline{\bz})$ can be expressed as
\begin{equation}\label{eq:pred31}
	\begin{split}
		& p(\bY_{\text{new}}\, | \,  \bY, \overline{\bz})=\frac{p(\bY_{\text{new}}, \bY\, | \,  \overline{\bz})}{p(\bY \, | \,  \bz)}\\
		&=\prod_{h=1}^H \prod_{h'=1}^h \frac{\mbox{B}(a,b)\mbox{B}(a+m_{hh'}+m^{\text{new}}_{hh'},b+\overline{m}_{hh'}+\overline{m}^{\text{new}}_{hh')}}{\mbox{B}(a+m_{hh'},b+\overline{m}_{hh'})\mbox{B}(a,b)}\\
		&\qquad \quad \times \prod_{h=H+1}^{H+H_k} \prod_{h'=1}^h \frac{\mbox{B}(a+m^{\text{new}}_{hh'},b+\overline{m}^{\text{new}}_{hh'})}{\mbox{B}(a,b)}\\
		&=\prod_{h=1}^H \prod_{h'=1}^h \frac{\mbox{B}(a_{hh',\bY}+m^{\text{new}}_{hh'},b_{hh',\bY}+\overline{m}^{\text{new}}_{hh')}}{\mbox{B}(a_{hh',\bY},b_{hh',\bY})}\\
		&\qquad \quad \times \prod_{h=H+1}^{H+H_k} \prod_{h'=1}^h \frac{\mbox{B}(a+m^{\text{new}}_{hh'},b+\overline{m}^{\text{new}}_{hh'})}{\mbox{B}(a,b)},
	\end{split}
\end{equation}
where $a_{hh',\bY}=a+m_{hh'}$ and $b_{hh',\bY}=b+\overline{m}_{hh'}$, while $m^{\text{new}}_{hh'}$ and $\overline{m}^{\text{new}}_{hh'}$ are the number of edges and non-edges among groups $h$ and $h'$ that involve at least a new node (if any) allocated to these clusters. In \eqref{eq:pred31}, the first factor corresponds to groups already occupied by the in-sample nodes, while the second refers to the $H_k$ potential new clusters created by the $k$ new incoming nodes. 

The combination of \eqref{eq:pred3}--\eqref{eq:pred31} allows to evaluate the predictive probability of any configuration $\bY_{\text{new}}$ via Monte Carlo as 
\begin{equation*}
\hat{p}(\bY_{\text{new}}\, | \,  \bY)=\frac{1}{n_{\text{tot}}}\sum_{s=n_{\text{burn}}+1}^{n_{\text{iter}}}p(\bY_{\text{new}}\!\mid \bY, \overline{\bz}^{(s)}),
\end{equation*}
where $p(\bY_{\text{new}}\, | \, \bY, \overline{\bz}^{(s)})$ is computed by evaluating \eqref{eq:pred31}  at the MCMC samples of  $\bz^{(s)}$ and $\bz^{(s)}_{\text{new}}$ from $p(\overline{\bz}\, | \, \bY)$. To obtain $\overline{\bz}^{(s)}$, note that $p(\overline{\bz}\, | \, \bY)=p(\bz_{\text{new}} \, | \, \bz, \bY)p(\bz \, | \,  \bY)$, where, by the pEx-SBM formulation, $p(\bz_{\text{new}} \, | \,  \bz, \bY)=p(\bz_{\text{new}} \, | \, \bz)$, and the samples $\bz^{(s)}$ from $p(\bz \, | \, \bY)$ are already available from the output of Algorithm~\ref{alg:gibbs}. Therefore, for each $\bz^{(s)}$ it suffices to simulate $\bz^{(s)}_{\text{new}}$ from $p(\bz_{\text{new}} \, | \,  \bz^{(s)})$ or, alternatively, from the augmented posterior, i.e.,  $p(\bz_{\text{new}}, \bw_{\text{new}} \, | \, \bz^{(s)}, \bw^{(s)})$, similarly to Algorithm~\ref{alg:gibbs}. In view of our construction, the entries in $\bz^{(s)}_{\text{new}}$ can be coherently generated in a sequential manner. This is achieved via the joint urn scheme in \eqref{eq:jointpred1} and \eqref{eq:jointpred2} from $p(z_{v_\text{new}}, w_{v_\text{new}} \, | \, \bz^{(s)}, \bw^{(s)},\bz_{v_\text{new}-1}^{(s)}, \bw_{v_\text{new}-1}^{(s)})$ for $v_{\text{new}} =V+1,\dots,V+k$. In the conditioning argument, the quantities  $\bz_{v_\text{new}-1}^{(s)}$ and $\bw_{v_\text{new}-1}^{(s)}$ are the  simulated group and subgroup allocations for the new incoming nodes $V+1, \ldots, v_\text{new}-1$. 

The above strategy allows to evaluate  the predictive probabilities for any possible configuration of new edges and non-edges within $\bY_{\text{new}}$. However, in practice, the set of all configurations is excessively large and difficult to summarize in a meaningful manner. Therefore, it is often convenient to complement a joint analysis with the predictive summaries for each new edge $y_{v_{\text{new}},u}$ with $v_{\text{new}} =V+1,\dots,V+k$ and $u=1, \ldots, v_{\text{new}}-1$, namely $\mathbb{P}(y_{v_{\text{new}},u}=1 \mid \bY)=1-\mathbb{P}(y_{v_{\text{new}},u}=0 \mid \bY)$. 

Let $\overline{z}_v:=\overline{z}_{v_{\text{new}}}$ be the allocation of $v_{\text{new}}$. Moreover, denote by $\mathcal{L}_{\bY}$ the posterior law of $\overline{\bz}$, and by  $\mathcal{L}_{\bY,\overline{\bz}}$ the conditional law of the block-probability $\psi_{\overline{z}_v,\overline{z}_u}$ between clusters $\overline{z}_v$ and $\overline{z}_u$ given the  network $\bY$ and all allocations $\overline{\bz}$. Then, $\mathbb{P}(y_{v_{\text{new}},u}=1 \, | \, \bY)$ can be equivalently expressed as
	\begin{equation}
	\begin{split}\label{eq:pred4}
		&\mathbb{P}(y_{v_{\text{new}},u}=1 \,|\, \bY)\\
		&\ =\int\int \cmean{y_{v_{\text{new}},u}}{\bY,\,\overline{\bz},\,\psi_{\overline{z}_v,\overline{z}_u}}\,\mathcal{L}_{\bY,\overline{\bz}}(\ddr \psi_{\overline{z}_{v},\overline{z}_{u}})\,\mathcal{L}_{\bY}(\ddr\overline{\bz})\\
		&\ =\int\frac{a+m_{\overline{z}_{v},\overline{z}_{u}}}{a+b+m_{\overline{z}_{v},\overline{z}_{u}}+
			\overline{m}_{\overline{z}_{v},\overline{z}_{u}}}\,\mathcal{L}_{\bY}(\ddr\overline{\bz}),
			\end{split}
	\end{equation}
after noticing that, due to beta-binomial conjugacy, we have 
\begin{equation*}
(\psi_{\overline{z}_v,\overline{z}_u}\,| \, \bY,\bz, \bz_{\text{new}}) \sim \mbox{beta}(a+m_{\overline{z}_{v},\overline{z}_{u}},b+\overline{m}_{\overline{z}_{v},\overline{z}_{u}})
\end{equation*}
where $m_{\overline{z}_{v},\overline{z}_{u}}$ and $\overline{m}_{\overline{z}_{v},\overline{z}_{u}}$ denote the total number of edges and non-edges between pairs of in-sample nodes allocated to $\overline{z}_v$ and $\overline{z}_u$. Therefore the last integral within \eqref{eq:pred4} can be evaluated again via Monte Carlo leveraging the samples $(\bz^{(s)},\bz^{(s)}_{\text{new}})$ generated to compute \eqref{eq:pred3}. This yields the estimate of $\mathbb{P}(y_{v_{\text{new}},u}=1 \, | \, \bY)$, for each $v_{\text{new}} =V+1,\dots,V+k$ and $u=1, \ldots, v_{\text{new}}-1$, defined as 
\begin{equation*}
\hat{\mathbb{P}}(y_{v_{\text{new}},u}=1 \, | \, \bY)=\frac{1}{n_{\text{tot}}}\sum_{s=n_{\text{burn}}+1}^{n_{\text{iter}}}\frac{(a+m^{(s)}_{\overline{z}_{v},\overline{z}_{u}})}{(a+m^{(s)}_{\overline{z}_{v},\overline{z}_{u}}+b+\overline{m}^{(s)}_{\overline{z}_{v},\overline{z}_{u}})},
\end{equation*}
where $m^{(s)}_{\overline{z}_{v},\overline{z}_{u}}$ and $\overline{m}^{(s)}_{\overline{z}_{v},\overline{z}_{u}}$ coincide with $m_{\overline{z}_{v},\overline{z}_{u}}$ and $\overline{m}_{\overline{z}_{v},\overline{z}_{u}}$ evaluated at  $(\bz^{(s)},\bz^{(s)}_{\text{new}})$.

\section{Simulation Studies}\label{sec_4}

\begin{figure}[b!]
\centering
	\includegraphics[trim=0cm 2.7cm 0cm 0.3cm,clip,width=0.5\textwidth]{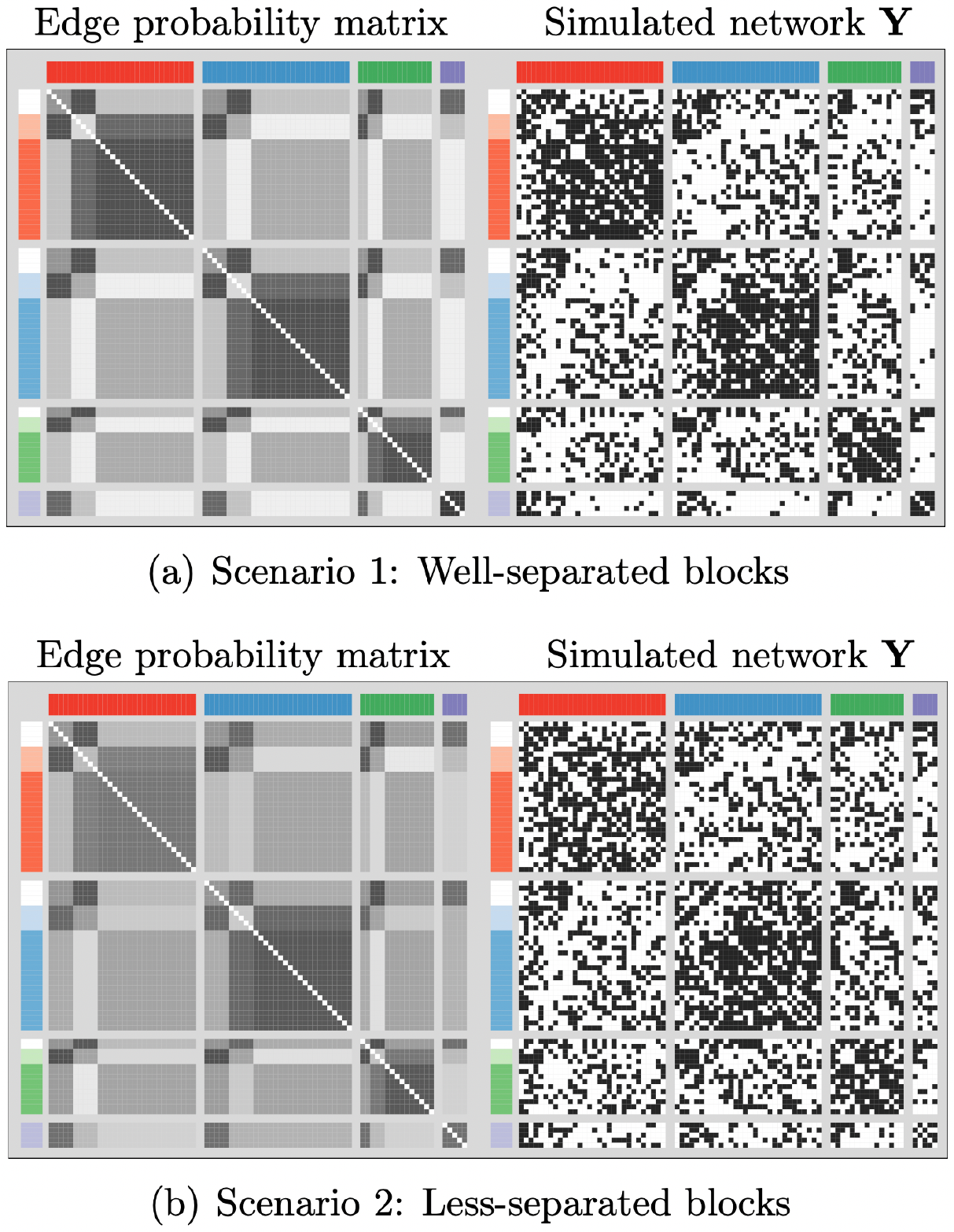}
	\vspace{1pt}
	\caption{\footnotesize For scenario 1 (panel (a)) and 2 (panel (b)), graphical representation of the true underlying edge probability matrix (left), and of one example of  \textit{supra}-adjacency matrix simulated from this edge probability matrix (right). The color annotation of columns refers to layers' division, while  the row colors display the true underlying grouping structure. In the edge probability matrices, the color of each entry  ranges from white to black as the corresponding edge probability goes from 0 to 1. In the \textit{supra}-adjacency matrices, the black and white entries indicate edges and non-edges, respectively. }
	\label{figure:2}
\end{figure}

To assess the performance of the proposed pEx-SBM and illustrate its merits compared to both state-of-the-art \citep{zhang2016community,binkiewicz2017covariate,legramanti2022extended} and routinely-implemented \citep{blondel2008fast,come2021hierarchical} competitors, we simulate complex binary undirected networks among $V=80$ nodes divided in $d=4$ layers, of size $V_1=V_2=30$, $V_3=15$ and $V_4=5$, whose block interactions mimic those expected for the criminal network in Section~\ref{sec_5}. As shown in Figure~\ref{figure:2}, the first three layers comprise two within-layer groups, and one across-layer cluster, identified by the white color, which contains nodes from layers 1, 2 and 3. Recalling the criminal network in Figure~\ref{figure:1}, the larger groups in each layer can be thought of as \textit{locale}-specific affiliates who mainly interact with peers in the same cluster and with an additional low-sized group of higher-level supervisors that administer activities in each  \textit{locale} and report  to an across-layer group of bosses from the different \textit{locali}. These individuals are the only ones entitled to establish dense connections with nodes in the fourth layer, which consists of a single group of bosses at the top of the criminal organization that connect with few, yet central, \textit{locale}-specific bosses. As illustrated in Figure~\ref{figure:2} the edge probabilities among the nodes in these $H_0=8$ groups are defined to be consistent with this pyramidal block-connectivity architecture, while accounting for two different scenarios in which separation among blocks is either more (Scenario 1) or less (Scenario 2) pronounced.

Based on the above probability matrices, we simulate edges in the \textit{supra}-adjacency matrix $\bY$ from independent Bernoulli variables for each scenario (see Figure~\ref{figure:2}), and then  perform the analysis under three key examples of pEx-SBMs in ten replicated experiments relying on different $\bY$ simulated from the edge probability matrices in Figure~\ref{figure:2}. Computation and inference for both scenarios resort to the algorithms and methods in Section~\ref{sec_3.1}, for three pEx-SBMs with the customary $\mbox{beta}(1,1)$ (i.e., uniform) prior for the block-probabilities and pExP priors on $\bz$ induced by an $\mbox{H-DP}(\theta=0.5, \theta_0=4)$, an $\mbox{H-NSP}(\sigma=0.2,\sigma_0=0.8)$, and an H-DP with $\mbox{gamma}(5,10)$ and $\mbox{gamma}(12,3)$ hyperpriors for $\theta$ and $\theta_0$, respectively. These settings are useful for assessing sensitivity and robustness with respect to misscalibrated priors. In fact, these choices enforce a prior expected number of clusters of  $\approx 5$ for all the three pEx-SBM examples, lower than the true $H_0=8$. Posterior inference relies on $8{,}000$ samples of $\bz$ from Algorithm~\ref{alg:gibbs}, after a conservative burn-in of $2{,}000$. As confirmed by the traceplots for the logarithm of the likelihood in \eqref{eq1} (see the supplementary materials), in practice, a few MCMC iterations suffice for convergence of the collapsed Gibbs samplers. A basic \texttt{R} implementation of Algorithm~\ref{alg:gibbs} requires $\approx$ 1.5 minutes to simulate $10{,}000$ values of $\bz$ on a MacBook Air (M1, 2020), CPU 8-core and 8GB RAM.

\begin{table*}[t]
	\renewcommand{\arraystretch}{1.3}
	\caption{\footnotesize{Performance comparison between three key examples of pEx-SBMs and relevant competitors for Scenarios~1--2 under several measures.}}
	\label{table_runtimes_MC}
	\begin{adjustbox}{width=1\textwidth,center=\textwidth}
	\setlength{\tabcolsep}{12pt}
		\begin{tabular}[c]{lcc|ccccccccc}
			& \multicolumn{2}{c}{$\mbox{VI}(\hat{\bz},\bz_0)$ ($\hat{H}$)} & \multicolumn{2}{c}{$\mbox{med}(H \mid \bY)$}  &  \multicolumn{2}{c}{$\mathbbm{E}[\mbox{VI}(\bz,\bz_0) \mid \bY]$}  & \multicolumn{2}{c}{$\mbox{VI}(\hat{\bz},\bz_b)$} &   \multicolumn{2}{c}{WAIC}    \\ 
			\midrule
			\textsc{Scenario} & 1 & 2 & 1 & 2 & 1 & 2&1&2&1& 2\\ 
			\midrule
			pEx-SBM (H-DP) &  0.00 (8.0)   &   0.45 (7.1) & 8.0 {\scriptsize[0.1]} &    7.5 {\scriptsize[0.7]}&   0.02 & 0.62 & 0.12 &  0.82&   3508 &  4046\\
			pEx-SBM (H-NSP) &  0.00 (8.0) &   0.39 (7.3) &  8.2 {\scriptsize[0.9]} &   8.2 {\scriptsize[1.5]}&  0.04 &  0.60 & 0.17 &  0.84 &  3510 & 4049\\
			pEx-SBM (H-DP hyp) &   0.00 (8.0) &   0.43 (7.0) &  8.0 {\scriptsize[0.2]} &   7.6 {\scriptsize[1.0]}&  0.03 &  0.63 &   0.14 &   0.85 &   3509 &4045\\
			\midrule
			Supervised ESBM (DP)&  0.02 (8.1)& 0.74 (6.3) &  8.1 \scriptsize[0.7] & 7.2 \scriptsize[1.6] & 0.06 & 0.98 & 0.18 & 1.14& 3512&4082\\
			Louvain & 1.21 (3.9)& 2.21 (3.8) & --- & --- & --- & --- & --- & --- & --- & --- \\
			SBM (\texttt{greed}) &0.06 (7.5)& 2.18 (2.6) & ---  & ---  & --- & --- & ---  & --- & --- & --- \\
			JCDC ($w_n=5$) &1.54 (7.7)&2.41 (8.0) & --- &  --- & --- & --- & --- & --- & --- & --- \\
			JCDC ($w_n=1.5$) &1.45 (7.1)&1.58 (7.0) & --- &  --- & --- &  --- & --- & ---  & --- & --- \\
			CASC (\texttt{rCASC}) & 0.83 (8.0)& 1.49 (8.0) & --- & --- & --- & --- & --- & --- & --- & --- \\
			\hline
 \vspace{0.1pt}
		\end{tabular}
	\end{adjustbox}
{\noindent \footnotesize{Note:} Performance is assessed under the following measures:  VI distance $\mbox{VI}(\hat{\bz},\bz_0)$ of the estimated $\hat{\bz}$ from the truth $\bz_0$ (number of unique clusters in $\hat{\bz}$ inside brackets), posterior median of the number of groups $H$ (interquartile range inside brackets), posterior mean $\mathbbm{E}[\mbox{VI}(\bz,\bz_0) \mid \bY]$ of the VI distance from the true $\bz_0$,  distance $\mbox{VI}(\hat{\bz},\bz_b)$ between the estimated partition $\hat{\bz}$ and the $95\%$ credible bound $\bz_b$, WAIC. Only pEx-SBMs and Supervised ESBM rely on a Bayesian approach and, hence, posterior quantities and WAIC are only available for them. For JCDC  and SBM (\texttt{greed}) a favorable implementation is considered with routines' initializations at the true number of groups $H_0=8$. For CASC, the number of clusters is set exactly equal to $H_0=8$. The results are averaged over 10 replicated experiments. Lower VI distances and WAIC indicate more accurate performance. \par}
 \end{table*}

\begin{figure*}[b!]
	\centering
	\includegraphics[trim=0cm 0cm 0cm 0cm,clip,width=1.02\textwidth]{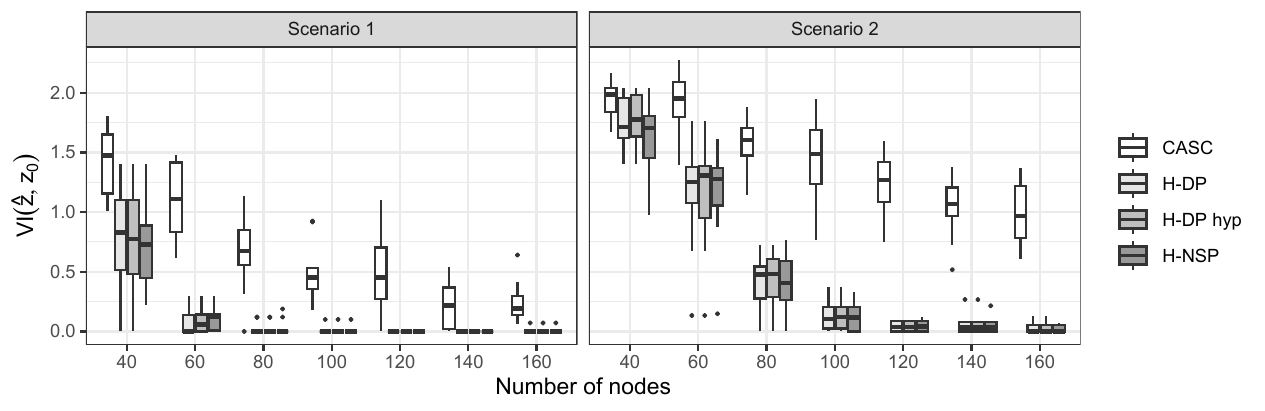}
	\caption{\footnotesize{Empirical assessment of frequentist posterior consistency for H-DP, H-DP hyperprior, and H-NSP monitored via the boxplots for $\mbox{VI}(\hat{\bz},\bz_0)$ from 10 replicated experiments on each network size from $V=40$ to $V=160$. As benchmark we consider CASC \citep{binkiewicz2017covariate} for which consistency in estimating $\bz_0$ has been proved theoretically. }
}
\label{figure:const}
\end{figure*}

As shown in the first three lines of Table~\ref{table_runtimes_MC}, all priors within the proposed pEx-SBM class learn the true underlying block structures in both Scenario 1 and 2 with a high accuracy, across replicated experiments. In Scenario 2, the reduced group separation leads, as expected, to a slight performance deterioration. However, even in this challenging context, pEx-SBMs are still accurate and systematically outperform relevant Bayesian and non-Bayesian competitors (see Table~\ref{table_runtimes_MC}). These results clarify that, regardless of the specific pExP prior considered, as long as such a prior incorporates the most suitable notion of exchangeability for the multilayer network analyzed, remarkable gains can be achieved over competitors, further motivating our broad focus on the general H-NRMI class.

The improvements w.r.t.\ the Louvain algorithm for community detection \citep{blondel2008fast} and   the \texttt{greed} implementation by \citet{come2021hierarchical} of classical single-layer SBM highlights the importance of including layer information for the analysis of node-colored networks. This can be included via state-of-the-art attribute-assisted methods in \citet{zhang2016community} (JCDC, setting either $w_n=5$ or $w_n=1.5$), \citet{binkiewicz2017covariate} (CASC leveraging the optimized parameter tuning in the \texttt{rCASC} library and a favorable implementation setting the number of clusters exactly equal to $H_0=8$) and \citet{legramanti2022extended} (supervised ESBM with a DP prior inducing the same expected number of groups as for pEx-SBM). As illustrated in Table~\ref{table_runtimes_MC}, both JCDC and CASC are not competitive with pEx-SBM. The former mainly searches for assortative blocks, and therefore, lacks sufficient flexibility, whereas the latter  inherits the difficulties of the underlying spectral clustering along with potential challenges in finding a proper balance between the exogenous layer division and the endogenous grouping structures in the network. The ESBM improves over JCDC and CASC, but it is still not competitive with pEx-SBM. Recall that ESBM employs exchangeable priors for $\bz$ supervised by layers in a way that does not yield a projective  formulation and, hence, cannot induce prediction rules as those derived for pEx-SBMs. 

The empirical evidence in Figure~\ref{figure:const} suggests that pEx-SBMs are also able  to recover the true partition $\bz_0$ as the network size $V$ grows, in replicated studies, thereby hinting at the validity of frequentist posterior consistency for our proposed methods, even in the challenging Scenario 2. This finding is further strengthened by the comparison with CASC for which consistency has been proved in  \citet{binkiewicz2017covariate}, though under a number of assumptions that rule out several data-generating processes of direct interest in practice. Relaxing these assumptions while extending the results to the node-colored network setting motivates future theoretical studies on pEx-SBM posterior consistency; see Section \ref{sec_6}.

As clarified in Section~\ref{sec_3.2}, pEx-SBM stands out also for its novel and principled $k$-step-ahead predictive strategies for both the edges and  the allocations of future incoming nodes. By evaluating these strategies on $10$ randomly-selected held-out nodes varying across the ten replicated studies in Scenario 1, yields an average mean squared error of $0.03$ among the true and predicted edge probabilities, and an average of  2.8 misallocated nodes out of 10 for all the three pEx-SBM examples. These are remarkable results when considering that only layer information is employed in these predictive tasks. 

Assessments on robustness to initialization, hyperparameter specification and inclusion of uninformative layers in the pEx-SBM examples analyzed can be found in the supplementary materials. Results suggest that all these three pEx-SBMs are robust. Among them, the H-DP  with hyperpriors on $\theta$ and $\theta_0$ stands out due to its ability of learning the parameters that control clustering properties together with the strength of layer information. As such, we will consider this prior in Section~\ref{sec_5}.

\section{Application to Criminal Networks}\label{sec_5}

We now showcase the potential of pEx-SBM on real-world data and consider the challenging criminal network application hinted at in Figure~\ref{figure:1}. The original data have been retrieved from the judicial acts of a law enforcement operation, named \textit{Operazione Infinito} \citep[e.g.,][]{calderoni2017communities}, that was conducted in Italy in order to disrupt a branch of the 'Ndrangheta Mafia that operates in the Milan area.

The original data are available in the UCINET software repository (covert networks). Here we consider the pre-processed version studied by \citet{legramanti2022extended}, which contains information on the presence or absence of at least one monitored co-attendance to a 'Ndrangheta summit for each pair of the $V=84$ criminals. Besides connectivity information, there is also knowledge on the role of each member (simple affiliate or boss) and on membership to the so-called \textit{locali}, namely, structural coordinated sub-units administering crime within specific territories. In analyzing this network, \citet{calderoni2017communities} focus on detecting simple communities, while \citet{legramanti2022extended} identify modular architectures via an ESBM supervised with both role and \textit{locale} affiliation. Nonetheless, obtaining reliable information on the roles of all criminals is often challenging, or even impossible, and, in general, is not available as prior information. Hence, an important goal is that of deducing the roles of criminals from the inferred group structures.

\begin{figure*}
	\centering
	\includegraphics[trim=2.2cm 2.8cm 1.2cm 2.2cm,clip,width=0.92\textwidth]{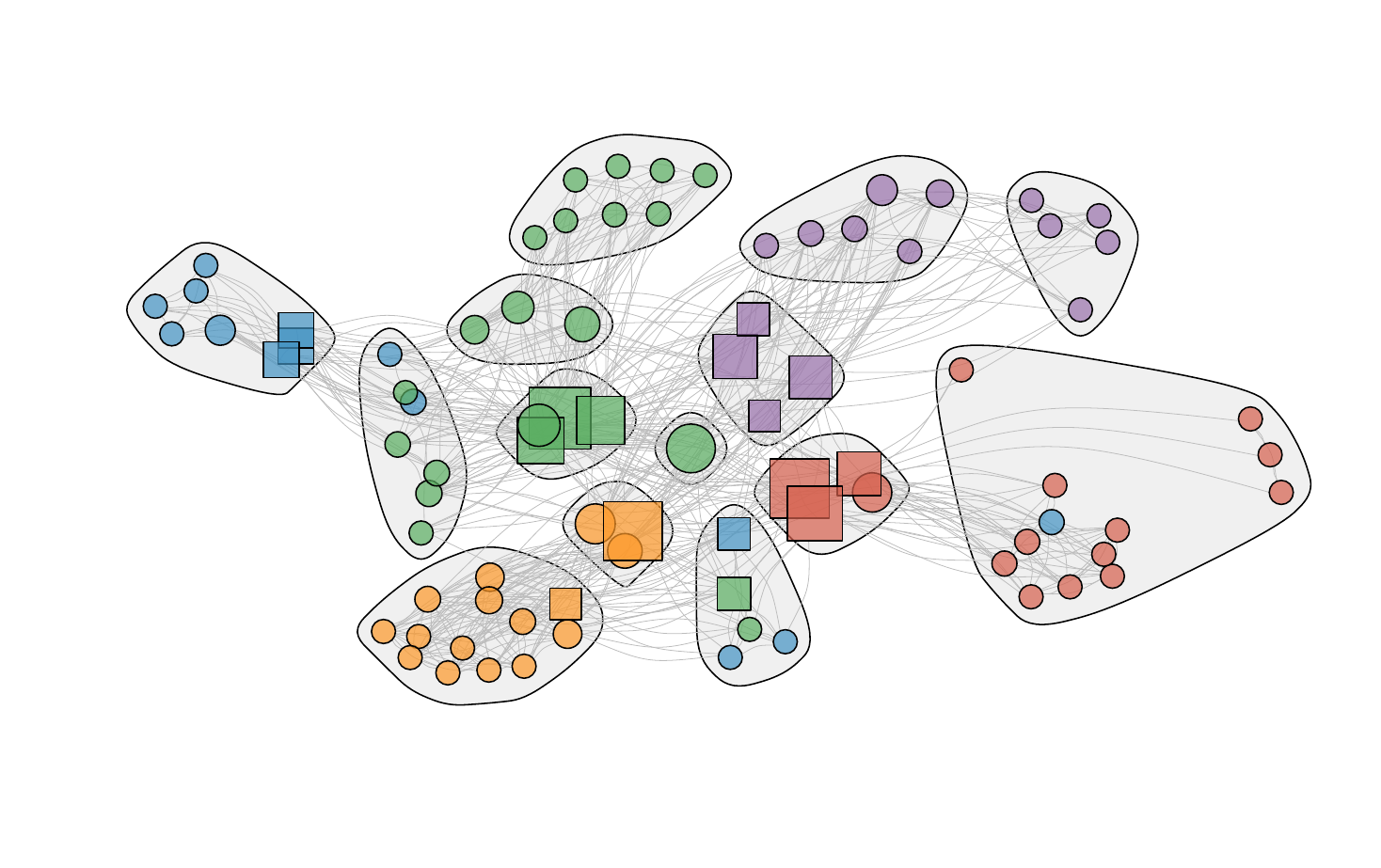}
	\caption{\footnotesize{Graphical representation of the \textit{Infinito} network along with the grouping structure estimated by the pEx-SBM. Node positions are obtained via force directed placement \citep{fru1991}, whereas colors indicate the division in layers corresponding to  \textit{locali} affiliation. The node size is proportional to the corresponding betweenness, while the shape indicates affiliates (circles) and bosses (squares). Gray areas highlight the groups inferred by pEx-SBM. }}
\label{figure:3}
\end{figure*}

Motivated by this remark, we only use \textit{locali}  to define the layers in the pEx-SBM, disregarding information on roles. Posterior inference is performed under the same settings of Section~\ref{sec_4} leveraging a H-DP with $\mbox{gamma}(10,2.5)$ and $\mbox{gamma}(5,0.45)$ hyperpriors for the parameters $\theta$ and $\theta_0$, so to induce a prior expected number of groups of $\approx 15$. Since there are 5 layers in total, with nodes possibly covering different roles, it is reasonable to expect, a priori,  the total number of groups to be at least double or triple the number of layers.  In fact, as illustrated in Figure~\ref{figure:3}, the minimum-VI point estimate $\hat{\bz}$ of $\bz$ points toward $\hat{H}=14$ groups, which is  in line with a number of network dynamics reported in the judicial acts. Such an accurate characterization of the block patterns in the observed network for pEx-SBM is also confirmed by a WAIC of $1282$, which improves the WAIC of $1299$ achieved by the most competitive alternative in the simulation studies in Section~\ref{sec_4}, i.e., ESBM with a DP prior and \textit{locali} supervision.

As is apparent from Figure \ref{figure:3}, pEx-SBM is able to disentangle core-periphery structures associated with boss-affiliate dynamics in each \textit{locale}. In fact, even if the role attribute is not incorporated within the model, for $4$ of the $5$ \textit{locali},  affiliates and bosses are grouped separately, hinted at a clear role separation in these \textit{locali}. Note that each small group of bosses also contains few affiliates. This suggests more nuanced roles beyond boss-affiliate separation, with some of the affiliates displaying connectivity behavior closer to those of bosses. This is further confirmed by the single-node group detected by pEx-SBM at the core of the network. Surprisingly, such a node is an affiliate that, however, displays a unique and central role in the connectivity architecture. According to the judicial acts, this node corresponds to a high-rank member who is in charge of coordinating the different \textit{locali}  and reporting to the leading Calabria-based families. Hence, its status appears to be even higher than that of \textit{locali} bosses. 

Across-layer groups are also inferred, as desired. These clusters reflect collaborative schemes between criminals in different \textit{locali} or more nuanced dynamics within the network, such as attempts of some members to change affiliation. For instance, the inferred group of red affiliates also includes a member of the blue \textit{locale} trying to create a new one by looking for affiliates in a different influence area.

A peculiar feature of this network is represented by the murder, during the investigations, of a high-rank member that destabilized the green \textit{locale}. As shown in Figure~\ref{figure:3}, the pEx-SBM estimate of $\bz$ unveils the consequences of this event, with the green \textit{locale} being the most fragmented in both within- and across-layer groups. Some of its affiliates are allocated to the group of nodes from the blue \textit{locale}. Moreover, even if a cluster of green bosses is detected, there is also a small group of three green affiliates seemingly forming a core-periphery structure with the more peripheral members of the same \textit{locale}. Interestingly, these three nodes were closely involved in the murder, according to the judicial documents.

The VI distance between $\hat{\bz}$ and the partition $\bz_b$ at the edge of the $95\%$ credible ball is $0.233$, much lower than the maximum achievable $\log_2 84 =6.392$ distance  among two generic partition of $V=84$ nodes. This concentration of the posterior for $\bz$ around $\hat{\bz}$ further strengthens the above claims. 

We conclude with an assessment of the predictive schemes in Section~\ref{sec_3.2}, focusing in particular on the out-of-sample accuracy in predicting the edges associated with $k=10$ randomly-selected held-out criminals. This test set comprises both affiliates and bosses from different \textit{locali}, thus allowing for a comprehensive evaluation. For these criminals we compute the predictive edge probabilities via~\eqref{eq:pred4}, conditioned only on the corresponding \textit{locale} affiliation and the network among the in-sample criminals. Despite the use of such a limited information, we obtain an area under the ROC curve of  $0.93$. This remarkable result highlights the practical effectiveness of the newly proposed predictive methods. These gains are confirmed by the posterior similarity matrix comprising in-sample and predictive co-clustering probabilities for estimating the allocations of out-of-sample nodes (see the supplementary materials).

\section{Conclusions and Future Research}\label{sec_6}

In this article, we proposed and developed a new class of SBMs for node-colored multilayer networks that infer complex block-connectivity structures both within and across layers. The layers’ division is accounted for in the formation of blocks via a rigorous and interpretable probabilistic construction. The proposed pEx-SBM allows for uncertainty quantification, derivation of clustering and co-clustering properties, and, crucially, prediction of the connectivity patterns for future nodes. Moreover, it is computationally tractable and, as shown in Sections \ref{sec_4}--\ref{sec_5}, yields remarkable practical improvements. To the best of our knowledge, current models for multilayer networks do not display all these  properties in a single formulation. 

While we focused on binary undirected networks, weighted edges are readily tackled by replacing the beta-binomial likelihood \eqref{eq1} with a Poisson-gamma for count edges, or a Gaussian-Gaussian for continuous ones. Directed and bipartite networks can be instead addressed by  modeling the row and column partitions of the non-symmetric or rectangular \textit{supra}-adjacency matrix via two distinct pExP priors. 

From a more general perspective, our contribution showcases in the node-colored setting the potential of a broadly-impactful and innovative idea. Namely that of designing models for multilayer network data that consist in matching the distinctive structures of a given multilayer network sub-class analyzed with the most suitable notion of probabilistic invariance encoded in a specific node-exchangeability assumption. For example, for edge-colored (multiplex) networks the natural pairing would be with \textit{separate exchangeability}, which, unlike for partial exchangeability, preserves the identity of nodes replicated across layers; see \citet{lin2021separate} for a review focusing on mixture models. A similar reasoning can be also applied to dynamic networks, where the time dimension demands for further restrictions to the invariance structure. 

Finally, proving frequentist posterior consistency of the proposed pEx-SBMs, as suggested by the empirical findings within Figure~\ref{figure:const}, is an important, yet challenging, direction of future research. Although it is possible to adapt the approach of \citet{geng_2019}  to a finite-dimensional version of  pEx-SBM under simplified settings, deriving a realistic and more comprehensive theory would first require overcoming two yet-unaddressed challenges.  First, the available Bayesian asymptotic theory for single-layered SBMs  \citep[e.g.,][]{geng_2019} has to be extended to node-colored multilayer networks and beyond  the restrictive  assumptions that are currently imposed in the literature (even for proving consistency of frequentist methods in single-layered settings). Second, one should also deal with the additional difficulty posed by the condition of partial exchangeability, for which the investigation of frequentist consistency properties is still limited \citep[e.g.,][]{catalano2022posterior} due to the challenges posed by the reduced mathematical tractability of the objects studied.

\section*{Code and Data}
Detailed code and data to reproduce the results can be found online within the GitHub repository: \url{https://github.com/francescogaffi/pexsbm}.
\onecolumn

\clearpage

\textwidth=18.4cm
\oddsidemargin=-1cm
\topmargin=-2.5cm
\textheight=22.9cm

\newenvironment{changemargin}[2]{%
\begin{list}{}{%
\setlength{\topsep}{0pt}%
\setlength{\leftmargin}{#1}%
\setlength{\rightmargin}{#2}%
\setlength{\listparindent}{\parindent}%
\setlength{\itemindent}{\parindent}%
\setlength{\parsep}{\parskip}%
}%
\item[]}{\end{list}}

\begin{center}
\vspace{70pt}
\huge Supplementary Materials for \\ ``Partially Exchangeable Stochastic Block Models for (Node-Colored) Multilayer Networks"
\end{center}

\setcounter{section}{0}
\setcounter{equation}{0}
\setcounter{table}{0}
\setcounter{figure}{0}

\renewcommand{\theequation}{S.\arabic{equation}}
\renewcommand{\thesection}{S\arabic{section}}  
\renewcommand{\thefigure}{S.\arabic{figure}}
\renewcommand{\thetable}{S.\arabic{table}}

\fontsize{11}{14.6}\selectfont
\vspace{90pt}

\begin{changemargin}{1.2cm}{1.2cm}

\section{Background Material}\label{sec_s_1}
Section~\ref{sec_s_1} reviews the core concepts underlying the proposed pEx-SBM model. More specifically, Section~\ref{sec_kivela} summarizes the general multilayer network framework presented in   \cite{kivela2014multilayer}~and clarifies how the node-colored networks we consider in our contribution represent a remarkable special case of such a framework. Sections~\ref{sec:exch}--\ref{sec:backpex},  discuss the connection between exchangeability~and~nonparametric priors defined via completely random measures, and then clarify the link between~partial~exchangeability and hierarchical compositions of such priors.

\vspace{17pt}
\subsection{Multilayer networks}\label{sec_kivela}
Recalling the comprehensive review by \cite{kivela2014multilayer}, the term \textit{multilayer networks} refers to  a general construction that encompasses most of the layered network structures treated in the literature. In the following, we summarize such a construction, and clarify how it includes, among others, the~node-colored networks we consider in our contribution. Refer to Section 2.1 in \cite{kivela2014multilayer}~for~a~more extensive and in-depth treatment.

Let $\mathbb{V}:=\{1,\dots,V\}$ denote a set of $V$ nodes in a network, for any $V\in\N$, and define with $\delta\in\N$~the number of \textit{aspects} of such a network, that is the number of directions of layerization. Depending on the type of multilayer network analyzed, each \textit{aspect} could denote, e.g., a division of the nodes~into~subpopulations (e.g., node-colored networks), the presence of different types of relationships monitored  (e.g., edge-colored networks), or a time dimension (e.g., dynamic networks), among~others. Within~this framework, any aspect $l\in\{1,\dots,\delta\}$ generates a set of \textit{elementary layers} $L_l=\{1,\dots,d_l\}$ for every $d_l\in\N$. For example, in node-colored networks  the elementary layers correspond to the different~subpopulations to which the nodes belong. Under these settings, a \textit{layer} is an element of the product~space $L_1\times\dots\times L_\delta$, (namely, a vector of coordinates identifying a location in the product space of elementary layers), whereas the set of \textit{existing nodes} is defined as $\mathbb{V}_M\subseteq \mathbb{V}\times L_1\times\dots\times L_\delta$. In symbols, node $i$ exists in layer $(j_1,\dots,j_\delta)$ whenever $(i,j_1,\dots,j_\delta)\in \mathbb{V}_M$, for $j_l\in L_l$ and $l\in\{1,\dots,\delta\}$. Consistent with these definitions, in the aforementioned node-colored network example each layer coincides with an elementary layer, whereas an existing node is a pair that defines the node index in $\mathbb{V}$ and the subpopulation~to which it belongs. If such a node-colored network were also observed at different time points,~this~would generate an additional aspect (i.e., time) with layers defined as pairs (subpopulation, time point) and existing nodes as triplets (node index, subpopulation,~time~point).~Finally,~denote~with~$\mathbb{E}_M\subseteq \mathbb{V}_M\times \mathbb{V}_M$ the set of edges. Such a set comprises all those couples $((i,\,j_1,\dots,j_\delta),(i',j_1',\dots,j_\delta'))$ of existing nodes  $(i,\,j_1,\dots,j_\delta)$ and $(i',j_1',\dots,j_\delta')$ among which an edge has been observed. With these settings one can give the following definition of multilayer network. 

\vspace{15pt}
\begin{defi}\label{def:multinet}
A $\delta$-multilayer network is a quadruplet $M:=(\mathbb{V}_M,\,\mathbb{E}_M,\,\mathbb{V},\,L)$ where $L=L_1\times\dots\times L_\delta$ is the product space of layers, $\mathbb{V}$ is the set of nodes, $\mathbb{V}_M$ corresponds to the set of coordinates of existing nodes, and $\mathbb{E}_M$ denotes the set of edges.
\end{defi}
\vspace{15pt}

Definition~\ref{def:multinet} allows for the existence of edges between nodes in the same layer, as well as across-layers, and even between copies of the same node existing in different layers. Moreover, notice that any multilayer network $M$, which is, in this formulation, a highly general object, can be flattened to obtain the node-labelled graph given by the couple $(\mathbb{V}_M,\mathbb{E}_M)$, which is called the \textit{supra}-graph of $M$. See, e.g., Figure 2 in \cite{kivela2014multilayer}. Hence, there is a close relationship between multilayer structures,~of~any kind that such general construction can encompass, and \textit{node-level covariate} structures. A $\delta$-multilayer network can indeed be represented by a \textit{supra}-adjacency matrix $\bY$, that is the adjacency matrix of~the \textit{supra}-graph. Once chosen an order in the product space of the layers $L$, e.g., the lexicographic order, based on a natural order of each entry, then it is easy to describe the matrix $\bY$ as a block matrix whose diagonal blocks are the adjacency matrices of each layer, and the off-diagonal blocks are the matrices of across-layers connections.

Under the above formulation,  network structures comprising different types of relationships --- with each layer encoding connections with respect to one of these types (e.g., friendship, kinship,~advice,~$\ldots$) --- can be represented through a block-diagonal \textit{supra}-adjacency matrix excluding the connections across layers. When $\delta=1$, these are called \textit{edge-colored} networks. If one considers the case in which each node exists within every layer, the network can be seen as a superposition of different sets of edges on the same collection of nodes. In such a case, the term \textit{multiplex network} is increasingly used in the literature on network data \citep[see, e.g., ][]{mucha2010community,barbillon2017stochastic,macdonald2022latent,pensky2021clustering,noroozi2022sparse,amini2019hierarchical}.

On the other hand, by forcing each node in the network to belong only to a single layer~via~the~disjointness~condition
\begin{equation}\label{eq:disjoint}
\forall\, i\in \mathbb{V} \quad\exists!\, (j_1,\dots,j_\delta)\in L \quad\text{ s.t. }\quad (i,\,j_1,\dots,j_\delta)\in \mathbb{V}_M,
\end{equation}
one can recover networks in which layers represent a division of nodes into distinct subpopulations,~so that a node cannot have copies in different layers. If $\delta=1$, these are called \textit{node-colored} networks. In~the main article, we propose a stochastic block model for such networks, based on the novel and general idea of matching the specific structure of each sub-class of multilayer network analyzed with the most suitable notion of probabilistic invariance encoded in a specific node-exchangeability assumption. As we clarify in the main article, in the context of node-colored networks the most natural invariance structure is the one encoded in the assumption of partial exchangeability.
\vspace{17pt}

\subsection{Exchangeable sequences and completely random measures}\label{sec:exch}
Let $\mathbb{X}$ be a complete and separable metric space equipped with the Borel $\sigma$-algebra $\mathscr X$. $\mathscr P$ is the space of probability distributions defined on $(\X,\Xcr)$ and endowed with the topology of weak convergence.~$\sigma(\mathscr P)$ is then the Borel $\sigma$-algebra of subsets of $\Pcr$. Moreover, denote with $[n]$ the set of the first $n$ integers $\{1,\dots,n\}$, for any $n\in\mathbb{N}$. For a sequence of $\X$-valued random elements $\bx=(x_{n})_{n\geq1}$, defined on some probability space $(\Omega,\mathscr{F},\mathbb{P})$ the following definition can be given.

\vspace{15pt}
\begin{defi}\label{def:exch}
$\bx$ is exchangeable if, for every $n\geq1$ and any permutation $\pi$ of the indices in $[n]$, it holds $(x_1,\dots,\,x_n)\overset{d}{=}(x_{\pi(1)},\dots,\,x_{\pi(n)})$.
\end{defi}
\vspace{15pt}

\noindent By virtue of the representation result in \citet{de1937prevision}, it is possible to state the following.
\vspace{15pt}

\begin{thm}[de Finetti]\label{thm:definetti}
A sequence $\bx$ is exchangeable if and only if there exists a probability measure $Q$ on the space of probability distributions $(\mathscr P, \sigma(\mathscr P))$ such that
 \begin{equation}\label{eq:defin}
  \prob{x_1\in A_1,\ldots,x_n\in A_n}=\int_{\mathscr P}\prod_{i=1}^n \, P(A_i)\, Q(\ddr P),
  \end{equation}
  
\noindent  for any $A_1,\ldots,A_n$ in $\Xcr$ and $n\ge 1$.
\end{thm}
\vspace{15pt}

\noindent The probability measure $Q$ directing the exchangeable sequence $(x_{n})_{n\geq1}$ is also termed \textit{de Finetti measure} and can be interpreted as a prior distribution in the Bayesian framework. The representation theorem in \eqref{eq:defin} can be equivalently rephrased by stating that, given an exchangeable sequence $(x_{n})_{n\geq1}$, there exists a random probability measure $\tilde{P}$, defined on $(\X,\Xcr)$ and taking values in $(\mathscr P, \sigma(\mathscr P))$, such that, for any $n\in\mathbb{N}$, it holds
    \begin{equation}\label{eq:definetti1}
x_1,\dots,\,x_n\mid\rpm\simiid\rpm, \qquad \mbox{with} \qquad \rpm\sim Q.
\end{equation}
The most popular instance of nonparametric prior $Q$ is the Dirichlet process (DP) prior, introduced in \cite{Fer(73)}, which selects discrete distributions with probability $1$. As shown in \cite{lp} most classes of discrete nonparametric priors, including the DP, can be seen as suitable transformations of \textit{completely random measures} (CRMs).
 
More specifically, let $\mathscr M$ be the set of boundedly finite measures on $\X$ equipped with the corresponding Borel $\sigma$-algebra $\sigma(\mathscr M)$. For details on the definition of this $\sigma$-algebra, see \cite{Dal(07)}. A CRM $\tilde \mu$ on $(\X,\Xcr)$ is a measurable function on $(\Omega, \Fcr,\P)$ taking values in $\mathscr M$ such that for any $k\geq2$ and pairwise disjoint sets $A_1,\dots,A_k$ in $\Xcr$ the random variables $\tilde \mu(A_1),\dots,\tilde \mu(A_k)$ are independent. CRMs have been introduced in \cite{Kin(67)}; see also  \cite{Kin(93)} and \cite{Dal(07)}. Any CRM $\tilde \mu$ fulfills the following representation. For $A\in\mathscr{X}$, we have
 \begin{equation}
 \tilde\mu(A)=\sum_{k\geq1}u_k\delta_{x_k}(A) +\beta(A)+\int_0^\infty s \,\tilde N(\ddr s,\,A),
 \end{equation}
 where $(x_k)_{k\geq1}$ is a sequence in $\mathbb{X}$, $(u_k)_{k\geq1}$ is a sequence of independent non-negative random variables, $\beta$ is a fixed non-atomic boundedly finite measure on $(\mathbb{X},\mathscr{X})$, and $\tilde N$ is a Poisson~process~on~$\R^+\times\mathbb{X}$ independent of $(u_k)_{k\geq1}$ and whose parameter measure $\nu$ satisfies $\int_{\mathbb{R}^+}\int_{B}\min\{s,1\}\:\nu(\ddr s,\ddr x)<\infty$
    for any bounded ${B}$ in $\mathscr{X}$. Intuitively, a CRM can be seen as a superposition of a random measure~with fixed~atoms, a deterministic non-atomic drift and a part characterized by random jumps and random locations, whose intensities, distribution and mutual dependence are governed by a Poisson process. Here, as customary in Bayesian nonparametrics, we focus on CRMs $\tilde \mu$ with no fixed atoms~and~no~drift. These~are~almost surely discrete and the corresponding Laplace functional admits the following \textit{L\'{e}vy--Khintchine} representation
  \begin{equation}
    \label{eq:lapl_funct}
    \E\left[\edr^{-\int_\X f(x)\,\tilde\mu(\ddr x)}\right]=
    \exp\biggl\{-\int_{\R^+\times\X}\left[1-\edr^{-sf(x)}\right]\,\nu(\ddr s,\ddr x)\biggr\},
    \end{equation}
where $f:\X\to\R$ is a measurable function such that $\int |f|\,\ddr\tilde\mu<\infty$ almost surely. The measure $\nu$ is known as the \textit{L\'evy intensity} of $\tilde\mu$ and regulates the intensity of the jumps of a CRM and their locations. By virtue of \eqref{eq:lapl_funct}, it characterizes the CRM $\tilde \mu$. Two noteworthy examples are given by the \textit{$\sigma$-stable~CRM} $\tilde\mu_\sigma$ with L\'evy intensity defined as
\begin{equation}\label{eq:stable_intens}
    \nu(\ddr s,\ddr x)=\frac{\sigma\, }{\Gamma(1-\sigma)}
    s^{-1-\sigma}\,\ddr s\,\alpha(\ddr x),
    \end{equation}
where $\alpha$ is a measure on $(\X,\Xcr)$ and $\sigma\in(0,1)$, and
by the \textit{gamma CRM} $\tilde \mu$, which corresponds to
 \begin{equation}\label{eq:gamma_intens}
    \nu(\ddr s,\ddr x)=e^{-s}s^{-1}\,\ddr s\,\alpha(\ddr x).
    \end{equation}
If we further impose the condition $0<\tilde\mu(\mathbb{X})<\infty$ almost surely, which is implied by $\nu(\R^+\times\mathbb{X})=\infty$ and $\int_{\R^+\times\X}\left[1-\edr^{-\lambda s}\right]\,\nu(\ddr s,\ddr x)<\infty$ for any $\lambda>0$, then, as proved in \cite{Reg(03)}, it is possible to define
\begin{equation}\label{eq:nrmi}
\rpm(A):=\frac{\tilde\mu(A)}{\tilde\mu(\mathbb{X})},
\end{equation}
for any $A\in\mathscr{X}$. Random probability measures defined as in \eqref{eq:nrmi} form the class of \textit{normalized random measures with independent increments} (NRMIs), introduced in \cite{Reg(03)}. As clarified in \eqref{eq:lapl_funct}, these measures are characterized by $\nu$. Moreover, $\tilde\mu$ is termed \textit{homogeneous CRM} if its L\'evy intensity factorizes as $\nu(\ddr s,\ddr x)=\rho(s)\,\ddr s\,\alpha(\ddr x)$ for some measurable positive function $\rho$ on $\R^+$~and~some measure $\alpha$ on $\mathbb{X}$; in this case $\tilde\mu$ is identified by $\rho$ and $\alpha$. Intensities in \eqref{eq:stable_intens} and \eqref{eq:gamma_intens} are in this class. Note that for homogeneous CRMs $\int_{\R^+\times\X}\left[1-\edr^{-\lambda s}\right]\,\nu(\ddr s,\ddr x)<\infty$ is equivalent to the finiteness of $\alpha$. 
Setting $c:=\alpha(\mathbb{X})$, we have 
\begin{equation}\label{eq:homo_crm}
\nu(\ddr s,\ddr x)=\rho(s)\,\ddr s \,c\, G(\ddr x),
\end{equation}
where $G(\cdot):=\alpha(\cdot)/\alpha(\mathbb{X})$ is now a probability measure on $\mathbb{X}$. Therefore, we can write
\begin{equation}\label{eq:pisnrmi}
\rpm\sim\text{NRMI}(\rho,\,c,\,G),
\end{equation}
for some measurable positive $\rho$,  $c>0$ and probability measure $G$. Note that $\mathbb{E}[\rpm(A)]=G(A)$ for any $A\in\mathscr{X}$, which means that, if $x\mid\rpm\sim\rpm$ then $\prob{x\in A}=G(A)$ for any $A\in\mathscr{X}$.
For particular choices of $\rho$ it is possible to recover noteworthy nonparametric priors. For example, taking $\rho(s)=e^{-s}s^{-1}$ as in \eqref{eq:gamma_intens}, we are normalizing a gamma CRM and, hence, the correspondent $\rpm$ in \eqref{eq:nrmi} is a DP. In this case the total mass $c$, also called \textit{concentration parameter}, is often denoted as $\theta>0$. Hence, we shall also write $\tilde P\sim \mbox{DP}(\theta, P_0)$, where $\theta=\alpha(\X)$ and $P_0=\E[\tilde P]$. If instead we choose $\rho$ as in \eqref{eq:stable_intens}, then $\rpm$ in \eqref{eq:nrmi} is called \textit{normalized stable process} (NSP) \citep{Kin(75)}.

\vspace{8pt}
\subsection{Partially exchangeable arrays and hierarchical processes}\label{sec:backpex}
As discussed within the main article, we consider a generalization of exchangeability in order to induce a random partition prior on the nodes which accounts for information from the layer division. Such~a generalization is known as \textit{partial exchangeability} \citep{deF(38)} and is a more natural hypothesis of dependence for random elements divided in a finite number $d$ of layers. In fact, as clarified~in~Definition~\ref{def:pex} of the main article, it assumes that the joint distribution of the entries in  a given infinite random array $\bX^\infty=\{(x_{ji})_{i\geq1}: j\in[d]\}$ is invariant with respect to within-layer permutations, but not necessarily with respect to across-layer ones.

Under the partial exchangeability assumption, the following extension of representation \eqref{eq:defin} holds.

\vspace{15pt}
\begin{thm}[de Finetti]\label{thm:repr}
The infinite random array $\bX^\infty$ is partially exchangeable if and only if there exists a measure $Q$ on the space of $d$-dimensional vectors of probability measures $\mathscr{P}^d$ endowed with the product $\sigma$-algebra $\bigotimes_{j=1}^d\sigma(\mathscr{P})$ such that
\begin{equation}\label{eq:defin1}
\begin{split}
 & \prob{x_{11}\in A_{11},\dots,x_{1V_{1}}\in A_{1V_{1}},\dots,x_{d1}\in A_{d1},\dots,x_{dV_{d}}\in A_{dV_{d}}}\\&\qquad \qquad \qquad \qquad \qquad=\int_{\mathscr P^d}\prod_{j=1}^d \prod_{i=1}^{V_j}\, P_j(A_{ji})\, Q(\ddr P_1,\dots,\ddr P_d),
 \end{split}
  \end{equation}
 for any $A_{11},\dots,A_{dV_{d}}\in\mathscr{X}$ and any $(V_1,\dots,V_d)\in\N^d$.
\end{thm}
\vspace{15pt}

Again the quantity $Q$ is called \textit{de Finetti measure}. By extending the representation in~\eqref{eq:definetti1} to the partial exchangeability setting, one obtains \eqref{eq:definetti} of the main article, which is equivalent to Theorem \ref{thm:repr}. Namely, given a partially exchangeable infinite array $\bX^\infty$, there exists a vector of random probability measures $(\rpm_1,\dots,\,\rpm_d)$, each defined on $(\X,\Xcr)$, and taking values in $(\mathscr P^d, \bigotimes_{j=1}^d\sigma(\mathscr P))$, such that, for any $j_1,\ldots,j_k\in[d]$ and $i_1,\ldots,i_k\ge 1$, one has
\begin{equation}\label{eq:definettis}
\begin{split}
		(x_{j_1i_1},\ldots,x_{j_ki_k})\,|\,(\rpm_1,\dots,\,\rpm_d)
			&\:\simiid\:\rpm_{j_1}\,\times\,\cdots\,\times\, \rpm_{j_k},\\
		(\rpm_1,\dots,\,\rpm_d)&\:\sim\: Q.
		\end{split}
\end{equation}
Partial exchangeability reduces to exchangeability if $\rpm_j=\rpm$ almost surely for any $j\in[d]$ and some random probability measure $\rpm$. In this case, \eqref{eq:definettis} reduces to \eqref{eq:definetti1}. Conversely, if $(\rpm_j)_{j\in[d]}$ are independent, that is $Q$ is a product law on $\mathscr{P}^d$, then the rows of the array $\bX^\infty$ are independent. These particular cases represent the extremes in the range of borrowing of information among nodes coming from different layers. Defining a partially exchangeable model for an array, then, amounts to choosing a de Finetti measure $Q$ for a vector of random probability measures which can allow flexible borrowing of information both within and across layers. To this end, and recalling the considerations for the classical exchangeability setting, a natural and routinely-employed solution is to rely on hierarchical compositions of the nonparametric priors presented in Section~\ref{sec:exch}. This yields the so-called \textit{hierarchical normalized random measures with independent increments} (H-NRMIs) priors; see Definition \ref{def:hnrmi} in the main article --- which extends  \eqref{eq:pisnrmi} to the partial exchangeability setting --- and \citet{camerlenghi2019distribution} for an in-depth treatment.

Example~\ref{exDPNSP} provides two popular instances of H-NRMIs, which further clarify the related construction via hierarchical compositions of the nonparametric priors.

\vspace{15pt}

\begin{exe}
\label{exDPNSP}
If for both levels of the hierarchy we consider a {\normalfont DP}, namely
 \setlength\abovedisplayskip{10pt}%
    \setlength\belowdisplayskip{10pt}%
\begin{equation}\label{eq:hdp}
\begin{split}
\rpm_1,\dots,\rpm_d\mid \rpm_0&\simiid \text{\normalfont{DP}}(\theta,\rpm_0),\\
\rpm_0&\sim\text{\normalfont{DP}}(\theta_0,\,P_0),
\end{split}
\end{equation}
then $(\rpm_1,\dots,\rpm_d)\sim\text{\normalfont H-DP}(\theta,\,\theta_0,\,P_0)$ \citep{Teh_06}.

\noindent If instead we consider two levels of {\normalfont NSP}, i.e.,
\begin{equation}\label{eq:hnsp}
\begin{split}
\rpm_1,\dots,\rpm_d\mid \rpm_0&\simiid \text{\normalfont{NSP}}(\sigma,\rpm_0),\\
\rpm_0&\sim\text{\normalfont{NSP}}(\sigma_0,\,P_0),
\end{split}
\end{equation}
then, $(\rpm_1,\dots,\rpm_d)\sim\text{\normalfont  H-NSP}(\sigma,\,\sigma_0,\,P_0)$ \citep{camerlenghi2019distribution}.
\end{exe}

Since H-NRMIs represent a particular specification of the de Finetti measure $Q$ in \eqref{eq:definettis}, 
if the $j$-th row, $(x_{j1},\dots,x_{jV_j})$, of a finite array $\bX\subset\bX^\infty$ is a conditionally iid sample from $\rpm_j$  in Definition~\ref{def:hnrmi}~of the main article, for any $V_j\ge 1$ and $j\in[d]$, then $\bX^\infty$ is partially exchangeable according to its rows. Moreover, because of the almost sure discreteness of each $\rpm_j$ and $\rpm_0$, the probability of a tie among the entries in the array $\bX$ is positive both within and across rows (i.e., layers), that is $\mathbb{P}(x_{ij}=x_{i'j'})>0$,
for any $i\in[V_j]$, $i'\in[V_{j'}]$ and $j,j'\in[d]$. As clarified in the main article, the proposed partially exchangeable partition prior for the node allocations to groups in the pEx-SBM model is directly obtained from the within- and across-layer clustering structures induced by these~ties.
\vspace{5pt}

\section{Results for the Hierarchical Normalized Stable Case}
\vspace{-5pt}
In the article, all the results obtained for the general class of H-NRMI are specialized to the H-DP~case. Here we  specialize these results also to the H-NSP. From Proposition \ref{prp:pred_hnrmi}, we can deduce the following.
\vspace{13pt}

\begin{cor}\label{cor:pred_nsp}
    Let $\bz\sim\mbox{\normalfont pExP}(\bV;\,\sigma,\,\sigma_0)$ where $\mbox{\normalfont pExP}(\bV;\,\sigma,\,\sigma_0)$ denotes the partition structure induced by a {\normalfont H-NSP} with parameters $\sigma,\,\sigma_0\in(0,1)$. Then
\begin{equation}
	\label{eq:pred_nsp}
	\mathbb{P}(z_v=h \mid \bz^{-v}, \bw^{-v})=
	\indic_{\{\ell^{-v}_{\cdot h}=0\}}\frac{H^{-v}\sigma_0}{|\bm{\ell}^{-v}|}\frac{\ell_{j\cdot}^{-v}\sigma}{V_j-1}
	+\indic_{\{\ell^{-v}_{\cdot h}\neq0\}}\left[\frac{\ell^{-v}_{\cdot h}-\sigma_0}{|\bm{\ell}^{-v}|}\frac{\ell_{j\cdot}^{-v}\sigma}{V_j-1}+\frac{n_{jh}^{-v}-\ell_{jh}^{-v}\sigma}{V_j-1}\right],
\end{equation}
for each $h\in[H^{-v}+1]$.
\end{cor}
\vspace{13pt}

\noindent Theorem \ref{coclust_hnrmi} yields, instead, the following expressions.
\vspace{13pt}

\begin{cor}\label{cor:coclust_nsp}
Let $\bz\sim \mbox{\normalfont pExP}(\bV;\,\sigma,\,\sigma_0)$. Then, if both $v$ and $u$ are in the same layer $j$, we have
\begin{equation}\label{eq:coclust_nsp1}
\begin{split}
&\mathbb{P}(z_v=z_u \mid \bz^{-vu}, \bw^{-vu})=
\frac{1}{(V_{j}-2)(V_j-1)}\left\{\sum_{h=1}^{H^{-vu}}\left(n_{jh}^{-vu}-\ell_{jh}^{-vu}\sigma\right)\left(n_{jh}^{-vu}-\ell_{jh}^{-vu}\sigma+1\right)\right.\\
&\qquad \qquad \qquad +\ell_{j\cdot}^{-vu}\sigma\left[(1-\sigma)+\frac{2}{|\bell^{-vu}|}\sum_{h=1}^{H^{-vu}}\left(n_{jh}^{-vu}-\ell_{jh}^{-vu}\sigma\right)\left(\ell_{\cdot h}^{-vu}-\sigma_0\right)\right]\\
&\qquad\qquad \qquad \left.+\frac{\ell_{j\cdot}^{-vu}\left(\ell_{j\cdot}^{-vu}+1\right)\sigma^2}{|\bell^{-vu}|\left(|\bell^{-vu}|+1\right)}\left[\sum_{h=1}^{H^{-vu}}\left(\ell_{\cdot h}^{-vu}-\sigma_0\right)\left(\ell_{\cdot h}^{-vu}-\sigma_0+1\right)+H^{-vu}\sigma_0(1-\sigma_0)\right]\right\},
\end{split}
\end{equation}
whereas, if node $v$ is in layer $j$ and node $u$ is in layer $j'$, with $j\neq j'$, it follows that
\begin{equation}\label{eq:coclust_nsp2}
\begin{split}
&\mathbb{P}(z_v=z_u \mid \bz^{-vu}, \bw^{-vu})=
\frac{1}{(V_{j}-1)(V_{j'}-1)}\left\{\sum_{h=1}^{H^{-vu}}\left(n_{jh}^{-vu}-\ell_{jh}^{-vu}\sigma\right)\left(n_{j'h}^{-vu}-\ell_{j'h}^{-vu}\sigma\right)\right.\\
&\qquad +\frac{\sigma}{|\bell^{-vu}|}\left[\ell_{j\cdot}^{-vu}\sum_{h=1}^{H^{-vu}}\left(n_{j'h}^{-vu}-\ell_{j'h}^{-vu}\sigma\right)\left(\ell_{\cdot h}^{-vu}-\sigma_0\right)+\ell_{j'\cdot}^{-vu}\sum_{h=1}^{H^{-vu}}\left(n_{jh}^{-vu}-\ell_{jh}^{-vu}\sigma\right)\left(\ell_{\cdot h}^{-vu}-\sigma_0\right)\right]\\
&\qquad \left.+\frac{\ell_{j\cdot}^{-vu}\left(\ell_{j'\cdot}^{-vu}\right)\sigma^2}{|\bell^{-vu}|\left(|\bell^{-vu}|+1\right)}\left[\sum_{h=1}^{H^{-vu}}\left(\ell_{\cdot h}^{-vu}-\sigma_0\right)\left(\ell_{\cdot h}^{-vu}-\sigma_0+1\right)+H^{-vu}\sigma_0(1-\sigma_0)\right]\right\}.
\end{split}
\end{equation}
\end{cor}
\vspace{13pt}

\noindent The following statement provides a prior elicitation result for the H-NSP, similar to that in Proposition~\ref{prp:cond_dir}.

\vspace{15pt}
\begin{prp}\label{prp:pelic_nsp}
    Let $\bz\sim \mbox{\normalfont pExP}(\bV;\,\sigma,\,\sigma_0)$. Then, for any generic node $v$ in layer $j$, we have
	\begin{equation}\label{eq:nspelicit1}
		\sigma\left(1+\frac{|\bH_j^{-v}|}{|\bell^{-v}|}\sigma_0\right)\leq \frac{V_j-1}{2\ell^{-v}_{j\cdot}}\Longrightarrow
		\mathbb{P}(z_v\in \bH_j^{-v}\mid \bz^{-v},\,\bw^{-v})\geq\mathbb{P}(z_v\notin \bH_j^{-v}\mid \bz^{-v},\,\bw^{-v}),
	\end{equation}
	for each $j\in[d]$. In particular, the left hand side of \eqref{eq:nspelicit1} is implied if $\sigma\leq 0.5 (1+(V_j-1)\sigma_0/d)^{-1}$. As discussed in Remark~\ref{rmk_ho} in the article,  this condition also guarantees $\mathbb{P}(z_v\in \bH_j^{-v}|\bz^{-v},\,\bw^{-v})\geq\mathbb{P}(z_v\in \bH^{-v}\smallsetminus  \bH_j^{-v}|\bz^{-v},\,\bw^{-v})$.
    \end{prp}
\vspace{15pt}

\noindent We conclude by tailoring the results on the conditional probabilities in Proposition \ref{prp:nuova} to the H-NSP. 

\vspace{15pt}

\begin{cor}\label{cor:jointpred_nsp}
If $\bz\sim\text{\normalfont pExP}(\bV;\,\sigma,\,\sigma_0)$, then, for any node $v$ in layer $j$, we have
\begin{eqnarray}
\begin{split}
\label{eq:jpred_hnsp}
&\cprob{z_v=h,w_v=\uptau}{\bz^{-v},\bw^{-v}}\\
&\qquad =\indic_{\left\{\uptau=\uptau^{-v}_{jht}\right\}}\frac{q_{jht}^{-v}-\sigma}{V_j-1}
+\indic_{\left\{\uptau=\ell_{j\cdot}^{-v}+1\right\}}\frac{\ell_{j\cdot}^{-v}\sigma}{V_j-1}\left[\indic_{\left\{h\leq H^{-v}\right\}}\frac{\ell^{-v}_{\cdot h}-\sigma_0}{|\bm{\ell}^{-v}|}+\indic_{\left\{h=H^{-v}+1\right\}}\frac{H^{-v}\sigma_0}{|\bm{\ell}^{-v}|}\right],
\end{split}
\end{eqnarray}
for any $h\in[H^{-v}+1]$, $\uptau\in[\ell^{-v}_{j\cdot}+1]$, where $\bm{\uptau}^{-v}_{jh}$ is defined as in Proposition \ref{prp:nuova} of the main article.
\end{cor}
\vspace{10pt}

\section{Proofs of Theorems, Propositions and Corollaries}\label{sec_s_2}
\vspace{-2pt}
Section~\ref{sec_s_2} provides the proofs of the theorems, propositions and corollaries stated in the main article and in these supplementary materials.

We start proving the expressions for the joint conditional probabilities given in Proposition \ref{prp:nuova} and its Corollaries \ref{exe:jpred_hdp} and \ref{cor:jointpred_nsp}, which we will also leverage on in the proofs of other results.

\vspace{8pt}

\begin{proof}[{\bf Proof of Proposition \ref{prp:nuova}}] By Bayes rule
\begin{equation}\label{eq:proof1}
\mathbb{P}(z_v=h,w_v=\uptau \mid \bz^{-v},\bw^{-v})= \frac{p(z_v=h,w_v=\uptau , \bz^{-v},\bw^{-v})}{p(\bz^{-v},\bw^{-v})},
\end{equation}
where both the numerator and the denominator can be directly retrieved from the pEPPF  in \eqref{eq:peppf1} of the main article. To this end, it suffices to notice that, having fixed a specific configuration for the~group~and subgroup allocations --- instead of just an array of frequencies --- the summation over all the configurations and the multiplication by the product of multinomial factors in \eqref{eq:peppf1} is not required since these operations yield the total mass assigned to equiprobable configurations with equal array of frequencies.  Hence, the denominator in \eqref{eq:proof1} can be analytically evaluated via
\begin{equation}\label{eq:proof2}
\begin{split}
&p(\bz^{-v},\bw^{-v})\\
&=\Phi_{H^{-v},0}^{(|\bell^{-v}|)}(\ell^{-v}_{\cdot 1},\dots,\ell^{-v}_{\cdot H^{-v}})
    \Phi_{\ell^{-v}_{j\cdot},\,j}^{(V_{j}-1)}(\bm{q}^{-v}_{j1},\dots,\bm{q}^{-v}_{jH^{-v}})\prod\nolimits_{j'=1,j'\neq j}^d\Phi_{\ell^{-v}_{j'\cdot},\,j'}^{(V_{j'})}(\bm{q}^{-v}_{j'1},\dots,\bm{q}^{-v}_{j'H^{-v}}),
    \end{split}
\end{equation}
where $j$ is the layer of node $v$. 

To derive a similar expression for the numerator, it is worth analyzing separately the two situations in which $\uptau$  corresponds either to an already-existing subgroup in layer $j$ or to a new one. In the former case, for any $\uptau\in[\ell_{j\cdot}^{-v}]$ the probability in \eqref{eq:proof1} is non-zero just when $h$ corresponds to the sociability profile associated to subgroup $\uptau$. It is easy to see that for any $h$, the set of such subgroup~indices~is~given~by~$\mathbbm{T}^{-v}_{jh}$ defined in \eqref{eq:taus} of the main article. Notice that $|\mathbbm{T}^{-v}_{jh}|=\ell_{jh}^{-v}$. Now, if  $\bm{\uptau}^{-v}_{jh}$ denotes the vector obtained ordering $\mathbbm{T}^{-v}_{jh}$, the subgroup corresponding to $\uptau^{-v}_{jht}$ is the $t$-th subgroup with sociability profile $h$ in layer $j$. Hence, to compute the numerator in \eqref{eq:proof1} for $\uptau=\uptau^{-v}_{jht}$, it suffices to increase the frequency $q^{-v}_{jht}$ by one, whereas $\bell^{-v}$ remains unchanged. As a result
\begin{equation}
\begin{split}
\label{eq:proof3}
&p(z_v=h,w_v=\uptau^{-v}_{jht} , \bz^{-v},\bw^{-v})=\Phi_{H^{-v},0}^{(|\bell^{-v}|)}(\ell^{-v}_{\cdot 1},\dots,\ell^{-v}_{\cdot H^{-v}})\\
&\qquad \quad \quad \times \Phi_{\ell_{j\cdot}^{-v},\,j}^{(V_j)}(\bm{q}_{j1}^{-v},\dots,\bm{q}^{-v}_{jh}+\e_t,\dots,\bm{q}^{-v}_{jH^{-v}})
{\prod\nolimits_{j'=1,j'\neq j}^d}\Phi_{\ell^{-v}_{j'\cdot},\,j'}^{(V_{j'})}(\bm{q}^{-v}_{j'1},\dots,\bm{q}^{-v}_{j'H^{-v}}),
\end{split}
\end{equation}
where $\e_t$ is a $\ell_{jh}^{-v}$-dimensional vector of $0$s, with a $1$ in the $t$-th entry. 

When, instead, $\uptau$ is a new subgroup in layer $j$, i.e., $\uptau=\ell_{j\cdot}^{-v}+1$, then node $v$ might be assigned~either a sociability profile already present in the network or, alternatively, a previously-unseen one, i.e., $h=H^{-v}+1$. This requires adding a new entry to $\q_{jh}^{-v}$ and increasing $\bell^{-v}$, for the creation of the new subgroup. Note that, when $h=H^{-v}+1$ then $(\q^{-v}_{jh},1)=1$.  Hence, in this case we have
\begin{equation}
\begin{split}
\label{eq:proof4}
&p(z_v=h,w_v=\ell_{j\cdot}^{-v}+1, \bz^{-v},\bw^{-v})=\Phi_{H_h^{-v},0}^{(|\bell^{-v}|+1)}(\ell_{\cdot 1}^{-v},\dots,\ell^{-v}_{\cdot h}+1,\dots,\ell^{-v}_{\cdot H_h^{-v}})\\
&\qquad \quad \quad	\times \Phi_{\ell_{j\cdot}^{-v}+1,j}^{(V_j)}(\q_{j1}^{-v},\dots,(\q^{-v}_{jh},1),\dots,\q^{-v}_{jH_h^{-v}})\prod\nolimits_{j'=1,j'\neq j}^d\Phi_{\ell^{-v}_{j'\cdot},\,j'}^{(V_{j'})}(\bm{q}^{-v}_{j'1},\dots,\bm{q}^{-v}_{j'H^{-v}}), 
\end{split}
\end{equation}
Replacing \eqref{eq:proof2}--\eqref{eq:proof4} in \eqref{eq:proof1} yields the expressions in \eqref{eq:jointpred1}--\eqref{eq:jointpred2} of the main article.
\end{proof}

\vspace{3pt}
\begin{proof}[{\bf Proof of Corollary \ref{exe:jpred_hdp}}]
Recall that the EPPF induced by a DP with concentration parameter $\bar{\theta}>0$ implies that the probability of any partition of $n$ objects in $K$ clusters is given by
\begin{equation}\label{eq:eppf_dp}
\Phi^{(n)}_{K}(n_1,\,\ldots,n_K)=\frac{\bar{\theta}^{K}}{[\,\bar{\theta}\,]_{n}}\prod\nolimits_{\kappa=1}^{K}(n_\kappa-1)!
\end{equation}
for any frequency vector $(n_1,\ldots,n_K)$ such that $n=\sum_{\kappa=1}^Kn_\kappa$, where $[\,\cdot\,]_n$ is the $n$-th ascending factorial.

When $\uptau\in\mathbbm{T}^{-v}_{jh}$, it suffices to substitute \eqref{eq:eppf_dp} in \eqref{eq:jointpred1} to get the first summand in \eqref{eq:jpred_hdp}. As for the case  $\uptau=\ell^{-v}_{j\cdot}+1$, if $h\in[H^{-v}]$ then $H^{-v}_h=H^{-v}$ and we have
\begin{equation}\label{eq:nonrichiesta1}
\frac{\Phi_{H^{-v},0}^{(|\bell^{-v}|+1)}(\ell_{\cdot 1}^{-v},\dots,\ell^{-v}_{\cdot h}+1,\dots,\ell^{-v}_{\cdot H^{-v}})}{\Phi_{H^{-v},0}^{(|\bell^{-v}|)}(\ell^{-v}_{\cdot 1},\dots,\ell^{-v}_{\cdot H^{-v}})}=\frac{\ell^{-v}_{\cdot h}}{\theta_0+|\bm{\ell}^{-v}|},
\end{equation}
while if $h=H^{-v}+1$, since $H^{-v}_h=H^{-v}+1$,
\begin{equation}\label{eq:nonrichiesta2}
\frac{\Phi_{H^{-v}+1,0}^{(|\bell^{-v}|+1)}(\ell_{\cdot 1}^{-v},\dots,\ell^{-v}_{\cdot H^{-v}},1)}{\Phi_{H^{-v},0}^{(|\bell^{-v}|)}(\ell^{-v}_{\cdot 1},\dots,\ell^{-v}_{\cdot H^{-v}})}=\frac{\theta_0}{\theta_0+|\bm{\ell}^{-v}|}.
\end{equation}
In both cases the second ratio in \eqref{eq:jointpred2} gives the common factor of the second summand in \eqref{eq:jpred_hdp}.
\end{proof}

\vspace{3pt}

\begin{proof}[{\bf Proof of Corollary \ref{cor:jointpred_nsp}}]
The EPPF induced by a NSP with parameter $\bar{\sigma}\in(0,1)$ gives
\begin{equation}\label{eq:eppf_nsp}
\Phi^{(n)}_{K}(n_1,\,\ldots,n_K)=\frac{(K-1)!\bar{\sigma}^{K-1}}{(n-1)!}\prod\nolimits_{\kappa=1}^{K}[\,1-\bar{\sigma}\,]_{n_\kappa-1},
\end{equation}
Substituting \eqref{eq:eppf_nsp} in \eqref{eq:jointpred1}--\eqref{eq:jointpred2}, with considerations as in proof of Corollary \ref{exe:jpred_hdp}, yields \eqref{eq:jpred_hnsp}.
\end{proof}
\vspace{3pt}

\noindent We now provide proofs of the remaining  results, in order of appearance.

\vspace{10pt}

\begin{proof}[{\bf Proof of Proposition \ref{prp:pred_hnrmi}}]
It suffices to note that 
\begin{equation}
\mathbb{P}(z_v=h \mid \bz^{-v}, \bw^{-v})=\sum\nolimits_{\uptau=1}^{\ell_{j\cdot}^{-v}+1} \mathbb{P}(z_v=h, w_v=\uptau \mid \bz^{-v}, \bw^{-v}),
\end{equation}
where the expressions of $\mathbb{P}(z_v=h, w_v=\uptau \mid \bz^{-v}, \bw^{-v})$ for any choice of $(h,\uptau)\in[H^{-v}+1]\times[\ell_{j\cdot}^{-v}+1]$ are given in Proposition \ref{prp:nuova}.
\end{proof}

\vspace{2pt}

\begin{proof}[{\bf Proof of Corollary \ref{ex1}}]
Equation \eqref{eq:pred_dir} in Corollary \ref{ex1} can be obtained from \eqref{eq:pred} by replacing the general $\Phi^{(\cdot)}_{\cdot}$ in \eqref{eq:pred1} with those specific to the DP case, given in \eqref{eq:eppf_dp}, or alternatively, by summing the H-DP joint conditional probabilities in \eqref{eq:jpred_hdp} over all the possible choices of $\uptau\in[\ell^{-v}_{j\cdot}+1]$.
\end{proof}

\vspace{2pt}

\begin{proof}[{\bf Proof of Theorem \ref{coclust_hnrmi}}]
 We can write 
 \begin{eqnarray}\label{eq:pproof}
 \begin{split}
&\mathbb{P}(\{z_v=z_u\},\,w_v=\uptau,\,w_u=\uptau'\mid \bz^{-vu},\bw^{-vu})
=\frac{p(\{z_v=z_u\},\,w_v=\uptau,\,w_u=\uptau', \bz^{-vu},\bw^{-vu})}{p(\bz^{-vu},\bw^{-vu})}\\
&\quad \quad \qquad \qquad  \qquad \qquad \qquad= \frac{\sum_{h=1}^{H^{-vu}+1}p(z_v=h,z_u=h,\,w_v=\uptau,\,w_u=\uptau',\, \bz^{-vu},\,\bw^{-vu})}{p(\bz^{-vu},\bw^{-vu})}.
\end{split}
 \end{eqnarray}
Now, if $v$ and $u$ belong to the same layer $j$, the probability within the summation at the numerator will include, for any $h,\uptau$ and $\uptau'$ the common factor 
\begin{eqnarray*}
 \prod\nolimits_{j'=1,j'\neq j}^d\Phi_{\ell^{-vu}_{j'\cdot},\,j'}^{(V_{j'})}(\bm{q}^{-vu}_{j'1},\dots,\bm{q}^{-vu}_{j'H^{-vu}}),
 \end{eqnarray*}
  given by the allocations in all the other layers. 
 
 The form of the remaining factor depends, instead, on $\uptau$ and $\uptau'$. If both $\uptau$ and $\uptau'$ are already-existing subgroups, then $\uptau= \uptau^{-vu}_{jht}$ and $\uptau'= \uptau^{-vu}_{jht'}$ for some (possibly equal) $t,t'\in[\ell^{-vu}_{jh}]$, where the vector $\bm{\uptau}^{-vu}_{jh}$ is defined as in Proposition \ref{prp:nuova}. In this case, the multiplicative factor for a generic $t,t'\in[\ell^{-vu}_{jh}]$ is equal~to 
\begin{eqnarray*}\Phi_{H^{-vu},0}^{(|\bell^{-vu}|)}(\ell^{-vu}_{\cdot 1},\dots,\ell^{-vu}_{\cdot H^{-vu}})\Phi_{\ell^{-vu}_{j\cdot},\,j}^{(V_j)}(\q_{j1}^{-vu},\dots,\q_{jh}^{-vu}+\e_t+\e_{t'},\dots,\q_{jH^{-vu}}^{-vu}).
\end{eqnarray*}
If, instead, $\uptau$ is already occupied and $\uptau'$ is new, that is $\uptau= \uptau^{-vu}_{jht}$ for some $t\in[\ell^{-vu}_{jh}]$ and $\uptau'=\ell^{-vu}_{j\cdot}+1$, then the factor is 
\begin{eqnarray*}
\Phi_{H^{-vu},0}^{(|\bell^{-vu}|+1)}(\ell^{-vu}_{\cdot 1},\dots,\ell_{\cdot h}^{-vu}+1,\dots,\ell^{-vu}_{\cdot H^{-vu}})\Phi_{\ell^{-vu}_{j\cdot}+1,\,j}^{(V_j)}(\q_{j1}^{-vu},\dots,(\q_{jh}^{-vu}+\e_t,1),\dots,\q_{jH^{-vu}}^{-vu}).
\end{eqnarray*}
 \setlength\abovedisplayskip{8pt}%
    \setlength\belowdisplayskip{8pt}%
When both $\uptau$ and $\uptau'$ correspond to the same new subgroup, i.e., $\uptau=\uptau'=\ell^{-vu}_{j\cdot}+1$, we have 
\begin{eqnarray*}
\Phi_{H_h^{-vu},0}^{(|\bell^{-vu}|+1)}(\ell^{-vu}_{\cdot 1},\dots,\ell_{\cdot h}^{-vu}+1,\dots,\ell^{-vu}_{\cdot H_h^{-vu}})\Phi_{\ell^{-vu}_{j\cdot}+1,\,j}^{(V_j)}(\q_{j1}^{-vu},\dots,(\q_{jh}^{-vu},2),\dots,\q_{jH_h^{-vu}}^{-vu}),
\end{eqnarray*}
 while if both are new but different, i.e., $\uptau=\ell^{-vu}_{j\cdot}+1,\,\uptau'=\ell^{-vu}_{j\cdot}+2$ (or vice versa), the multiplicative~factor is equal to
 \begin{eqnarray*}
 \Phi_{H_h^{-vu},0}^{(|\bell^{-vu}|+2)}(\ell^{-vu}_{\cdot 1},\dots,\ell_{\cdot h}^{-vu}+2,\dots,\ell^{-vu}_{\cdot H_h^{-vu}})\Phi_{\ell^{-vu}_{j\cdot}+2,\,j}^{(V_j)}(\q_{j1}^{-vu},\dots,(\q_{jh}^{-vu},1,1),\dots,\q_{jH_h^{-vu}}^{-vu}).
 \end{eqnarray*}
As in \eqref{eq:proof2}, it is also easy to check that the denominator in \eqref{eq:pproof} is equal to 
\begin{eqnarray*}
\Phi_{H^{-vu},0}^{(|\bell^{-vu}|)}(\ell^{-vu}_{\cdot 1},\dots,\ell^{-vu}_{\cdot H^{-vu}})\Phi_{\ell^{-vu}_{j\cdot},\,j}^{(V_j-2)}(\q_{j1}^{-vu},\dots,\q_{jH^{-vu}}^{-vu})\prod\nolimits_{j'=1, j'\neq j}^d\Phi_{\ell^{-vu}_{j'\cdot},\,j'}^{(V_{j'})}(\bm{q}^{-vu}_{j'1},\dots,\bm{q}^{-vu}_{j'H^{-vu}}).
\end{eqnarray*}
Replacing the above factors within the ratio in \eqref{eq:pproof} and summing over all the possible choices of $\uptau$ and $\uptau'$, yields expressions \eqref{eq:monster1}--\eqref{eq:monster11}. The proof of \eqref{eq:monster2}--\eqref{eq:monster22} in Theorem \ref{coclust_hnrmi} follows along the same line of reasoning. It suffices to note that when $v$ and $u$ are not in the same layer it is not possible for $v$ and $u$ to be allocated to the same subgroup, neither new nor already-occupied.
\end{proof}

\vspace{10pt}

\begin{proof}[{\bf Proof of Corollary \ref{cor:coclust_hdp}}]
Similarly to the proof of Corollary \ref{ex1}, \eqref{eq:coclust_hdp1} and \eqref{eq:coclust_hdp2} in Corollary \ref{cor:coclust_hdp} can be derived from \eqref{eq:monster1} and \eqref{eq:monster2}, respectively, by replacing the general functions $\Phi^{(\cdot)}_{\cdot}$ in \eqref{eq:monster11} and \eqref{eq:monster22} with those specific to the DP case in \eqref{eq:eppf_dp}.
\end{proof}

\vspace{10pt}

\begin{proof}[{\bf Proof of Proposition \ref{prp:cond_dir}}]
Summing the predictive probabilities in \eqref{eq:pred_dir} over $h\in \bH_j^{-v}$ and $h\in\bH^{-v}\setminus \bH_j^{-v}$, we have that the inequality in the right hand side of \eqref{eq:elicit1} is satisfied if and only if

\begin{equation}\label{prooff1}
\frac{|\bell^{-v}|p_v}{\theta_0+|\bell^{-v}|}\frac{\theta}{\theta+V_j-1}+\frac{V_j-1}{\theta+V_j-1}\geq\frac{|\bell^{-v}|(1-p_v)}{\theta_0+|\bell^{-v}|}\frac{\theta}{\theta+V_j-1},
\end{equation}
where $p_v=(1/|\bm{\ell}^{-v}|)\sum_{h\in \bH_j^{-v}}\ell_{\cdot h}^{-v}$. This is equivalent to
\begin{equation}\label{eq:prooff}
p_v\geq\frac{1}{2}\left(1-\frac{|\bell^{-v}|+\theta_0}{|\bell^{-v}|}\frac{V_j-1}{\theta}\right).
\end{equation}
Therefore, whenever the right hand side of \eqref{eq:prooff} is negative, regardless of the value of the proportion $p_v$ of subgroups with sociability profiles already observed in layer $j$, the inequality is always satisfied. This is implied by the left hand side of \eqref{eq:elicit1}. 

Summing to the right hand side of \eqref{prooff1} the probability of node $v$ being assigned a new sociability profile, i.e., 
\begin{equation}
\frac{\theta_0}{\theta_0+|\bell^{-v}|}\frac{\theta}{\theta+V_j-1}
\end{equation}
we obtain the inequality in the right hand side of \eqref{eq:elicit2}. With the previous strategy we can again retrieve the sufficient condition.
\end{proof}

\vspace{6pt}

\begin{proof}[{\bf Proof of Proposition \ref{prp:pelic_nsp}}]
    Summing the predictive probabilities in \eqref{eq:pred_nsp} over $h\in \bH_j^{-v}$ and $h\notin \bH_j^{-v}$, we have that the inequality in the right hand side of \eqref{eq:nspelicit1} is satisfied if and only if
\begin{equation}\label{proofff1}
p_v - \frac{|\bH^{-v}_j|\sigma_0}{|\bell^{-v}|}+\frac{V_j-1}{\sigma\ell^{-v}_{j\cdot}}-1\geq(1-p_v)+\frac{|\bH^{-v}_j|\sigma_0}{|\bell^{-v}|},
\end{equation}
where $p_v=(1/|\bm{\ell}^{-v}|)\sum_{h\in \bH_j^{-v}}\ell_{\cdot h}^{-v}$. This is equivalent to
\begin{equation}\label{eq:proofff}
1-p_v\leq\frac{V_j-1}{2\sigma\ell_{j\cdot}^{-v}}-\frac{|\bH_j^{-v}|\sigma_0}{|\bell^{-v}|}.
\end{equation}
Therefore, whenever the right hand side of \eqref{eq:proofff} is greater than $1$, regardless of the value of the proportion $p_v$ of subgroups with sociability profiles already observed in layer $j$, the inequality is always satisfied. This is implied by the left hand side of \eqref{eq:nspelicit1}. In particular, since $|\bH^{-v}_j|\leq V_j-1$, $|\bell|\geq d$, and $\ell_{j\cdot}^{-v}\leq V_j-1$, we have that \eqref{eq:proofff} is implied by $\sigma\leq 0.5 (1+(V_j-1)\sigma_0/d)^{-1}$.
\end{proof}

\vspace{4pt}

\begin{proof}[{\bf Proof of Corollary \ref{cor:suff}}]
Corollary \ref{cor:suff} can be directly deduced from \eqref{eq:jointpred1}--\eqref{eq:jointpred2}.
\end{proof}

\vspace{4pt}

\begin{proof}[{\bf Proof of Corollary \ref{cor:pred_nsp}}]
Equation \eqref{eq:pred_nsp} in Corollary \ref{cor:pred_nsp} can be also obtained from \eqref{eq:pred} by replacing the general functions $\Phi^{(\cdot)}_{\cdot}$ in \eqref{eq:pred1} with those specific to the NSP, given in \eqref{eq:eppf_nsp}, or alternatively,~by~summing the H-NSP joint conditional probabilities in \eqref{eq:jpred_hnsp} over all possible choices of $\uptau\in[\ell^{-v}_{j\cdot}+1]$.
 \end{proof}

\vspace{4pt}

\begin{proof}[{\bf Proof of Corollary \ref{cor:coclust_nsp}}]
Similarly to the proof of Corollary \ref{cor:pred_nsp}, \eqref{eq:coclust_nsp1} and \eqref{eq:coclust_nsp2} in Corollary \ref{cor:coclust_nsp} can be derived from \eqref{eq:monster1} and \eqref{eq:monster2}, respectively, by replacing the general functions $\Phi^{(\cdot)}_{\cdot}$ in \eqref{eq:monster11} and \eqref{eq:monster22} with those specific to the NSP case in \eqref{eq:eppf_nsp}.
\end{proof}

\vspace{5pt}

\section{Additional Methodological and Empirical Results}\label{sec_s_3}
\vspace{-4pt}
We provide below details on $\text{\normalfont pExP}(\bV;\,\theta,\,\theta_0)$ with hypepriors on $\theta$ and $\theta_0$, along with additional  empirical results which complement the analyses in Sections~\ref{sec_4} and \ref{sec_5} of the main article. 
\vspace{-3pt}

\subsection{ pExP$(\bV;\,\theta,\,\theta_0)$ with hypepriors on $\theta$ and $\theta_0$}
\vspace{-2pt}
As discussed in Section~\ref{sec_3.1} of the main article it is possible to include gamma hyperpriors for the parameters $\theta$ and $\theta_0$ of $\text{\normalfont pExP}(\bV;\,\theta,\,\theta_0)$ through a tractable construction that yields closed-form conjugate full conditionals. This allows for a direct modification of   Algorithm~\ref{alg:gibbs} that includes an additional step to sample from these full-conditionals of the  $\text{\normalfont pExP}(\bV;\,\theta,\,\theta_0)$  parameters. More specifically, extending and adapting the results of \citet{escobar1995bayesian} from the DP to the H-DP setting within our~formulation, we have that, when $\theta \sim \mbox{gamma}(\alpha,\beta)$ and $\theta_0 \sim \mbox{gamma}(\alpha_0,\beta_0)$, the two full-conditionals for $\theta$ and $\theta_0$~are 
\begin{eqnarray*}
p(\theta_0\mid\bz,\bw,\bY,\theta)\propto \frac{\theta_0^{H}}{[\theta_0]_{|\bell|}}\times p(\theta_0) \qquad \mbox{and} \qquad p(\theta\mid\bz,\bw,\bY,\theta_0)\propto \prod_{j=1}^d\frac{\theta^{\ell_{j\cdot}}}{[\theta]_{V_j}}\times p(\theta),
\end{eqnarray*}
where $p(\theta_0)$ and $p(\theta)$ are the densities of the gamma hypepriors for $\theta_0$ and $\theta$, respectively, whereas $[\theta_0]_{|\bell|}$ and $[\theta]_{V_j}$ denote the two ascending factorials that can be also expressed as $\Gamma(\theta_0+|\bell|)/\Gamma(\theta_0)$ and $\Gamma(\theta+V_j)/\Gamma(\theta)$. Thus, $1/[\theta_0]_{|\bell|} \propto \mbox{B}(\theta_0, |\bell|)=\smash{ \int_0^1 \eta^{\theta_0-1}_0(1-\eta_0)^{|\bell|-1} d \eta_0}$ and $1/[\theta]_{V_j} \propto \mbox{B}(\theta, V_j)= \smash{\int_0^1 \eta_j^{\theta-1}(1-\eta_j)^{V_j-1} d \eta_j}$, for every layer $j=1, \ldots, d$. Therefore, introducing augmented variables $\eta_0$ and $\eta_1,\dots,\eta_d$ such that
\begin{equation}
(\eta_0\mid\theta_0,\bz,\bw)\sim \text{beta}(\theta_0,|\bell|) \qquad \mbox{and}  \qquad  (\eta_j\mid\theta)\overset{\mbox{\footnotesize ind}}{\sim} \text{beta}(\theta,V_j), \ j=1, \ldots, d,
\label{eta_post_sampl}
\end{equation}

\noindent and defining $\nu_0:= \log (\eta_0) \leq 0$ and $\nu:=\sum_{j=1}^d\log(\eta_j) \leq 0$, we have
\begin{equation}
(\theta_0\mid\eta_0,\bz,\bw)\sim \mbox{gamma}(\alpha_0+H,\beta_0-\nu_0)  \quad \mbox{and}  \quad  (\theta\mid\eta_1,\dots,\eta_d,\bz,\bw)\sim\mbox{gamma}(\alpha+|\bell|,\beta-\nu).
\label{theta_ga_post_sampl}
\end{equation}
Hence, the inclusion of an additional step in Algorithm~\ref{alg:gibbs} sampling first from \eqref{eta_post_sampl} and then from \eqref{theta_ga_post_sampl} allows for tractable posterior computation also when including hyperpriors in $\text{\normalfont pExP}(\bV;\,\theta,\,\theta_0)$.

\vspace{10pt}
\begin{figure}[t!]
\centering
    \includegraphics[trim=0cm 0cm 0cm 0cm,clip,width=1\textwidth]{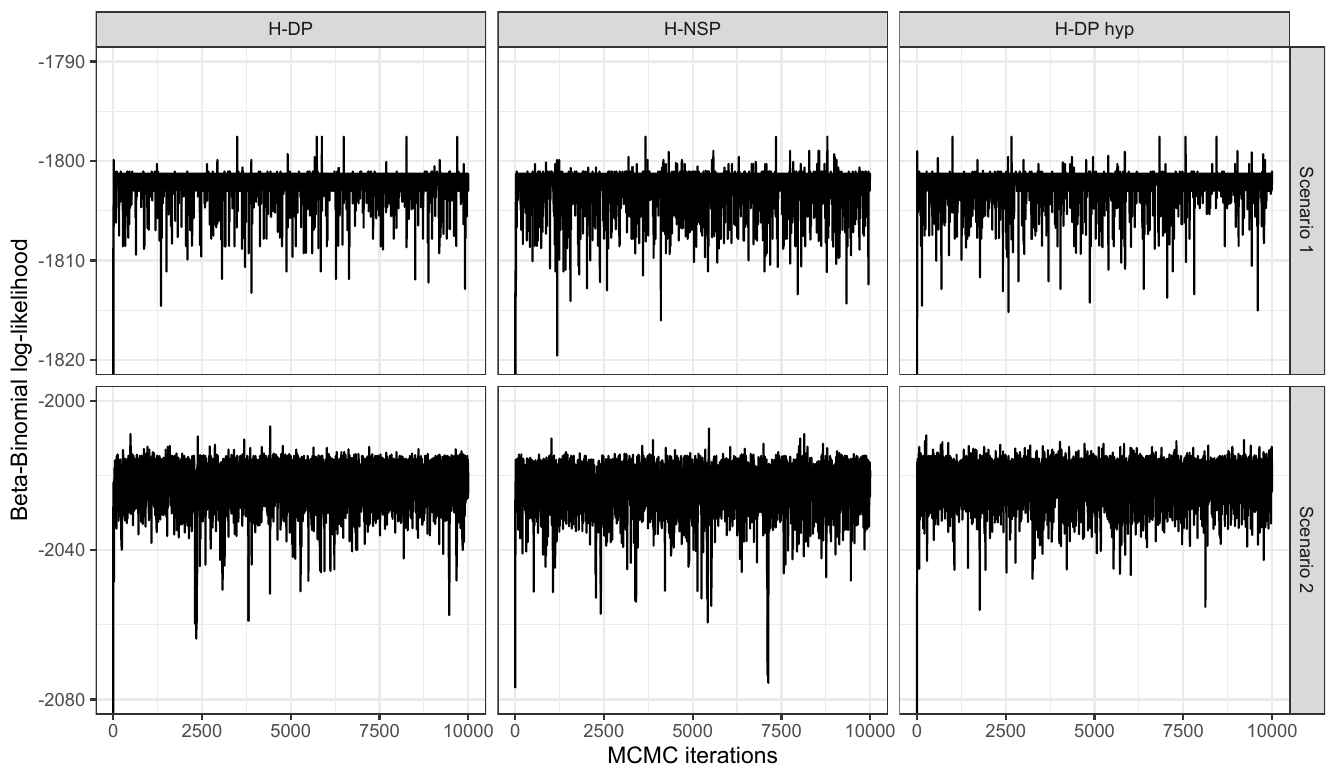}
      \vspace{-15pt}
    \caption{\footnotesize{Traceplots of $\log p(\bY \mid \bz^{(s)})$, $s=1, \ldots, n_{\mbox{\tiny iter}}$ for H-DP, H-NSP, and H-DP with hyperpriors on $\theta$ and $\theta_0$, in one of the ten replicated experiments considered in Section~\ref{sec_4}, under both Scenario 1 and 2.}}
    \label{figure:1_suppl}
\end{figure}

\subsection{Section~\ref{sec_4} (additional empirical results) }
As discussed within Section~\ref{sec_4}, convergence of the Gibbs sampling schemes for  $\mbox{H-DP}(\theta=0.5, \theta_0=4)$, $\mbox{H-NSP}(\sigma=0.2,\sigma_0=0.8)$, and H-DP with $\mbox{gamma}(5,10)$ and $\mbox{gamma}(12,3)$ hyperpriors for $\theta$ and $\theta_0$ is assessed from the analysis of the traceplots  for the logarithm of the likelihood in \eqref{eq1}. Figure~\ref{figure:1_suppl} provides a graphical representation of these traceplots for one of the  ten replicated experiments considered within Section~\ref{sec_4}, under both Scenario 1 and 2. The results in Figure~\ref{figure:1_suppl} suggest rapid convergence~and~adequate mixing in both Scenario 1 and 2, for the three pEx-SBM examples under analysis. The higher variability of the traceplots for Scenario 2 correctly unveils the increased posterior uncertainty on $\bz$ arising from such a more challenging scenario with reduced separation among the different groups.

\begin{table*}[b]
	\renewcommand{\arraystretch}{1}
	\centering
	\caption{\footnotesize{Robustness assessments for three key examples of pEx-SBMs in Scenarios~1--2: Posterior~mean~$\mathbbm{E}[\mbox{VI}(\bz,\bz_0) \mid \bY]$ of the VI distance from the true $\bz_0$, when changing the initialization of the collapsed Gibbs samplers and the~hyperparameter values w.r.t.~the default settings in Section~\ref{sec_4}. Values within brackets are the posterior means $\mathbbm{E}[\mbox{VI}(\bz,\bz_0) \mid \bY]$  obtained under~the default settings in the main article.  Results are averaged over ten replicated experiments. }}
	\vspace{5pt}
	\label{table_runtimes_MC_supp}
	\begin{adjustbox}{width=0.9\textwidth,center=\textwidth}
	\setlength{\tabcolsep}{5pt}
		\begin{tabular}{l@{\hskip 0.5in}c@{\hskip 0.5in}c@{\hskip 1in}cc}
			& \multicolumn{2}{l}{initialization $\mathbbm{E}[\mbox{VI}(\bz,\bz_0) \mid \bY]$}  &  \multicolumn{2}{l}{hyperparameters $\mathbbm{E}[\mbox{VI}(\bz,\bz_0) \mid \bY]$}    \\ 
			\midrule
			\textsc{Scenario} & 1 & 2 & 1 & 2 \\ 
			\midrule
			pEx-SBM (H-DP) \quad &  0.03 (0.02)   &   0.62 (0.62) & 0.07 (0.02)   &   0.67 (0.62) \\
			pEx-SBM (H-NSP) &  0.04 (0.04) &   0.61 (0.60) &    0.07 (0.04) &   0.61 (0.60)\\
			pEx-SBM (H-DP hyp) $\qquad$   &   0.03 (0.03) &   0.62 (0.63) &   0.05 (0.03) &   0.63 (0.63)   \\
			\hline
		\end{tabular}
	\end{adjustbox}
\end{table*}

Table~\ref{table_runtimes_MC_supp} quantifies the robustness of the three  pEx-SBMs examples analyzed in Section~\ref{sec_4}, with a focus on the initialization of the Gibbs sampling routine in Section~\ref{sec_3.1} and hyperparameters specification. Recalling Algorithm~\ref{alg:gibbs}, as a default setting for the initialization of $\bz$ we consider $V$ different groups, each comprising a single node. Table~\ref{table_runtimes_MC_supp} clarifies that the inferred posterior concentration~around~$\bz_0$ in replicated studies does not change when initializing Algorithm~\ref{alg:gibbs} to the other extreme setting characterized by a single group containing all the $V$ nodes in the network. Similar robustness is observed~also~when varying the hyperparameters of the three priors under analysis. Within Table~\ref{table_runtimes_MC_supp} we assess,~in~particular, an $\mbox{H-DP}(\theta=2, \theta_0=7)$, an $\mbox{H-NSP}(\sigma=0.35,\sigma_0=0.9)$, and an H-DP with $\mbox{gamma}(12,6)$ and $\mbox{gamma}(14,2)$ hyperpriors for $\theta$ and $\theta_0$, respectively. Although these alternative hyperparameter settings~induce~a~prior expected number of groups of $\approx 10$ (i.e., twice the one obtained under the prior specifications in Section~\ref{sec_4}), as shown in  Table~\ref{table_runtimes_MC_supp} the overall posterior concentration~around $\bz_0$ in replicated studies remains almost the same as the one achieved under the original settings,~with~H-DP~based on hyperpriors showcasing the improved robustness. 

The slightly improved robustness of H-DP with hypeprior can be explained by its ability to learn the parameters that control the clustering properties and the strength of the layer information in informing inference on $\bz$. Such an advantage is also observed when studying the performance of the three~pEx-SBMs examples in Scenario 1 under a non-informative layer division obtained by considering a random permutation of the original layer labels in Section~\ref{sec_4} (see also Figure~\ref{figure:2}). By performing~posterior~inference in this context under the default settings from Section~\ref{sec_4} for the three  pEx-SBMs examples yields a $\mbox{VI}(\hat{\bz},\bz_0)$ distance, averaged over ten replicated studies, of $0.16$ for H-DP and H-NSP, while H-DP with hyperprior achieves a slightly improved robustness with a value of $0.14$.~As~expected, when layers are non-informative, performance slightly deteriorates relative to the results displayed in Table~\ref{table_runtimes_MC} within the main article. Nonetheless, a closer inspection of Table~\ref{table_runtimes_MC} shows that, even in this  challenging setting, pEx-SBMs still display higher accuracy than the one achieved by  state-of-the-art  and routinely-implemented methods (e.g., Louvain \citep{blondel2008fast}, JCDC \citep{zhang2016community} and CASC \citep{binkiewicz2017covariate}) supervised by the original informative layer division. Considering non-informative layers for CASC yields a $\mbox{VI}(\hat{\bz},\bz_0)$ distance, averaged over ten replicated studies, of $2.93$, which is orders of magnitude higher than the one achieved under the three  pEx-SBMs. This suggest that  pEx-SBM is less sensitive to non-informative layers than state-of-the-art   methods.

\begin{figure}[t]
\centering
    \includegraphics[trim=0cm 0cm 0cm 0cm,clip,width=0.42\textwidth]{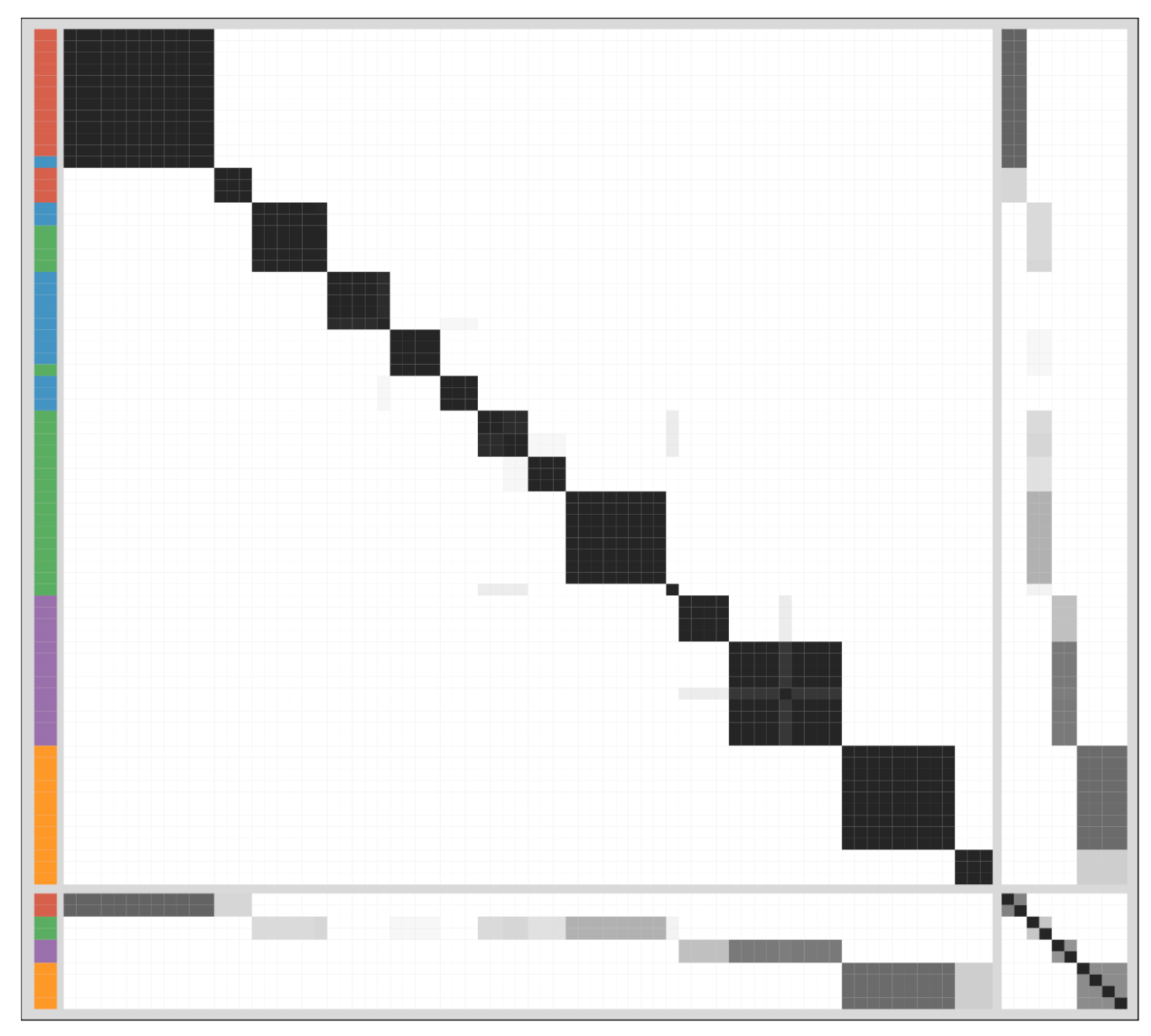}
    \caption{\footnotesize{For the criminal network application in Section~\ref{sec_5}, graphical representation of the posterior similarity~matrix comprising in-sample and predictive co-clustering probabilities. The two gray lines separate~in-sample nodes (larger group) and out-of-sample ones (smaller group). Side colors correspond to \textit{locale} affiliation (i.e., layers), whereas~the~color of~each entry in the matrix ranges from white to black as the estimated co-clustering probability goes from 0 to 1.}}
    \label{figure:2_suppl}
          \vspace{-5pt}
\end{figure}

\vspace{5pt}
\subsection{Section~\ref{sec_5} (additional empirical results) }
Figure~\ref{figure:2_suppl} complements the predictive studies on the criminal network in Section~\ref{sec_5} of the main~article with  a focus on the estimated posterior similarity matrix~comprising both in-sample and predictive co-clustering probabilities. The results nicely illustrate one of the important advantages of the proposed pEx-SBM. Namely its ability to properly quantify both in-sample and predictive clustering uncertainty. This is evident in the higher co-clustering uncertainty for pairs of nodes involving at least an out-of-sample one, whereas those comprising both in-sample nodes display, as expected, a lower variability. Interestingly, the higher heterogeneity is observed for the co-clustering probabilities involving out-of-sample nodes from the green \textit{locale}. As discussed within Section~\ref{sec_5} (see also Figure~\ref{figure:3}), such a \textit{locale} is the most fragmented in different groups. This structure properly translates into a higher co-clustering uncertainty in Figure~\ref{figure:2_suppl}  for out-of-sample nodes from the green \textit{locale}.

\vspace{4pt}

\section{Glossary}\label{sec_gloss}
\vspace{-5pt}
The table below provides a glossary of the main quantities involved in our adaptation of the \textit{Chinese restaurant franchise} (CRF) metaphor to the pEx-SBM construction.

\begin{table*}[h]
	\renewcommand{\arraystretch}{1}
	\centering
	\vspace{5pt}
	\begin{adjustbox}{width=0.9\textwidth,center=\textwidth}
	\setlength{\tabcolsep}{5pt}
		\begin{tabular}{ll}
			\midrule
			\textsc{Quantity} \qquad \qquad & \textsc{Description} \\ 
			\midrule
			$V$ \quad & total number of  nodes in the network \\
			$d$ \quad & total number of  layers in the network \\
			$H$ \quad & total number of sociability profiles (i.e., clusters) in the network \\
			$V_j$ \quad & number of  nodes in layer $j$ \\
			$w_{ji}$ \quad & label of the subgroup to which node $i$ in layer $j$ has been assigned \\
			$\ell_{jh}$ \quad & number of subgroups in layer $j$ with sociability profile $h$\\
			$q_{jht}$ \quad & number of nodes in layer $j$ assigned to the $t$-th subgroup with sociability profile $h$\\
			$\ell_{j\cdot}$ \quad & number of subgroups in layer $j$\\
			$\ell_{\cdot h}$ \quad & total number of subgroups with sociability profile $h$ \\ 
			$n_{jh}$ \quad & number of nodes in layer $j$ with sociability profile $h$ \\
			$\Phi_{\ell_{j\cdot},\,j}^{(V_j)}$ & EPPF regulating the distribution of the division in subgroups within layer $j$\\
			$\Phi_{H,0}^{(|\bell|)}$ &  EPPF driving the sociability profile assignment to the subgroups in the different layers\\
						\hline
		\end{tabular}
	\end{adjustbox}
\end{table*}

\end{changemargin}

\vspace{1pt}
\begingroup
\fontsize{11pt}{12.8pt}\selectfont

\endgroup

\end{document}